\documentclass[review,authoryear]{elsart}

\usepackage{graphicx}
\usepackage{epstopdf, epsfig}
\usepackage{color} 
\graphicspath{{./figures/}} 

\usepackage{amssymb}
\usepackage{amsmath}
\usepackage{natbib}
\usepackage{subfloat}
\usepackage{subcaption}
\usepackage{bm}
\usepackage{lscape}
\usepackage [autostyle, english = american]{csquotes}
\MakeOuterQuote{"}

\newsavebox{\largestimage}
\def\vec#1{\mbox{\boldmath $#1$}}
\newcommand{\bu}{\mathbf{u}}

\def\Otf{\Omega^\mathrm{f}(t)}
\newcommand{\xx}{\mbox{$\mathbf{x}$}}

\def\stress{{\vec \sigma}}
\def\div{\vec \nabla}

\usepackage{environ}
\NewEnviron{myequation}{%
	\begin{equation}
	\scalebox{0.7}{$\BODY$}
	\end{equation}
}

\def\vec#1{\mbox{\boldmath $#1$}}

\usepackage[english]{babel}
\usepackage{algorithm}

\def\dd{\partial}

\def\vec#1{\mbox{\boldmath $#1$}}

\usepackage[english]{babel}

\def\dd{\partial}
\def\bI{{\bf I}}

\def\bu{{\vec u}}
\def\force{{\vec f}}

\def\G{\Gamma}

\def\stress{{\vec \sigma}}
\def\Otf{\Omega^\mathrm{f}(t)}

\def\Otnm1f{\Omega^\mathrm{f}(t^{\mathrm{n}-1})}
\def\Otnm12f{\Omega^\mathrm{f}(t^{\mathrm{n}-\frac{1}{2}})}

\def\Otnm1s{\Omega^\mathrm{s}(t^{\mathrm{n}-1})}
\def\Otnm12s{\Omega^\mathrm{s}(t^{\mathrm{n}-\frac{1}{2}})}

\def\div{\vec \nabla}

\def\testf{\vec \phi^\mathrm{f}}
\def\tests{\vec \phi^\mathrm{s}}

\def\dO{\mathrm{d}{\Omega}}
\def\dG{\mathrm{d}{\Gamma}}

\def\vphi{\vec{\varphi}^\mathrm{s}}
\def\uf{{\vec u}^\mathrm{f}}
\def\us{{\vec u}^\mathrm{s}}
\def\usn{{\vec u}^\mathrm{s,n}}
\def\usnn{{\vec u}^\mathrm{s,n-1}}
\def\usnnn{{\vec u}^\mathrm{s,n-2}}

\def\vphi{\vec{\varphi}^\mathrm{s}}
\def\vphin{\vec{\varphi}^\mathrm{s,n}}

\def\grad{\bm \nabla}
\def\Os{\Omega^\mathrm{s}}
\def\Of{\Omega^\mathrm{f}}

\newcommand{\f}{\mathrm{f}}
\newcommand{\h}{\mathrm{h}}

\renewcommand{\i}{\mathrm{i}}
\renewcommand{\j}{\mathrm{j}}

\newcommand{\n}{\mathrm{n}}

\newcommand{\ww}{\mbox{\boldmath $w$}}
\newcommand{\zz}{\mbox{$\mathbf{z}$}}

\newcommand{\TT}{\mathcal{T}}

\def\ufn{{\vec u}^\mathrm{f,n}}
\def\ufnn{{\vec u}^\mathrm{f,n-1}}
\def\ufnnn{{\vec u}^\mathrm{f,n-2}}

\def\us{{\vec u}^\mathrm{s}}

\def\usn{{\vec u}^\mathrm{s,n}}
\def\usnn{{\vec u}^\mathrm{s,n-1}}
\def\usnnn{{\vec u}^\mathrm{s,n-2}}

\def\cuf{\check{{\vec u}}^\mathrm{f}}

\begin{document}
	\journal{Journal of Fluids and Structures}	
	
	\begin{frontmatter}
		
		\title{
			Numerical Study on Flapping Dynamics of a Flexible Two-Layered Plate in a Uniform Flow}
		
		\author{Aditya Karthik S.},
		\author{Supradeepan K.} and
		\author{P. S. Gurugubelli\corauthref{cor}}
		
		\address{Computing Lab, Department of Mechanical Engineering, \\Birla Institute of Technology and Science - Pilani, Hyderabad, India.}
		\corauth[cor]{Corresponding author.; phone: +91 040 6630 3618.\\
			\textit{E-mail address:} pardhasg@hyderabad.bits-pilani.ac.in (P.S. Gurugubelli).}
		
		\begin{abstract}
			Over the past few decades, the prospect of energy generation from an oscillating piezoelectric patch has gained attention. A typical setup of this kind would be a piezoelectric patch mounted on a flexible flat plate that is exhibiting self-sustained flapping motion. Such piezoelectric patches are generally multilayered consisting of piezoelectric, substrate and electrode layers placed on top of each other. Although the flapping dynamics of single-layered structures have been extensively studied, the investigation into the flapping dynamics of multilayered structures is minimal. In this paper, we first propose a robust, numerically stable, quasi-monolithic formulation with exact interface tracking to simulate the fluid-multilayered structure interactions. We validate the proposed formulation, using an in-house solver, by considering a simple two-layered plate-like structure with identical material properties against a single-layered plate. We then use this formulation to study a two-layered flexible plate, wherein each layer of the plate has independent material properties. 
			Systematic parametric simulations are conducted to understand the effect of difference in material properties between the two layers of a plate on the self-sustained flapping dynamics. The parametric simulations are performed at a Reynolds number $ Re=1000 $ by selecting different values of Young's modulus and density for each of the layers such that the average structure to fluid mass ratio $ \left(m^*\right)_\mathrm{avg}=0.1 $ and average non-dimensional bending stiffness $ \left(K_B\right)_\mathrm{avg}=0.0005 $. Firstly, the effects of difference in elasticity between the two layers on the flapping amplitude, frequency, forces and vortex shedding patterns are investigated. Following this, the effect of such difference in elastic properties on the onset of flapping is investigated for a case with $ Re=1000 $, $ \left(m^*\right)_\mathrm{avg}=0.1 $ and $ \left(K_B\right)_\mathrm{avg}=0.0008 $, for which, a single-layered plate does not undergo self-sustained flapping. Two distinct response regimes are observed depending on the difference in elastic properties between the two layers: (I) fixed-point stable; and (II) periodic limit cycle oscillations. Finally, we look into the effects due to the difference in the structural density between the two layers on the flapping dynamics of the plate.
		\end{abstract}
		
		\begin{keyword}
			Fluid-structure interaction (FSI)\sep Two-layered plate \sep Flapping dynamics \sep CFEI formulation
			\MSC[2020] 00-01\sep  99-00
		\end{keyword}
		
	\end{frontmatter}
	
	\section{Introduction}
	
	Thin, flexible, plate-like structures, when placed in an external flow field along its length with the leading edge clamped, can exhibit a self-sustained flapping motion above a critical flow speed due to coupled fluid-elastic instability. The value of critical flow speed depends on the destabilizing inertial effects and stabilizing elastic effects. The problem of a single isotropic flexible plate with leading-edge fixed is extensively studied for its relevance in the development of energy harvesting devices \citep{allen2001,tang2009c,deniz2012,michelin_2013,michelin_2015} and flexible propulsive devices \citep{heaving_zhu_2013,quinn_2014,quinn_2014_1,heavingPropulsor}. 
	Piezoelectric plates/patches convert the structural strain energy into electric current through piezoelectric effect and are therefore widely used for harvesting fluid kinetic energy either by allowing the piezoelectric patches to undergo flow-induced flapping motion or by attaching the patches onto a structure that is undergoing flow-induced vibration. A typical Piezoelectric patch consists of multiple layers, including the piezoelectric, substrate and electrode layers.
	
	Extensive experimental works have been conducted to understand the flapping instabilities of plate-like structures over the years. \cite{allen2001} were one of the first to show that the piezoelectric materials such as polyvinylidene fluoride (PVDF) and polyurethane (PU) can exhibit flow-induced flapping motion when placed in the wake of a bluff body. Around the same time, \cite{taylor_eel} conducted similar experiments by placing a long piezoelectric strip in the wake of bluff bodies immersed in water to demonstrate its energy harvesting potential. Later, \cite{akaydin_2010} reported a combination of experimental and numerical studies regarding the use of piezoelectric material to harness energy from unsteady airflows, where they have also discussed the power output and efficiency of the system. Following this, an interesting experimental study was carried out by \cite{lipson_2011}, who investigated a bio-inspired piezo-leaf undergoing self-induced aerodynamics instability. Notably, this study featured a cross-flow arrangement which is in contrast to the traditional set-up where the plate is in parallel with the flow direction. Recent experiments by \cite{kim2013} have shown that compared to the traditional flag orientation, wherein the leading edge is clamped and trailing edge left free, the orientation where the trailing edge is clamped, and the leading edge is left free (known as the inverted flag orientation) might be better suited for energy harvesting purposes. These observations have also been verified numerically by \cite{gurugubelli_JFM} and \cite{mittal_2016}. In \cite{mittal_piezo_inverted}, experiments were conducted considering a piezoelectric membrane in the inverted flag orientation, and the results demonstrate that the orientation is indeed better suited for energy harvesting than the traditional orientation. Apart from these studies, more specific studies such as the influence of electrode position in a piezoelectric energy harvesting flag \citep{piezo_electrode_position} and the effect of the number of piezoelectric patches on the plate \citep{tang2019} have also been investigated experimentally. 
	
	Along with experimental studies, many numerical studies have also been carried out to enhance the understanding of flapping dynamics for a single layered plate over the last few decades. A variety of numerical techniques have been employed to simulate interactions between a viscous fluid flow, i.e. the Navier-Stokes equations, and a flexible structure. 
	In one of the earliest numerical works, \cite{zhu2002} have numerically investigated flexible filament flapping within a flowing soap film. Here a finite difference method was used for both the structural and flow equations, and the coupling was enforced using the immersed boundary method. Later, two-dimensional flapping dynamics of a flag structure in a uniform stream was studied by \cite{connell2007} using a strongly coupled finite difference based numerical solver. This work presented an extensive analysis on the effects of various non-dimensional parameters, i.e. structure to fluid mass ratio $( m^*)$, the non-dimensional flexural rigidity $ (K_B)$, and the Reynolds number $ (Re)$, on the flapping response dynamics. Using an immersed boundary technique, \cite{huang_3d} numerically investigated three dimensional flapping dynamics to understand the vortex structures and three dimensional deformation of the plate. \cite{lbm_flapping} developed a lattice Boltzmann based immersed boundary method to investigate the two dimensional flapping dynamics involving or more than one flexible structures. \cite{jie_CFEI} presented a two-dimensional finite element based formulation that is numerically stable for very low structure to fluid mass ratio with a quasi-monolithic coupling and exact interface tracking. This formulation, has been used by \cite{gurugubelli_JFM} and \cite{gurugubelli_jfs_inverted_LAF} to investigate the two and three dimensional flapping dynamics, respectively, of a flexible plate in the inverted orientation. \cite{gurugubelli_ijhff_parallel} extended the quasi-monolithic formulation of \cite{jie_CFEI} for investigating the flapping dynamics of multiple flexible structures. \cite{gilmanov2015} has used a coupled finite element and finite difference approach to solve the structural and flow equations respectively to simulate the three dimensional flapping of an inverted flag in a uniform flow. Most recently, three dimensional simulations of an inverted flag in linear shear flows have been studied in \cite{invertedPlate_ShearFlow} using the lattice Boltzmann method for the fluid equations, the finite element method for the structural equations and immersed boundary approach to tackling the coupling. A comprehensive summary of all the numerical and experimental research on the fluid dynamics of flapping foils can be found in \cite{xia_review_2019}	
	
	Even though piezoelectric energy harvesting devices are multilayered, the existing understanding of non-linear flapping dynamics is primarily based on single-layered plate studies. In the present work, in order to gain better insight into the behaviour of these multilayered plates, we consider a two-layered flexible plate, with each layer having independent material properties (density, Young's modulus and Poisson's ratio). A robust, in-house FEM solver that is stable for mass ratios of the order 0.1 is used to solve the problem of a two-layered flexible plate placed in a uniform stream. The quasi-monolithic formulation proposed in \cite{jie_CFEI} has been extended to solve the fluid-multilayered structure interactions of an elastic body consisting of multiple materials.  Systematic investigations are then carried out to study the effects of differences in the material properties within a plate on the flapping dynamics. Initially, the effect of different elastic properties between the two layers on the flapping amplitude, frequency and forces are studied. Following this, we look into the effect of such variation in the elastic properties between layers on the onset of flapping in certain cases. Finally, a similar investigation into the effect of having layers with different structure-to-fluid mass ratios on the flapping dynamics is carried out.
	
	In Section~\ref{formulation}, we describe the quasi-monolithic formulation along with the time-discretization employed. Then in Section~\ref{prob_state}, the problem statement, along with the relevant non-dimensional parameters, are defined. Following this, in Section~\ref{results}, the validation of the in-house solver along with the results of the parametric investigations are detailed. Finally, a summary of all the findings is provided in Section~\ref{conc}. 
	
	\section{Governing Equations \& Numerical Methodology} \label{formulation}
	Let us consider a two-dimensional flexible structure $\Os$ made-up of $n$ different materials, i.e. $\Os = \Os_\mathrm{1}\cup\Os_\mathrm{2}\cup\Os_\mathrm{3}\cup\cdots\Os_\mathrm{n}$, interacting with the surrounding two-dimensional incompressible viscous fluid flow $\Of$. Let, $\Gamma_\mathrm{i}$ represents the interface between fluid and a part of the flexible structure made-up of $\mathrm{i}^\mathrm{th}$ material. Similarly, $\Gamma_\mathrm{i,j}$ denotes the interface between two parts of the flexible structure made-up of the $\mathrm{i}^\mathrm{th}$ and $\mathrm{j}^\mathrm{th}$ materials respectively, provided $\i\ne\j$.  
	
	In an arbitrary Lagrangian-Eulerian (ALE) reference frame, the Navier-Stokes equations for an incompressible flow are given as
	\begin{equation}
	\begin{aligned}\label{Eq:NavierStokes}
	\rho^\mathrm{f}\left[{\dd \bu^\mathrm{f} \over \dd t} +	\left(\bu^\mathrm{f} 
	-\vec{w}\right)
	\cdot \div \bu^\mathrm{f} \right]= \div
	\cdot \stress^\mathrm{f}+ \rho^\mathrm{f} \force^\mathrm{f} \quad &\mbox{on$\quad \Otf,$} \\
	\div \cdot \bu^\mathrm{f} = 0 \quad &\mbox{on$\quad \Otf,$}
	\end{aligned}
	\end{equation}
	where $\rho^\mathrm{f}$ is the fluid density and {$\bu^\mathrm{f}=\left(u^\mathrm{f},v^\mathrm{f}\right)=\left(u^\mathrm{f}\left(\xx,t\right),v^\mathrm{f}\left(\xx,t\right)\right)$} represents the fluid velocity with $x$ and $y$-velocity components $u^\mathrm{f}$ and $v^\mathrm{f}$, respectively, defined at the spatial points $\xx \in \Otf$. $\vec{w}=\vec{w}(\xx,t)$ denotes the velocity of the spatial points $\xx$ in the fluid domain $\Otf$. $\boldsymbol{f}^\mathrm{f}$ is the
	fluid body force and $\boldsymbol{\sigma}^\mathrm{f}$ represents the 
	Cauchy stress tensor for a Newtonian fluids given as
	\begin{equation}
	\begin{aligned}
	\stress^\mathrm{f} &= -p \bI + \mu^\mathrm{f} \left[\div \bu^\mathrm{f}+
	\left(\div \bu^\mathrm{f}\right)^T\right],
	\label{eq:cauchyStress}
	\end{aligned}
	\end{equation}
	where $\mu^\mathrm{f}$ is the fluid's dynamic viscosity and $p\vec{\mathrm{I}}$ denotes the isotropic fluid pressure tensor. 
	The dynamics of a flexible structure consisting of $n$ different materials should satisfy the Navier's equation over each $\Os_\i$, which is given as
	\begin{equation}
	\begin{aligned}\label{Eq:StructuralDynamics}
	\rho^\mathrm{s}_\mathrm{i} \frac{\partial
		\vec{u}^\mathrm{s}}{\partial t} =\bm{\nabla} \cdot
	\bm{\sigma}_\mathrm{i}^\mathrm{s} + \rho^\mathrm{s}_\mathrm{i} \vec{f}_\i^\mathrm{s} \qquad
	\mbox{in}\quad\Omega^\mathrm{s}_\mathrm{i}\quad\forall\quad\i=\{1,2,3,\cdots,n\}.
	\end{aligned}
	\end{equation}
	Here $\rho^\mathrm{s}_\mathrm{i}$ represents density of the flexible structure's $\mathrm{i}^\mathrm{th}$ material, $\bu^\mathrm{s}=\bu^\mathrm{s}(\zz_\i,t)$ is the velocity defined 
	at a Lagrangian point $\zz_\i \in \Os_\i$, $\boldsymbol{f}_\i^\mathrm{s}$ denotes the body force and $\boldsymbol{\sigma}^\mathrm{s}_\mathrm{i}$ represents the first Piola-Kirchhoff stress tensor for a Saint Venant Kirchhoff linearly hyper-elastic material, which is given as	
	\begin{align}
	\stress^\mathrm{s}_\i = 2 \mu^\mathrm{s}_\i \vec{F} \vec{E}
	+ \lambda^\mathrm{s}_\i \left[\mathrm{tr} \left(\vec{E}\right)\right] \vec{F},
	\label{eq:SVKModel}
	\end{align}
	where $\mathrm{tr}(\cdot)$ denotes the trace operator applied over a tensor, $\vec{F}$ is the deformation gradient and 
	$\vec{E}$ represents the Green-Lagrangian strain tensor.
	
	In addition to the governing Eqs.~(\ref{Eq:NavierStokes})~and~(\ref{Eq:StructuralDynamics}), the fluid multi-material structure interactions should also satisfy the following interface conditions along $\Gamma_\mathrm{i}=\Of \cap \Os_\mathrm{i}$ and $\Gamma_\mathrm{i,j}=\Os_\mathrm{i}\cap\Os_\mathrm{j}~|~\i\ne\j$
	\begin{eqnarray}
	&\int_{\vphi(\gamma,t)}\stress^\mathrm{f}(\vec{\vphi}(\zz,t),t)\cdot\vec{\mathrm{n}}^\mathrm{f} \mathrm{d}\Gamma+ \int_{\gamma}\stress^\mathrm{s}_\mathrm{i}(\zz,t)\cdot \vec{\mathrm{n}}^\mathrm{s} \mathrm{d}\Gamma=0
	\quad \forall \quad \zz\in\gamma , \gamma \subset \G_\mathrm{i},\label{Eq:TractionCont}\\
	&\uf(\vphi(\zz_\i,t),t)=\us(\zz_\i,t) \quad \forall \quad \zz_\i\in \G_\mathrm{i}, \label{Eq:VelocityCont}\\
	&\int_{\gamma}\stress^\mathrm{s}_\mathrm{i}(\zz,t)\cdot\vec{\mathrm{n}}^\mathrm{s}_\mathrm{i} \mathrm{d}\Gamma+ \int_{\gamma}\stress^\mathrm{s}_\mathrm{j}(\zz,t)\cdot \vec{\mathrm{n}}^\mathrm{s}_\mathrm{j} \mathrm{d}\Gamma=0
	\qquad \forall \quad \zz\in\gamma, \gamma \subset \G_\mathrm{i,j},
	\label{Eq:TractionCont_solid_solid}\\
	&\us(\zz_\i)=\us(\zz_\j) \qquad \vphi(\zz_\i,t) = \vphi(\zz_\j,t) \qquad \forall \quad \zz_\i,\zz_\j\in \G_\mathrm{i,j}~\mbox{and}~\zz_\i=\zz_\j, \label{Eq:VelocityCont_solid_solid}
	\end{eqnarray}
	Here, Eq.~(\ref{Eq:TractionCont}) represents the traction continuity along a small element $\gamma$ on the fluid-structure interface $\Gamma_\mathrm{i}$, wherein $\vec{\mathrm{n}}^\mathrm{f}$ and $\vec{\mathrm{n}}^\mathrm{s}$ denote the element $\gamma\mbox{'s}$ unit normal vectors in the direction pointing away from the fluid and structural domains respectively. $\boldsymbol{\varphi}^\mathrm{s}\left(\xx,t\right)$ is a mapping function that maps the Lagrangian point $\xx \in \Os_i$ to its deformed state at a time instance $t$. Equation~(\ref{Eq:VelocityCont}) denotes the velocity continuity condition, i.e. velocity of the fluid should be equal to the velocity of solid at the point of contact.
	Similar to Eq.~(\ref{Eq:TractionCont}) for the fluid-structure interface, Eq.~(\ref{Eq:TractionCont_solid_solid}) represents the traction continuity along the structure-structure interface $\Gamma_\mathrm{i,j}=\Os_\mathrm{i}\cap\Os_\mathrm{j}$ for any element $\gamma$ on it. In Eq.~(\ref{Eq:TractionCont_solid_solid}), $\vec{\mathrm{n}}^\mathrm{s}_\mathrm{i}$ and $\vec{\mathrm{n}}^\mathrm{s}_\mathrm{j}$ represent the unit normal vectors for the element $\gamma\mbox{'s}$ in the outward direction to $\Os_\mathrm{i}$ and $\Os_\mathrm{j}$ respectively. Finally, Eq.~(\ref{Eq:VelocityCont_solid_solid}) represents the continuity in position along the interface $\Gamma_\mathrm{i,j}$. It should be noted that the current formulation does not take the delamination effects into account. 
	
	\subsection{Quasi-Monolithic Formulation for MultiLayered Structures}\label{sec:numMethod}
	In this section, we propose a high-order finite element based formulation for
	simulating the 
	fluid-structure interaction of a viscous 
	fluid 
	flow interacting with
	a structure consisting of two or more materials. The proposed formulation is
	an extension of the combined field with explicit interface (CFEI) formulation
	proposed for an elastic body made up of a single material in \cite{jie_CFEI}. 
	The CFEI formulation is stable for low structure-to-fluid mass ratios and has been 
	extensively used for investigating the flow-induced flapping instability of
	thin flexible structures \citep{bourlet2015,gurugubelli_JFM,gurugubelli_ijhff_parallel,gurugubelli_jfs_inverted_LAF} 
	both for its numerical stability and its ability to capture the boundary layer
	induced effects accurately due to the exact interface tracking feature of
	the formulation.
	
	To construct the weak form of the Navier-Stokes equations (Eq.~\ref{Eq:NavierStokes}),
	consider a test function  for the fluid velocities,
	$\testf(\xx)\in\mathcal{H}^\mathrm{1}$, which is sufficiently smooth such that 
	$\testf(\xx)=0~\forall~\xx\in\G^\mathrm{f}_\mathrm{d}$, where $\Gamma^\mathrm{f}_\mathrm{d}$ represents
	the Dirichlet boundary of $\Of$. Similarly, we also introduce a test function $q\in\mathcal{L}_\mathrm{2}$ for the fluid pressure. 
	Hence, the weak form for Eq.~(\ref{Eq:NavierStokes}) has been written as
	\begin{align}
	\int_{\Otf} \rho^\mathrm{f}\left(\dd_t\bu^\mathrm{f}
	+\left(\bu^\mathrm{f}-\vec{w}\right)\cdot\div\bu^\mathrm{f}\right)\cdot \testf \dO + \int_{\Otf}\stress^\mathrm{f} : \div\testf\dO =\notag\label{eq:weakFormNavierStokes_1}\\ 
	\int_{\Otf}
	\force^\mathrm{f} \cdot \testf \dO+\int_{\G^\mathrm{f}_\mathrm{n}(t)} \vec{T}^\mathrm{f} \cdot \testf \dG+
	\displaystyle\sum\limits_{\i} \int_{\G_\mathrm{i}\left(t\right)} \left(\stress^\mathrm{f} \cdot \vec{\mathrm{n}}^\mathrm{f}\right) \cdot \testf  \mathrm{\dG}, \\
	\int_{\Omega^\mathrm{f}(t)} q \div \cdot \bu^\mathrm{f}  \dO = 0.
	\label{eq:WeakFormNavierStokes}
	\end{align}
	Here $\partial_t$ denotes partial time derivative operator ${\partial \left( \boldsymbol{\cdot} \right) }/{\partial t}$ and
	$\G^\mathrm{f}_\mathrm{n}$ represents the Neumann boundary of $\Of$ where $\stress^\mathrm{f}\cdot \vec{\mathrm{n}}^\mathrm{f}=\vec{T}^\mathrm{f}$.
	The weak form  of the structural dynamics equation (Eq.~\ref{Eq:StructuralDynamics})  can be constructed by introducing 
	a set of sufficiently smooth test functions defined by $\tests_\i(\zz_\i)\in\mathcal{H}^\mathrm{1}~|~\tests_\i(\zz_\i)=0~\forall~\zz_\i\in\G^\mathrm{s}_\mathrm{i,d}$ for the structural velocities, where
	$\G^\mathrm{s}_\mathrm{i,d}$ represents the Dirichlet boundary of the structural domain $\Os_\i$. The weak form of Eq.~(\ref{Eq:StructuralDynamics}) can be expressed as
	\begin{align}
	\int_{\Omega^\mathrm{s}_\i} \rho^\mathrm{s}_\i\dd_t\bu^\mathrm{s}\cdot\tests_\i\dO+\int_{\Omega^\mathrm{s}_\i}\stress^\mathrm{s}_\i:\div\tests_\i\dO = \int_{\Omega^\mathrm{s}_\i}
	\force^\mathrm{s}_\i\cdot \tests_\i \dO+ \notag\\ \int_{\G^\mathrm{s}_{\mathrm{i},\mathrm{n}}} \vec{T}^{\mathrm{s}}_{\mathrm{i}} \cdot \tests_\i \dG+\int_{\G_\i}
	\left(\stress^\mathrm{s}_\i   {\cdot}\vec{\mathrm{n}}^\mathrm{s}_\i\right) \cdot \tests_\i  + \displaystyle\sum\limits_{\j}\int_{\G_{\i\j}}
	\left(\stress^\mathrm{s}_\i   {\cdot}\vec{\mathrm{n}}^\mathrm{s}_\i\right) \cdot \tests_\i \mathrm{\dG}.\label{Eq:WeakFormStructuralDynamics}
	\end{align}
	Here $\G^\mathrm{s}_\mathrm{i,n}$ is the Neumann boundary of $\Os_\i$ where $\stress^\mathrm{s}_\i\cdot \vec{\mathrm{n}}^\mathrm{s}_\i=\vec{T}^\mathrm{s}_\i$.
	Finally, the weak form for the traction continuity Eqs.~(\ref{Eq:TractionCont}~and~\ref{Eq:TractionCont_solid_solid}) can be written as 
	\begin{equation}\begin{aligned}
	&\int_{\vphi(\gamma,t)}(\stress^\mathrm{f}(\vec{\vphi}(\zz,t),t)\cdot\vec{\mathrm{n}}^\mathrm{f})\cdot\testf \mathrm{d}\Gamma+ \int_{\gamma}(\stress^\mathrm{s}_\mathrm{i}(\zz,t)\cdot \vec{\mathrm{n}}_\i^\mathrm{s})\cdot\tests_\i \mathrm{d}\Gamma=0, \\
	&\int_{\gamma}(\stress^\mathrm{s}_\mathrm{i}(\zz,t)\cdot\vec{\mathrm{n}}^\mathrm{s}_\mathrm{i})\cdot\tests_\i \mathrm{d}\Gamma+ \int_{\gamma}(\stress^\mathrm{s}_\mathrm{j}(\zz,t)\cdot \vec{\mathrm{n}}^\mathrm{s}_\mathrm{j})\cdot\tests_{\j} \mathrm{d}\Gamma=0.
	\label{eq:WeakFormTraction}
	\end{aligned}
	\end{equation}	
	 The above weak forms of the traction continuity conditions are constructed by considering a conforming mesh along
	 the interfaces $\G_\i$ and $\G_{\i\j}$, thereby resulting in 
	\begin{equation}
		\testf(\vphi(\zz_\i,t))=\tests(\zz_\i) \quad \forall \zz_\i \in \Gamma_\i \quad \mbox{and} \quad \tests(\zz_\i)=\tests(\zz_\j) \quad \forall \zz_\i,\zz_\j \in \G_{\i\j}~|~\zz_\i=\zz_\j.
	\end{equation}
	Therefore, the weak form of the combined field formulation for the fluid-structure interactions for the structure consisting of multiple materials can be obtained by combining Eqs. (\ref{eq:weakFormNavierStokes_1})-(\ref{eq:WeakFormTraction}) 
	and written as
	\begin{align}\label{varOF}
	\int_{\Otf} \rho^\mathrm{f}\left(\dd_t\bu^\mathrm{f}(\xx,t)
	+ \left(\bu^\mathrm{f}-\vec{w}\right)\cdot\grad\bu^\mathrm{f}\right)\cdot \testf(\xx) \dO+ \int_{\Otf}\stress^\mathrm{f} : \grad\testf \dO \notag \\
	-\int_{\Omega^\mathrm{f}(t)}q \grad \cdot \bu^\mathrm{f} \dO \notag\\
	+\displaystyle\sum\limits_{\i}\int_{\Os_\i} \rho^\mathrm{s}_\i\dd_t\bu^\mathrm{s}\cdot \tests_\i \dO +\displaystyle\sum\limits_{\i}\int_{\Os_\i}\stress^\mathrm{s}_\i:\grad\tests_\i \dO= \notag\\
	\int_{\Otf} \force^\mathrm{f}\cdot \testf \dO
	+\int_{\Gamma_\mathrm{n}^\mathrm{f}(t)} \vec{T}^{\mathrm{f}} \cdot \testf \mathrm{\dG} + \sum\limits_{\i} \int_{\Os_\i} \force^\mathrm{s}_\i\cdot\tests_\i
	\dO+ \sum\limits_{\i}\int_{\Gamma_\mathrm{i,n}^\mathrm{s}} \vec{T}^{\mathrm{s}}_{\mathrm{i}} \cdot \tests_\i \mathrm{\dG}.
	\end{align}
	The above weak form is the generic form of the combined field formulations presented in \cite{jie_CFEI} and \cite{gurugubelli_ijhff_parallel}. For $n=1$, the weak form in Eq.~(\ref{varOF}) reduces into the formulation proposed in 
	\cite{jie_CFEI} for the problem of an incompressible fluid interacting	with a single flexible structure. On the other hand,
	for $\G_{\i\j}=\emptyset$ the weak form would represent the case of the fluid interacting with multiple flexible structures with no
	common interface between them  presented in \cite{gurugubelli_ijhff_parallel}. 
	
%
	The weak variation form in Eq.~(\ref{varOF}) is discretized to approximate the fluid velocity, 
	pressure and structural velocity using $\mathbb{P}_2/\mathbb{P}_{1}/\mathbb{P}_2$ isoparametric 
	elements. The mixed $\mathbb{P}_2/\mathbb{P}_{1}$ finite element discretization for the 
	fluid velocity and pressure would satisfy the inf-sup condition for well-posedness of the system of linear equations. 
	Let, $\TT^\mathrm{
		f}_\mathrm{h}(t)$ represents the finite element mesh defined on the domain $\Of(t)$ for the fluid velocity and pressure. Similarly, let $\TT^\mathrm{s}_\mathrm{i,h}$ is the triangular mesh defined on domains $\Os_{\i}$ for the flexible structure with $\i^\mathrm{th}$ material properties. For better computational efficiency, the edges of the isoparametric elements are assumed to be straight unless they represent	either the interface or curved boundaries.	
	
	
	A typical fluid-structure interaction formulation requires solving of three fields i.e. fluid, structure and the fluid mesh. However, the weak form presented in Eq.~(\ref{varOF}) accounts only for the fluid and structural fields. This is where the novelty of the quasi-monolithic solver comes into the picture. Unlike traditional monolithic solvers where all three fields are solved together, in the quasi-monolithic formulation the equations corresponding to the movement of fluid mesh is decoupled from the other two fields, i.e. fluid and structure. 
	A second-order extrapolation scheme is considered to describe the displacement vector $\vec{\eta}^\mathrm{s,n}_\mathrm{h}$ of the flexible structure to decouple the fluid mesh movement
	\begin{align}
	\vec{\eta}^\mathrm{s,n}_\mathrm{h}(\zz_\i)&= \vec{\eta}^\mathrm{s,n-1}_\mathrm{h}(\zz_\i)+\frac{3\Delta t}{2}\usnn_\mathrm{h}(\zz_\i)- \frac{\Delta t}{2} \usnnn_\mathrm{h}(\zz_\i) \qquad \forall \zz_\i \in \TT_\mathrm{i,h}^\mathrm{s},
	\label{vphihn}
	\end{align}
	where $\Delta t$ is the time step. Subscript $\h$ above represents the approximation due to the mesh discretization. 
	
	Let, $\vec{\eta}^\mathrm{f,n}_\h$ and $\xx^\mathrm{n}_\h$ denote displacement and the corresponding updated position of a {fluid} mesh node on $\TT^\mathrm{
		f}_\mathrm{h}(t^\n)$ due to the deformation/movement of the structure respectively. Therefore, the fluid mesh $\TT^\mathrm{
		f}_\mathrm{h}(t^\n)$ should satisfy the displacement continuity with the structure along the interfaces, which is given as
	\begin{align}
	\vec{\eta}^\mathrm{f,n}_\h(\vphin(\zz_\i))	= \vec{\eta}^\mathrm{s,n}_\mathrm{h}(\zz_\i) \quad \forall \zz_\i\in (\TT_\mathrm{i,h}^\mathrm{s} \cap \TT_\mathrm{h}^\mathrm{f}). \label{meshDispContinuity}
	\end{align} 
	Given $\vec{\eta}^\mathrm{s,n}_\mathrm{h}(\zz_\i)~\forall~\zz_\i\in\G_\i$, the internal fluid nodes can be updated using pseudo-elastic material model \citep{tezduyar_ALE} given by,
	\begin{align}
	\label{MeshEq}
	\div \cdot \stress^\mathrm{m} = \vec{0},
	\end{align} 
	and satisfying Eq.~(\ref{meshDispContinuity}) as the boundary condition. In the above equation, $\stress^\mathrm{m}$ represents the stress experienced by the {fluid} mesh due to the displacement of the interface solid Lagrangian nodes. Considering the {fluid} mesh as a linearly elastic material, the stress experienced by the fluid mesh can be given by
	\begin{align}
	\label{meshEq}
	\qquad \stress^\mathrm{m} = (1+\tau_\mathrm{m})\left[\left(\div \vec{\eta}^\mathrm{f}+\left(\div
	\vec{\eta}^\mathrm{f}\right)^T\right)+\left(\div \cdot \vec{\eta}^\mathrm{f}\right)\vec{\mathrm{I}} \right],
	\end{align}
	where $\tau_\mathrm{m}$ is a mesh stiffness variable chosen as a function of the element size to 
	limit the distortion of the small elements located in the immediate vicinity of the fluid-structure 
	interface. The mesh stiffness variable $\tau_\mathrm{m}$ for the $\j^\mathrm{th}$ element is defined as $\frac{\max |T|-\min|T|}{|T_\mathrm{j}|}$, where $T_\mathrm{j}$ represents the area of $\mathrm{j}^\mathrm{th}$ element and $T$ is the set of areas of all the elements on the mesh $\tau^\mathrm{f}$ respectively. 
	
	Using the updated mesh positions, $\xx^\n$, a second order approximation of the fluid mesh velocities can be given as
	\begin{align}
	\vec{w}^\mathrm{n}(\xx^\n) = \frac{1}{\Delta t}\left[\left(\xx^\mathrm{n}-\xx^\mathrm{n-1}\right)+\frac{1}{2}
	\left(\xx^\mathrm{n-1}-\xx^\mathrm{n-2}\right)  
	-\frac{1}{2} \left(\xx^\mathrm{n-2}-\xx^\mathrm{n-3}\right)\right].
	\label{wn}
	\end{align}

	Finally, without loss of accuracy the nonlinear convective term in the weak form Eq.~(\ref{varOF}) can be 
	linearized by using a second-order explicit time-accurate extrapolation for the fluid velocity, which is given by
	\begin{equation}\label{convVel}
	\cuf_\h(\xx^\n) = 2\ufnn_\h (\xx^\mathrm{n-1})-\ufnnn_\h (\xx^\mathrm{n-2}).
	\end{equation}
	Therefore, the fully discretized finite element form of the quasi-monolithic formulation for the interaction of a fluid with the structure made up of multiple material will be
	\begin{align}
	\left.
	\begin{aligned}
	\int_{\TT^\mathrm{
			f}_\mathrm{h}(t^\n)}
	\left[\frac{\rho^\mathrm{f}}{\Delta t}\left(\frac32\ufn_\h(\xx^\n) - 2 \ufnn_\h(\xx^\mathrm{n-1})+\frac12 \ufnnn_\h(\xx^\mathrm{n-2})\right)
	\right. \nonumber \\
	\left.+\left(\cuf_\h(\xx^\n)-\ww_\h^\n(\xx^\n)\right)\cdot\grad \ufn_\h(\xx^\n) \right]\cdot\testf d\TT
	\nonumber \\
	+\int_{\TT^\mathrm{
			f}}
	\rho^\mathrm{f}\nu^{\f} \left(\grad \ufn_\h(\xx^\n) + \left(\grad \ufn_\h\left(\xx^\n\right)\right)^T\right):\grad \testf d\xx
	\end{aligned}
	\right\} A \notag \\
	\left.
	\begin{aligned}
	-\int_{\Of_{\h,t^\n}}
	p^{\f,\n}_\h (\xx^\n)  (\grad\cdot\testf) d\xx
	\end{aligned}
	\right\} B \notag \\
	\left.
	\begin{aligned}
	-\int_{\Of_{\h,t^\n}}
	q^\f  (\grad\cdot\ufn_\h) d\xx
	\end{aligned}
	\right\} C \notag \\
	\left.
	\begin{aligned}
	+\displaystyle\sum\limits_{\i}\int_{\Os_{\i,\h}}
	\frac{\rho^\mathrm{s}}{\Delta t}\left(\frac32\usn_\h - 2 \usnn_\h+\frac12 \usnnn_\h\right) \cdot\tests_\i d\xx 
	\end{aligned}
	\right\} D \notag \\
	=
	\left.
	\int_{\Of_{\h,t^\n}}\force^\mathrm{f}\cdot\testf d\Omega
	+\int_{(\G^\mathrm{f}_\mathrm{n})_\h}\vec{T}^{\mathrm{f}} \cdot\testf d\G
	+\displaystyle\sum\limits_{\i}\int_{\Os_{\i,\h}}\force_\i^\mathrm{s}\cdot\tests_\i d\Omega
	+ \displaystyle\sum\limits_{\i}\int_{(\G^\mathrm{s}_\mathrm{i,n})_\h}\vec{T}^{\mathrm{s}}_{\mathrm{i}} \cdot\tests_\i d\G,
	\right\} E
	\label{2ndscheme}
	\end{align}
	where $A$ contains the transient, convective and diffusive contributions, 
	$B$ and $C$ are the pressure and continuity terms respectively,
	$D$ represents the solid momentum equation and $E$ consists of the external body force and boundary conditions.

	\section{Problem Statement}\label{prob_state}
\begin{figure}
		\begin{center}
			\includegraphics[width=1\columnwidth]{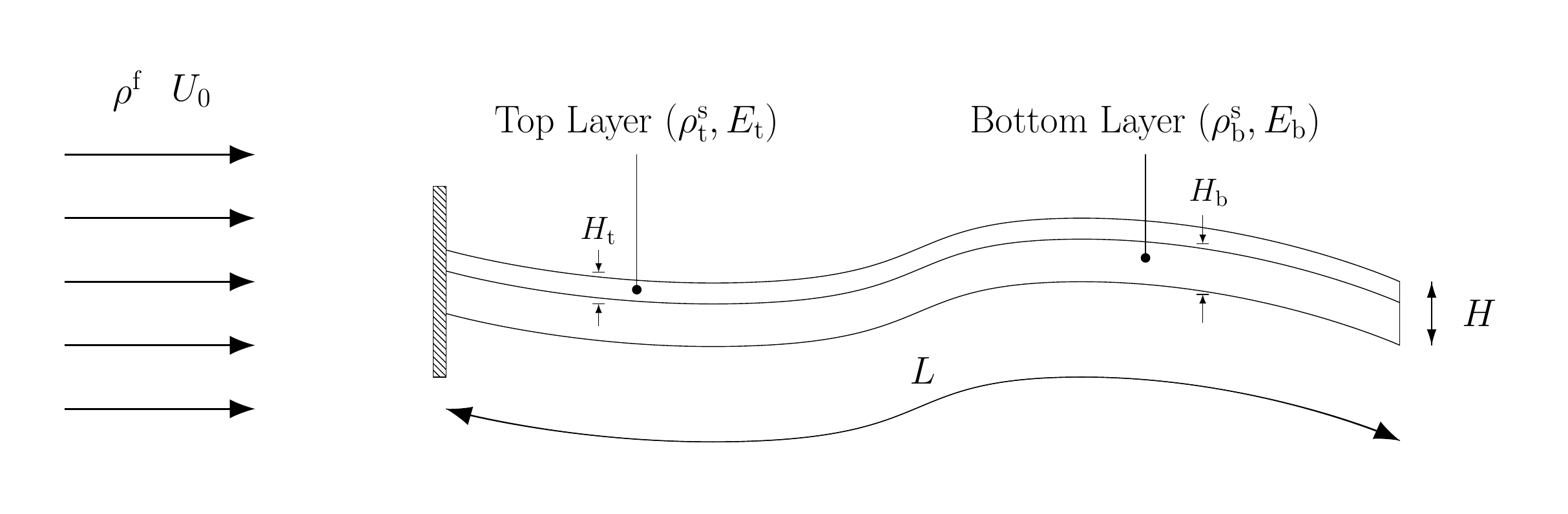}
			\caption{A schematic illustration of a two-layered flexible plate, with its leading edge fixed, exhibiting self-sustained flapping in a uniform flow.}
			\label{Problem_Statement}
		\end{center}
	\end{figure}

	In the present work, flapping dynamics of a flexible plate made-up of two layers stacked on top of each other having different material properties (i.e. density, Young's modulus, Poisson's ratio) and thicknesses, is studied in a two-dimensional framework using the numerical methodology described in section \ref{formulation}. Figure~\ref{Problem_Statement} presents a representative schematic for the passive flapping of a two-layered flexible plate of length $L$ and thickness $H$ with the leading edges fixed, and interacting with an incompressible viscous uniform flow $U_0$ {in the direction of the plate's length} and density $\rho^\mathrm{f}$. 	 
	For this study  the density, Young's modulus, Poisson's ratio and thickness of the bottom layer are represented as $\rho^\mathrm{s}_\mathrm{b}, E_\mathrm{b}, \nu_{\mathrm{b}},$ and $H_\mathrm{b}$, respectively. Similarly, the material properties and thickness for the top layer are considered as $\rho^\mathrm{s}_\mathrm{t}, E_\mathrm{t}, \nu_{\mathrm{t}},$ and $H_\mathrm{t}$. The total thickness ($H$) of the two-layered plate is given by $H=H_\mathrm{t}+H_\mathrm{b}$.
	
	The flapping dynamic dynamics of a single-layered flexible plate is relatively well understood and is known to depend on structure to fluid mass ratio ($m^*$), non-dimensional flexural rigidity ($K_B$) and Reynolds number ($Re$) \citep{shelley05,connell2007,jaiman2013}. Similar non-dimensional parameters for a two-layered plate can be defined based on the average elastic modulus $D^\mathrm{s}_\mathrm{avg}=\left(\frac{E^\mathrm{s}_\mathrm{t}}{(1-\nu_{\mathrm{t}}^2)}H_\mathrm{t}+\frac{E^\mathrm{s}_\mathrm{b}}{(1-\nu_{\mathrm{b}}^2)}H_\mathrm{b}\right)/H$, and average plate density $\rho^\mathrm{s}_\mathrm{avg}= \left(\rho^\mathrm{s}_\mathrm{t}H_\mathrm{t}+\rho^\mathrm{s}_\mathrm{b}H_\mathrm{b}\right)/H$ can be defined as
	\begin{equation}
	(m^*)_\mathrm{avg}=\alpha h \qquad (K_{B})_\mathrm{avg}=\frac{\beta_\mathrm{avg} h^3}{12} \qquad Re = \frac{\rho^\mathrm{f}U_0L}{\mu^\mathrm{f}}, \label{numbers1}
	\end{equation}
	where $\alpha=\rho^\mathrm{s}_\mathrm{avg}/\rho^\mathrm{f}$ is the non-dimensional density ratio, $\beta_\mathrm{avg}=D^\mathrm{s}_\mathrm{avg}/\rho^\mathrm{f}U_0^2$ denotes the non-dimensional elastic modulus and $h=H/L$ represents the non-dimensional thickness of the two-layered plate. Here we have expressed $(m^*)_\mathrm{avg}$ and $(K_B)_\mathrm{avg}$ has a function of two non-dimensional parameters, i.e. $\alpha$ and $h$ for $(m^*)_\mathrm{avg}$, and $\beta$ and $h$ in the case of $(K_B)_\mathrm{avg}$. This representation enables us to isolate the density and elasticity effects from the thickness. At the first instant one might feel this representation is redundant, however in the case two-layered plate $(m^*)_\mathrm{avg}$ and $(K_B)_\mathrm{avg}$ defined based on $\rho^\mathrm{s}_\mathrm{avg}$ and  $D^\mathrm{s}_\mathrm{avg}$ respectively can have same values of $(m^*)_\mathrm{avg}$ and $(K_B)_\mathrm{avg}$ for different combinations of $\rho^\mathrm{s}_\mathrm{t}, E_\mathrm{t}, \nu_{\mathrm{t}},$ $H_\mathrm{t}$, $\rho^\mathrm{s}_\mathrm{b}, E_\mathrm{b}, \nu_{\mathrm{b}},$ and $H_\mathrm{b}$.
	
	\begin{figure}
		\centering
		\includegraphics[trim={0mm 0mm 0mm 0mm},clip,width=\textwidth]{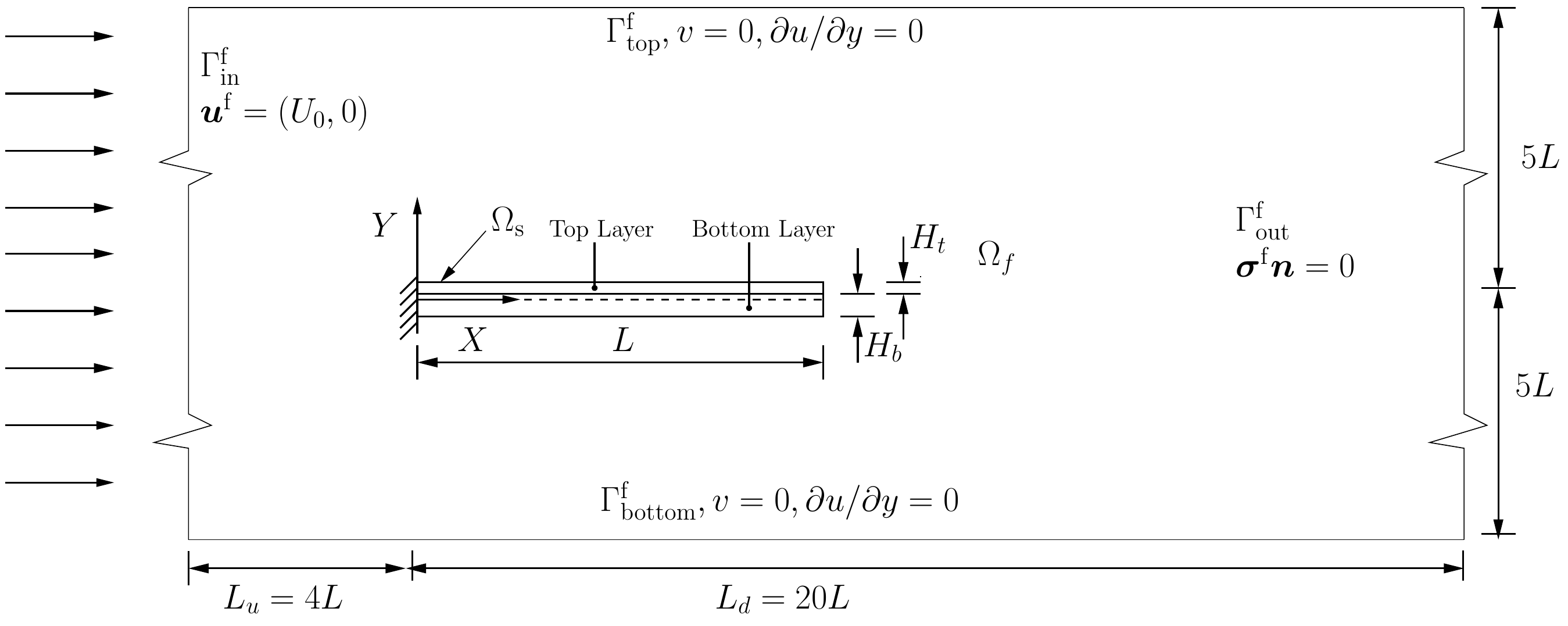}
		\caption{Computational setup corresponding to a two-layered flexible plate in a uniform flow of velocity $U_{0}$ along with the boundary conditions and interface separating fluid-solid domain. The dashed line $\left(---\right)$ indicates the neutral surface of a given two-layered flexible plate.}\label{double_layered}
	\end{figure}
	
	Figure \ref{double_layered} shows the computational setup along with the numerical boundary conditions considered for simulating the interactions between a two-layered flexible plate that is clamped at its leading edge and a uniform axial flow. The computational domain is of size $24L \times 10L$, where $L$ is the length of the flexible plate. The thickness of the two-layered flexible plate is considered as $H=0.01L$ so that $H\ll L$. For simplicity, in this study it has been assumed that $H_\mathrm{t}=H_\mathrm{b}$, i.e. $H_\mathrm{t}=H_\mathrm{b}=0.005L$. The fluid is considered to be flowing from left to right with a free steam velocity $U_0$, i.e. it enters and leaves the computational domain through $\Gamma_\mathrm{in}^\mathrm{f}$ and $\Gamma_\mathrm{out}^\mathrm{f}$, respectively. Hence, a traction free boundary condition is imposed at the outlet $(\Gamma_\mathrm{out}^\mathrm{f})$ of the computational domain. A free-slip boundary condition is applied along the top $(\Gamma_\mathrm{top}^\mathrm{f})$ and the bottom $(\Gamma_\mathrm{b}^\mathrm{f})$ boundaries of the computational domain. A no-slip boundary condition is enforced along the plate-fluid interface. 
	
	
	\section{Results and Discussion}\label{results}
	Before we start investigating the flapping dynamics of a two-layered flexible plate, we first present the verification of the proposed numerical methodology described in section~\ref{sec:numMethod} and its implementation. To verify the implementation, we construct a special case by providing identical material properties for both the layers and compare the flapping response against the results published in \cite{bourlet2015} for a single-layered foil. The identical material properties considered for both the layers would correspond to the single-layer non-dimensional parameters of $m^*=0.1$, $K_B=0.0001$ and $Re=1000$. 	
	
\begin{figure}
		\centering
		\includegraphics[width=0.49\columnwidth]{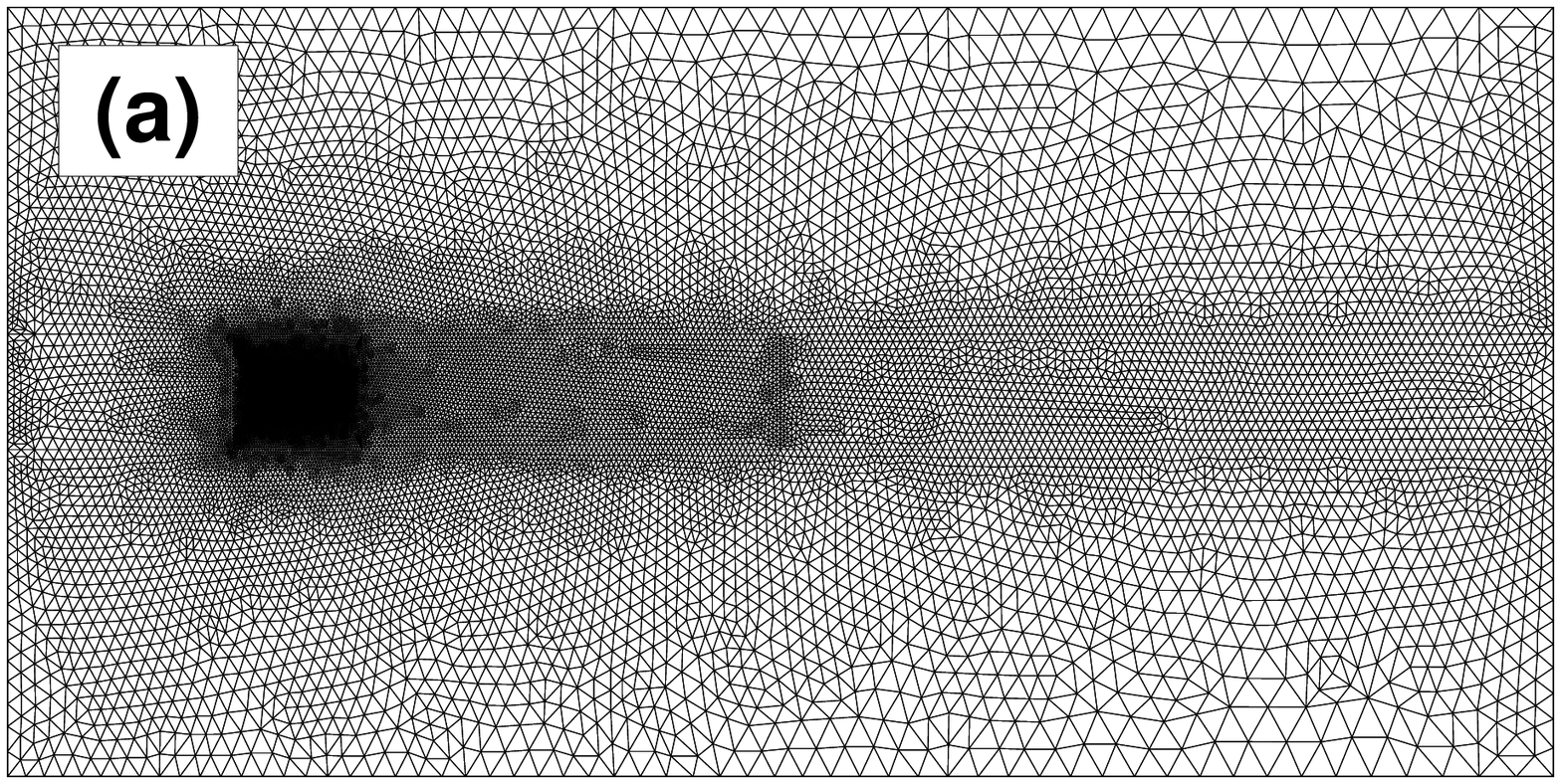}
		\includegraphics[width=0.49\columnwidth]{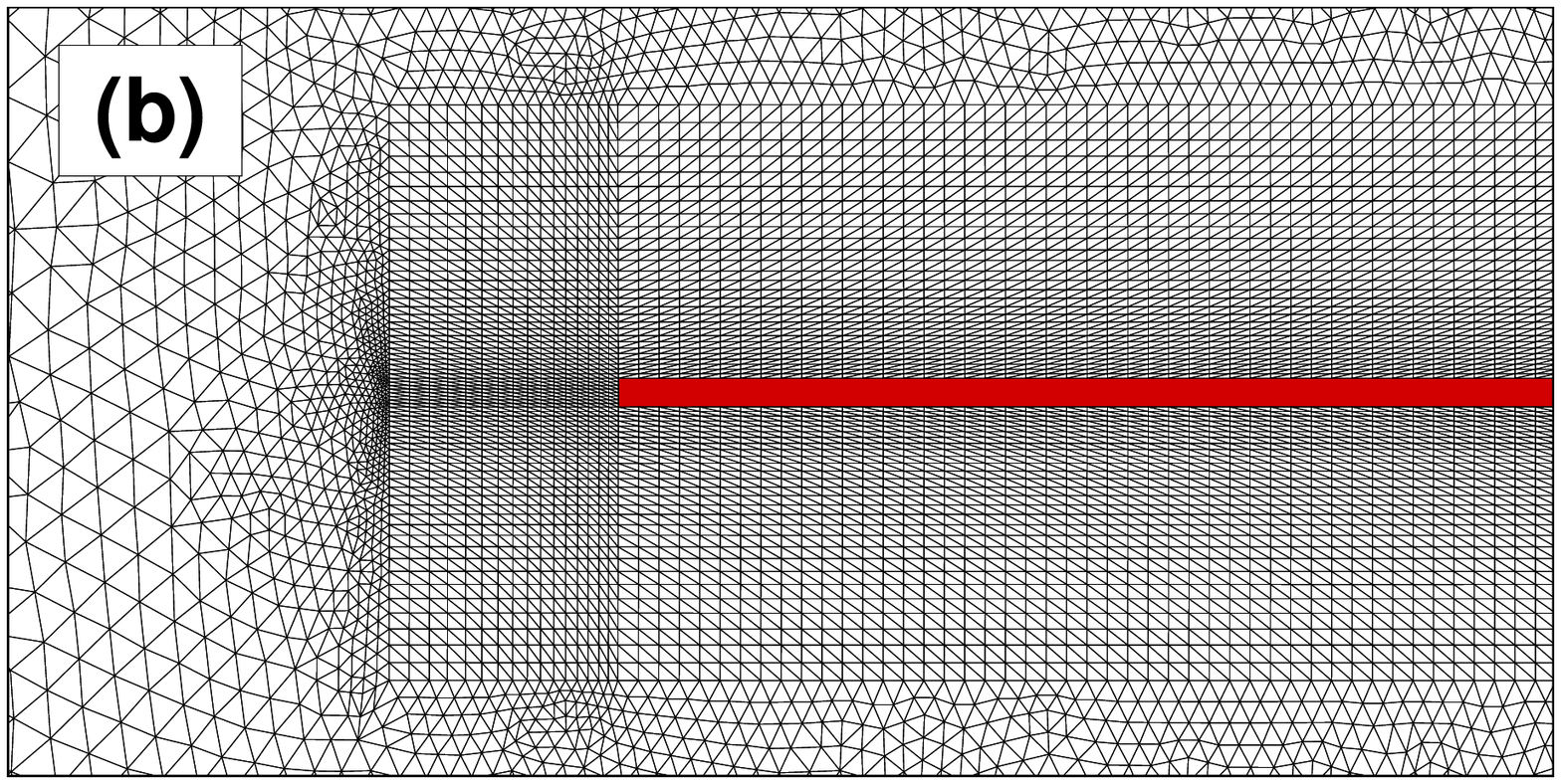}		
		\caption{(a) Distribution of finite element mesh in the fluid domain ($\Omega^\mathrm{f}(t)$) $-$ consisting of 7446 nodes and 14657 elements. This mesh distribution corresponds to the initial configuration of the fluid domain (see Fig.\ref{double_layered}). (b) Magnified view of finite element mesh in the vicinity of the leading-edge.}  
		\label{validation}
	\end{figure}

	Figure~\ref{validation}~(a) presents the high-order $\mathbb{P}_2$ finite-element computational mesh that has been selected based on the mesh sensitivity study. The mesh consists of $29,549$ nodes and $14,657$ $\mathbb{P}_2$ elements. Figure~\ref{validation}~(b) shows the zoomed-in view of the boundary-layer mesh around the flexible two-layered plate near its leading edge. Table~\ref{validationtable} presents a comparison of the trailing edge flapping amplitude and flapping Strouhal number for the numerical simulations performed using the solver based on the proposed formulation and the result from \cite{bourlet2015}. The maximum deviation in the trailing edge response is about $0.04\%$ and the small variation in the solution can be attributed to the computational mesh considered for this study, i.e. the computational mesh used in \cite{bourlet2015} consists of approximately 32,000 nodes against approximately 30,000 nodes considered here. 
	
	\begin{table}
		\centering
		\resizebox{0.49\textwidth}{!}{
			\begin{tabular}{lll}
				& \cite{bourlet2015} & Present  \\ \hline
				Flapping amplitude $\delta_y/L$      & $0.2435$               & $0.2434$ \\
				\% Change               & $-$                    & $-0.04$  \\
				Strouhal number $\delta_y f/U_0$        & $0.2341$               & $0.2340$ \\
				\% Change               & $-$                    & $-0.04$  
			\end{tabular}%
		}
		\caption{Verification of the proposed quasi-monolithic solver for the fluid-structure interaction of a two-layered flexible plate made up of identical material properties for both the layers whose equivalent single layer non-dimensional parameters correspond to $Re=1000$, $m^*=0.1$, and $K_B=0.0001$ with the numerical result of \cite{bourlet2015}.}
		\label{validationtable}
	\end{table}
	
	\subsection{Effect of Elastic Modulus on the Flapping Dynamics of Plate} \label{Delta_study}
	Our consideration of a two-layered flexible plate in this work is a simplified case of one or more flexible plate like structures having different elastic properties mounted on top of each other. An interesting case arises when the average non-dimensional elastic modulus $\beta_{\mathrm{avg}} $ of the two-layered plate is kept constant and the independent elastic properties of the two layers are varied. This asymmetry in contribution from the layers to the elastic modulus of the plate has a definite impact on the flapping dynamics of the plate. To study this, we introduce a parameter $ \Delta=\beta_{\mathrm{t}}-\beta_{\mathrm{b}}$ which represents the non-dimensional difference in the elastic modulus between the two layers, where 
	\begin{equation}\label{Beta_b_t}
	\beta_{\mathrm{b}}=\frac{E_{\mathrm{b}}}{\rho^{\mathrm{f}}U_0^2\left(1-\nu_{\mathrm{b}}^2\right)}
	\qquad
	\beta_{\mathrm{t}}=\frac{E_{\mathrm{t}}}{\rho^{\mathrm{f}}U_0^2\left(1-\nu_{\mathrm{t}}^2\right)}.
	\end{equation}

	\begin{figure}
		\centering
		\begin{subfigure}{0.49\textwidth}
			\includegraphics[width=0.97\columnwidth,trim=0mm 0mm 0mm 5mm,clip]{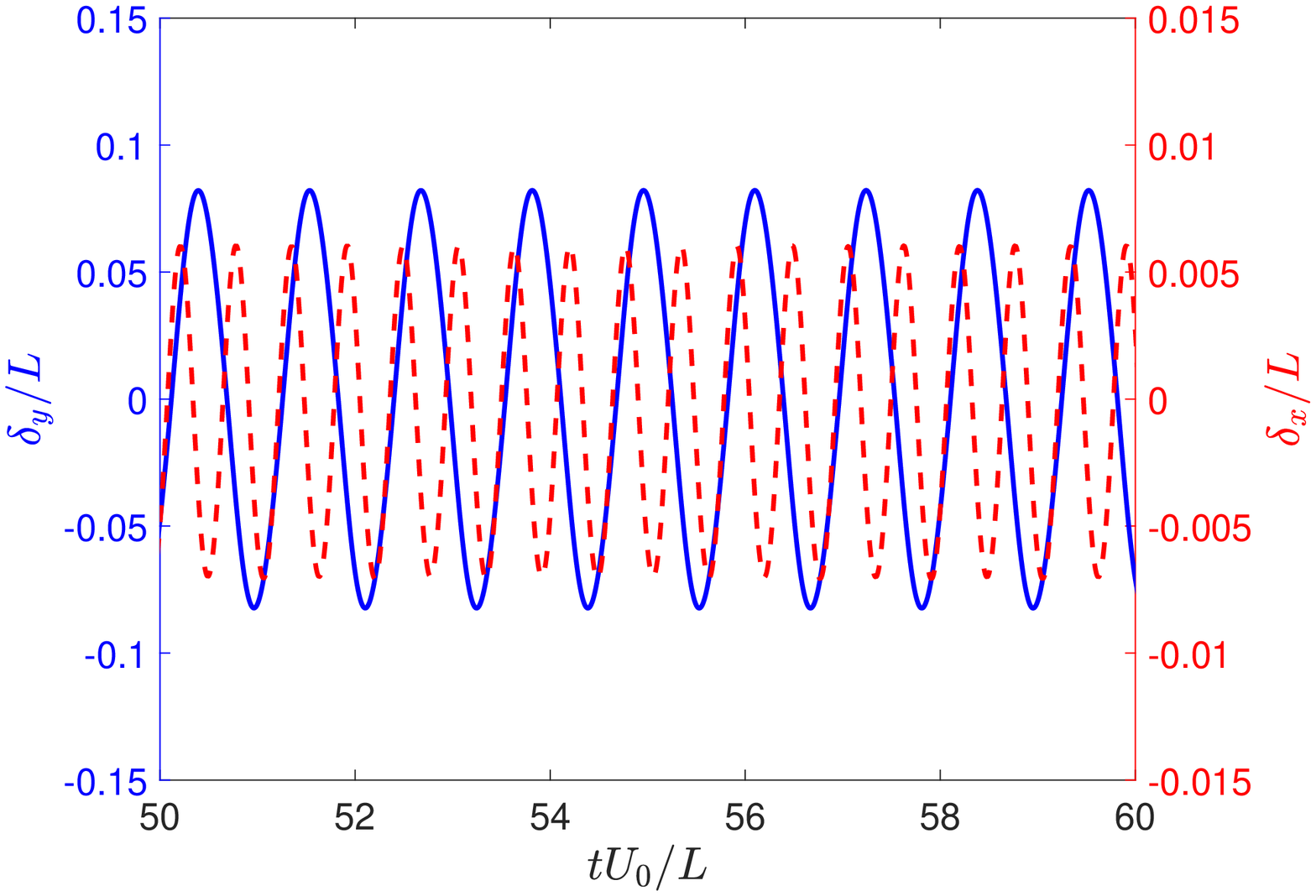}
		\end{subfigure}
		\begin{subfigure}{0.49\textwidth}
			\includegraphics[width=0.99\columnwidth,trim=0mm 0mm 10mm 5mm,clip]{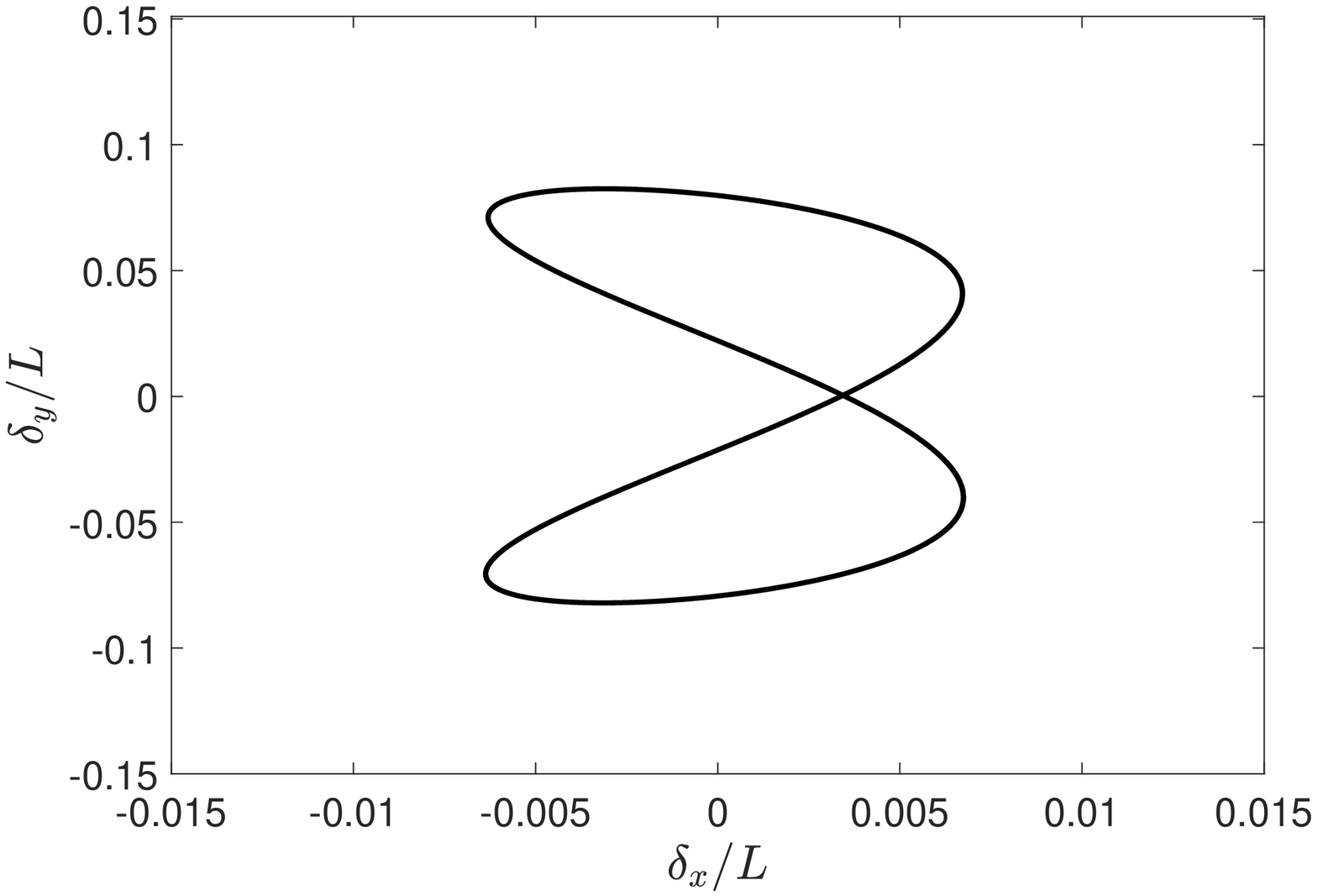}
		\end{subfigure}\\
		\begin{subfigure}{0.99\textwidth}
			\centering
			\caption{$\Delta=0$}	
		\end{subfigure}\\
		\begin{subfigure}{0.49\textwidth}
			\includegraphics[width=0.99\columnwidth,trim=0mm 0mm 0mm 5mm,clip]{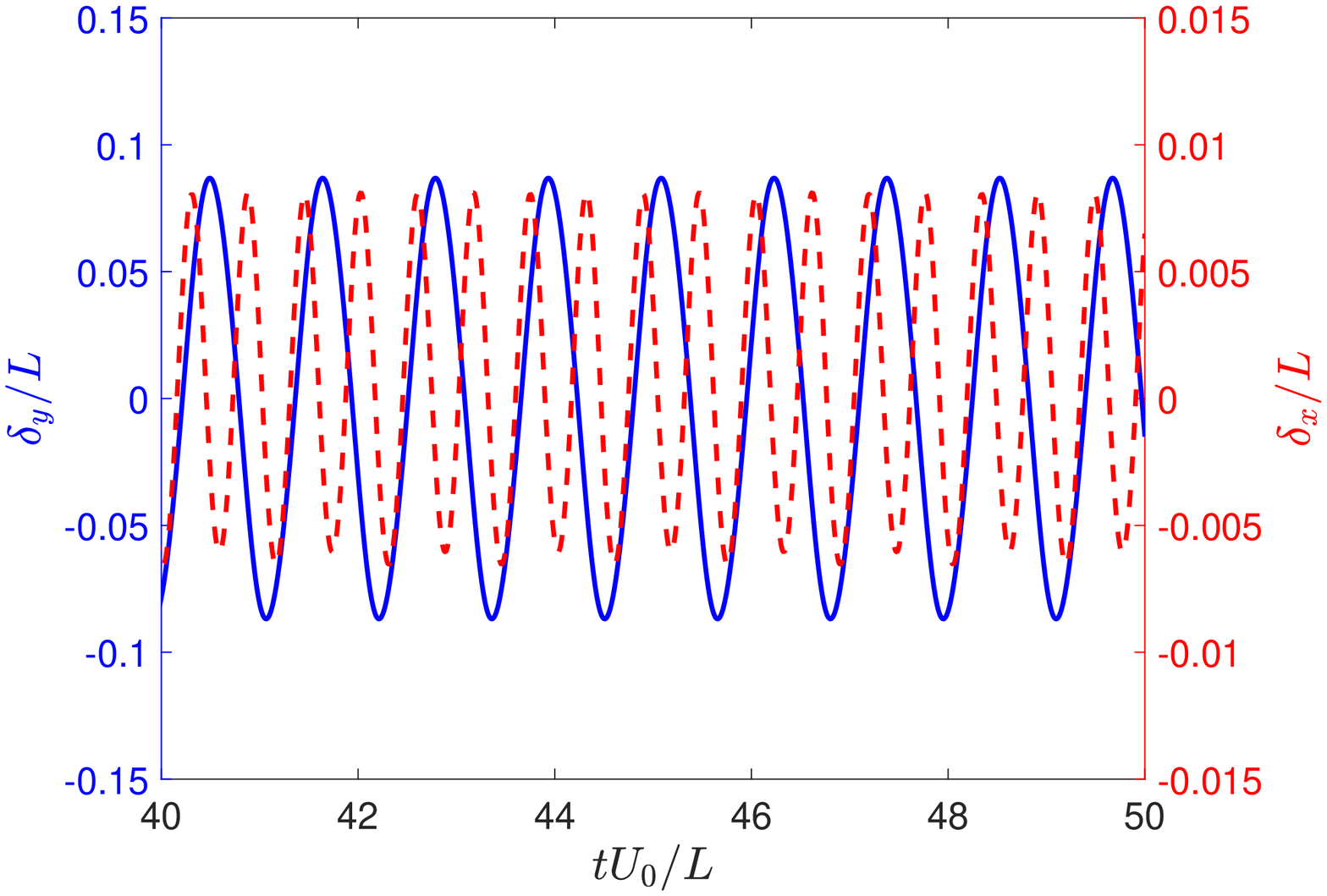}
		\end{subfigure}
		\begin{subfigure}{0.49\textwidth}
			\includegraphics[width=0.99\columnwidth,trim=0mm 0mm 10mm 5mm,clip]{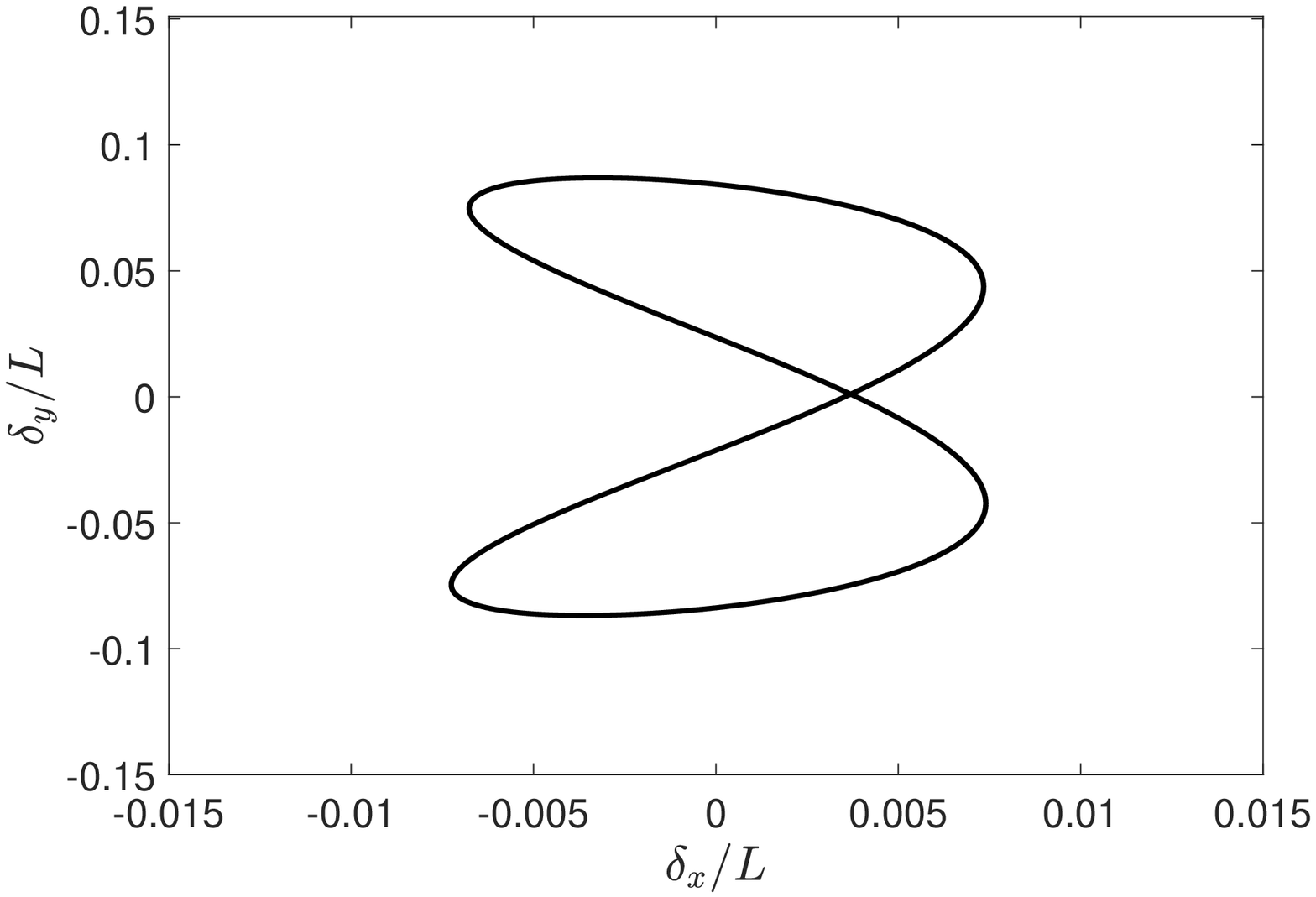}
		\end{subfigure}\\
		\begin{subfigure}{0.99\textwidth}
			\centering
			\caption{$\Delta=2400$}	
		\end{subfigure}\\
		\begin{subfigure}{0.49\textwidth}
			\includegraphics[width=0.99\columnwidth,trim=0mm 0mm 0mm 5mm,clip]{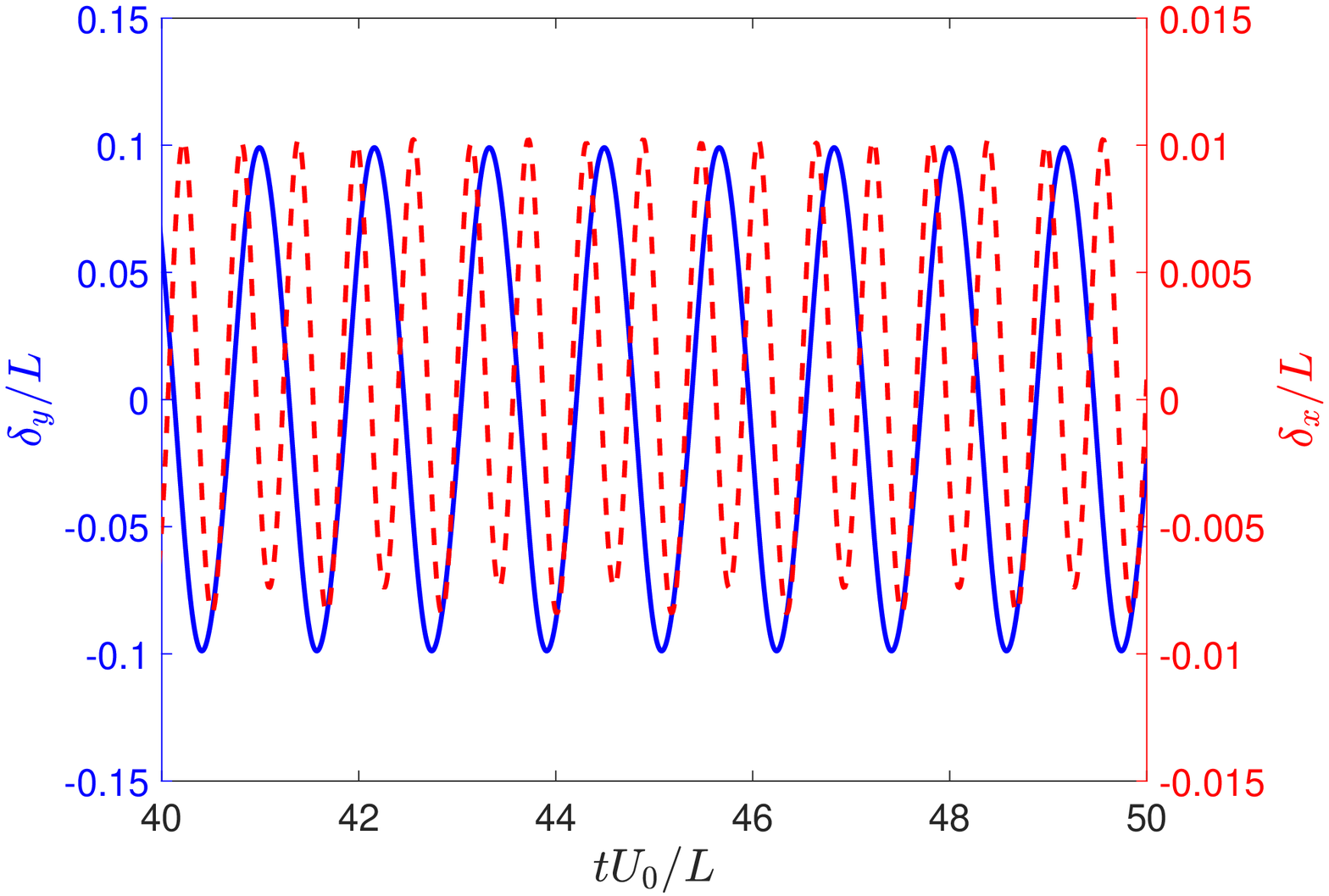}
		\end{subfigure}
		\begin{subfigure}{0.49\textwidth}
			\includegraphics[width=0.99\columnwidth,trim=0mm 0mm 10mm 5mm,clip]{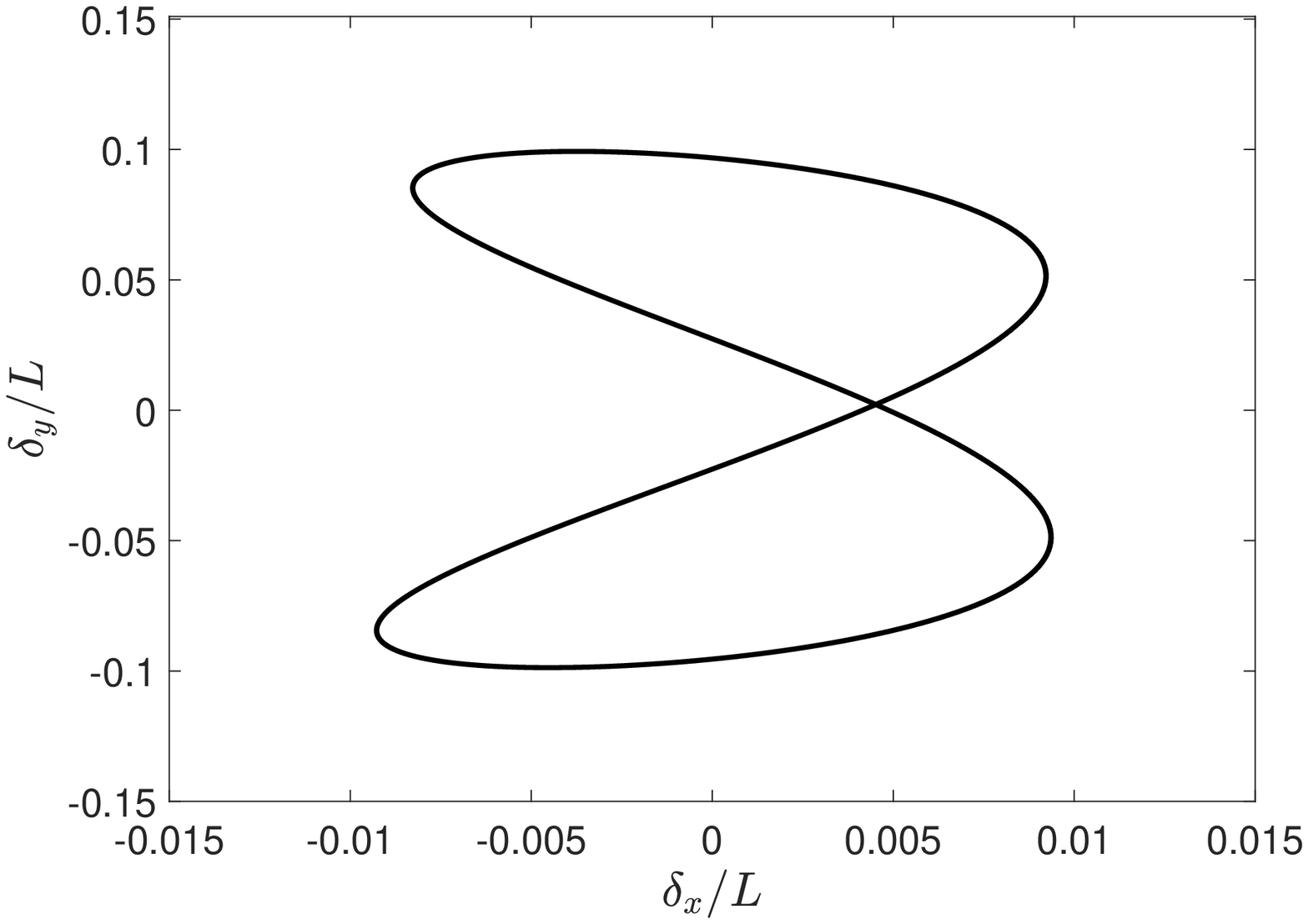}
		\end{subfigure}\\
		\begin{subfigure}{0.99\textwidth}
			\centering
			\caption{$\Delta=4800$}	
		\end{subfigure}\\
		\begin{subfigure}{0.49\textwidth}
			\includegraphics[width=0.99\columnwidth,trim=0mm 0mm 0mm 5mm,clip]{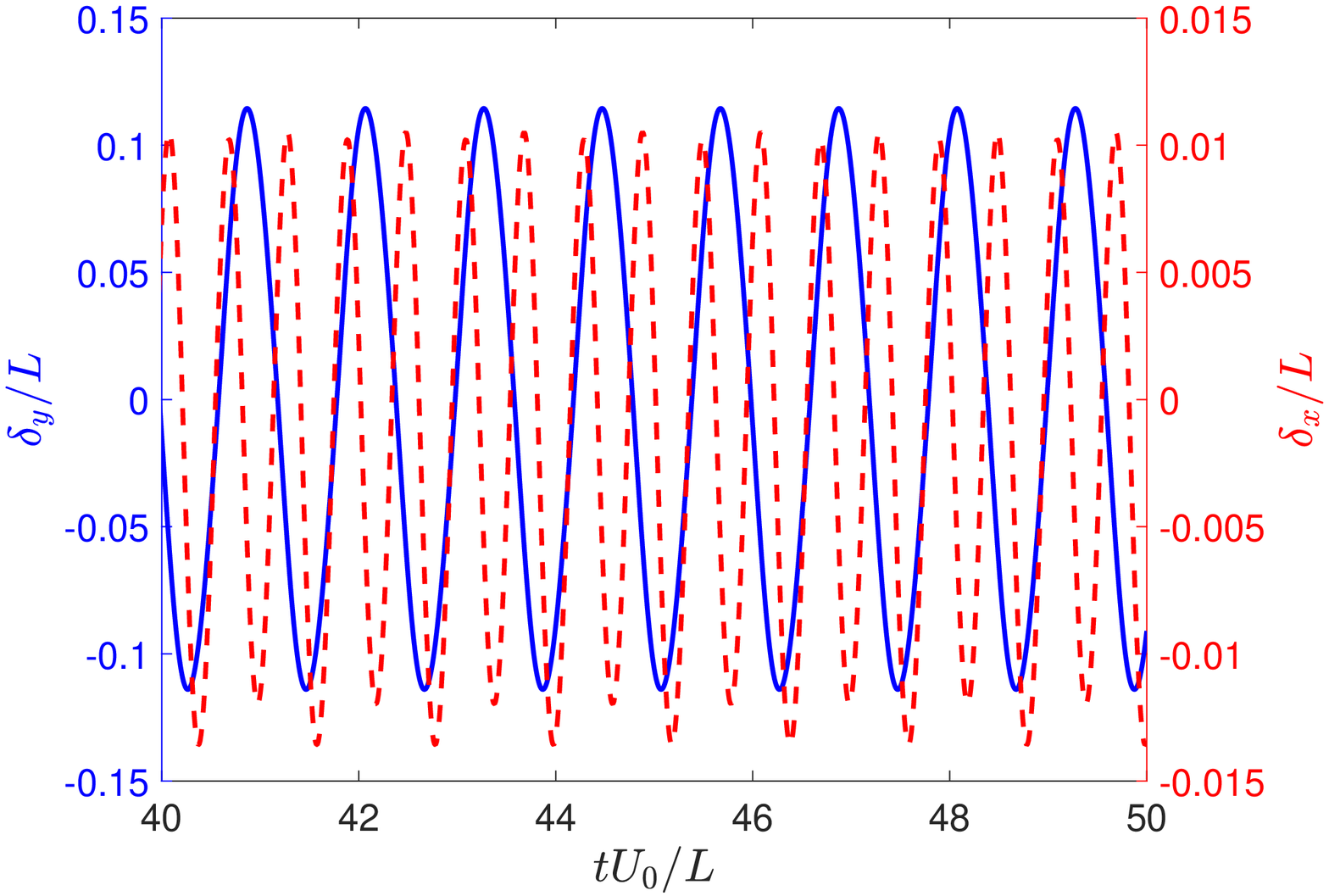}
		\end{subfigure}
		\begin{subfigure}{0.49\textwidth}
			\includegraphics[width=0.99\columnwidth,trim=0mm 0mm 10mm 5mm,clip]{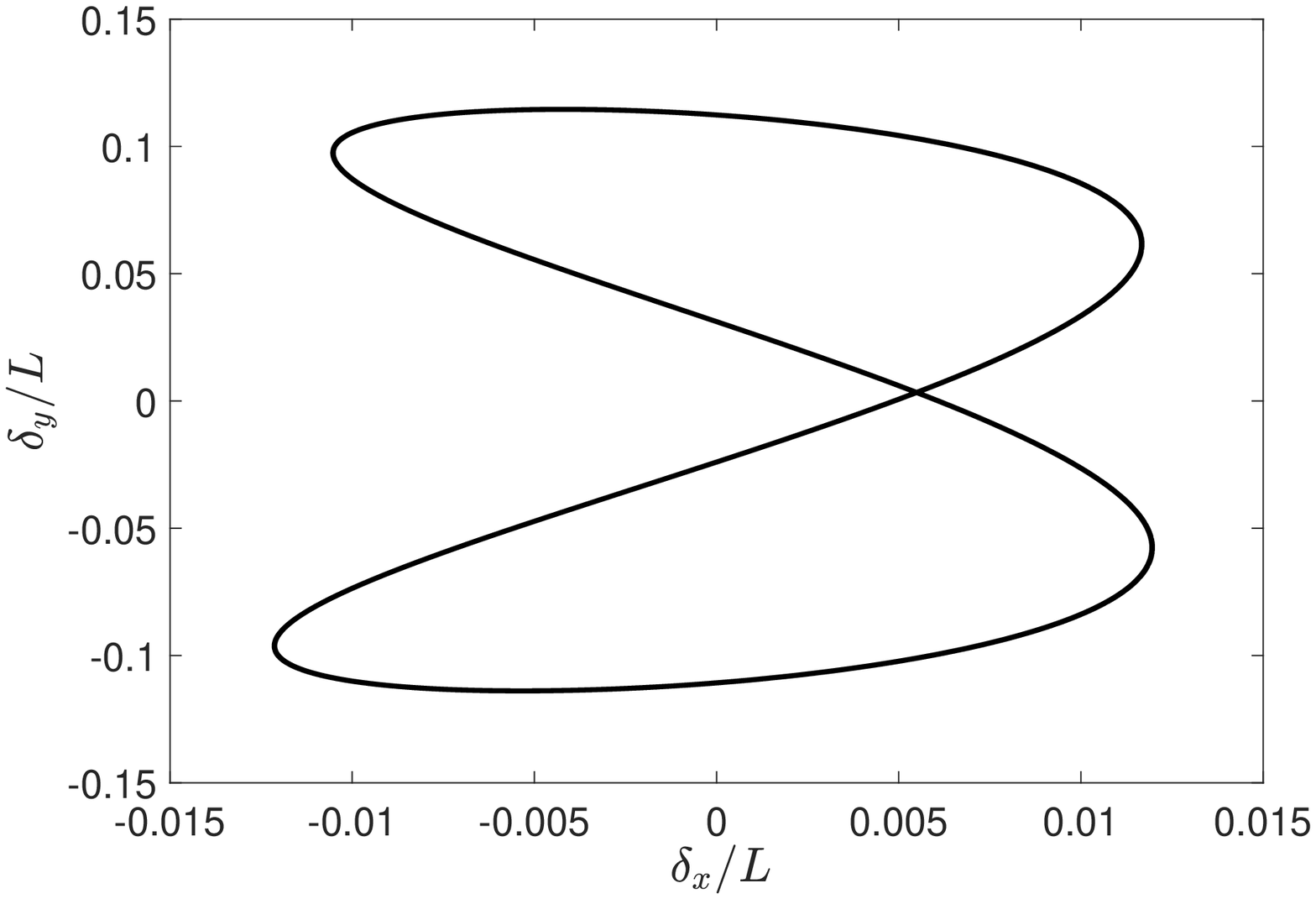}
		\end{subfigure}\\
		\begin{subfigure}{0.99\textwidth}
			\centering
			\caption{$\Delta=7200$}	
		\end{subfigure}\\
		\caption{Time history of the cross-stream (--------) and stream-wise $(---)$ displacement of the trailing edge (left), and the Lissajous curves (right) for: (a) $\Delta=0$, (b) $\Delta=2400$, (c) $\Delta=4800$ and (d) $\Delta=7200$ at a constant $Re=1000$, $m^*=0.1$, $H_\mathrm{t}/H=H_\mathrm{b}/H=0.5$ and $\beta_{\mathrm{avg}}=6000 $ ($ K_B=0.0005 $)}
		\label{tip_disp_6000}
	\end{figure}
	
	Four cases of $ \Delta =\left\{0,2400,4800,7200\right\}$ with $\beta_{\mathrm{avg}} $ = 6000 (corresponding to $ K_B=0.0005 $) for a constant Reynolds number of $ \mathrm{Re}=1000$, mass ratio $ \left(m^*\right)_{\mathrm{avg}}=0.1$ and $H_\mathrm{t}/H=H_\mathrm{b}/H=0.5$ are considered.
	Figure \ref{tip_disp_6000} presents the time history of the trailing edge displacement for the two-layered flexible plate and the corresponding Lissajous curves over a duration of ten non-dimensional time units. The first observation from these figures is that the amplitude of flapping in the cross-stream direction increases with an increase in $\Delta$, which is also clearly observed from the Lissajous curves (Fig.~\ref{tip_disp_6000},~right). Looking at the stream-wise displacement of the trailing edge, in addition to an increase in amplitude with increasing $ \Delta $, an asymmetry between consecutive oscillations is noted. A closer look at the Lissajous curves, in Fig.~\ref{tip_disp_6000}, for the cases corresponding to $ \Delta=\left\{4800,7200\right\} $, clearly reveals that the stream-wise displacement during the up-stroke is lesser than it is during the down-stroke. This difference can be attributed to the fact that the top layer is considerably more elastic in these two cases than the bottom layer, which allows for a greater degree of bending in the down-stroke compared to the up-stroke. 
	Along with information about the amplitude of displacements, the Lissajous curves in Fig.~\ref{tip_disp_6000} can also be used to infer the frequency ratio and the phase difference between the cross-stream and stream-wise displacements of the tip. In all the four cases, the shape of the curves points to a phase difference close to $ \pi/8 $ or $ 3\pi/8 $. Furthermore, as expected, the curves' shape also signifies a frequency ratio of 1:2 between the cross-stream and stream-wise displacements. 
	
	\begin{figure}
		\centering
		\begin{subfigure}{0.49\textwidth}
			\includegraphics[width=0.99\columnwidth]{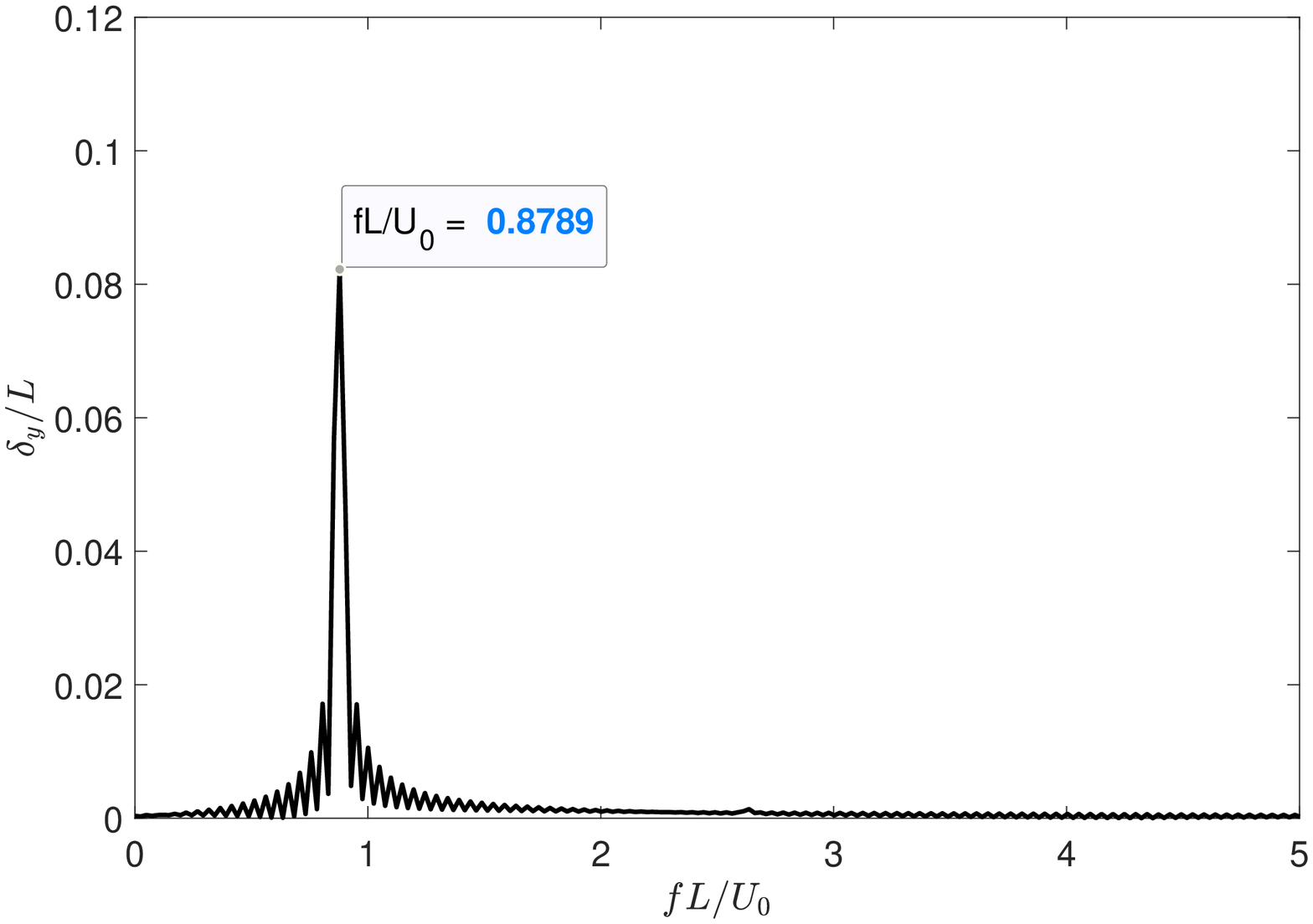}
			\caption{$ \Delta=0 $}	
		\end{subfigure}
		\begin{subfigure}{0.49\textwidth}
			\includegraphics[width=0.99\columnwidth]{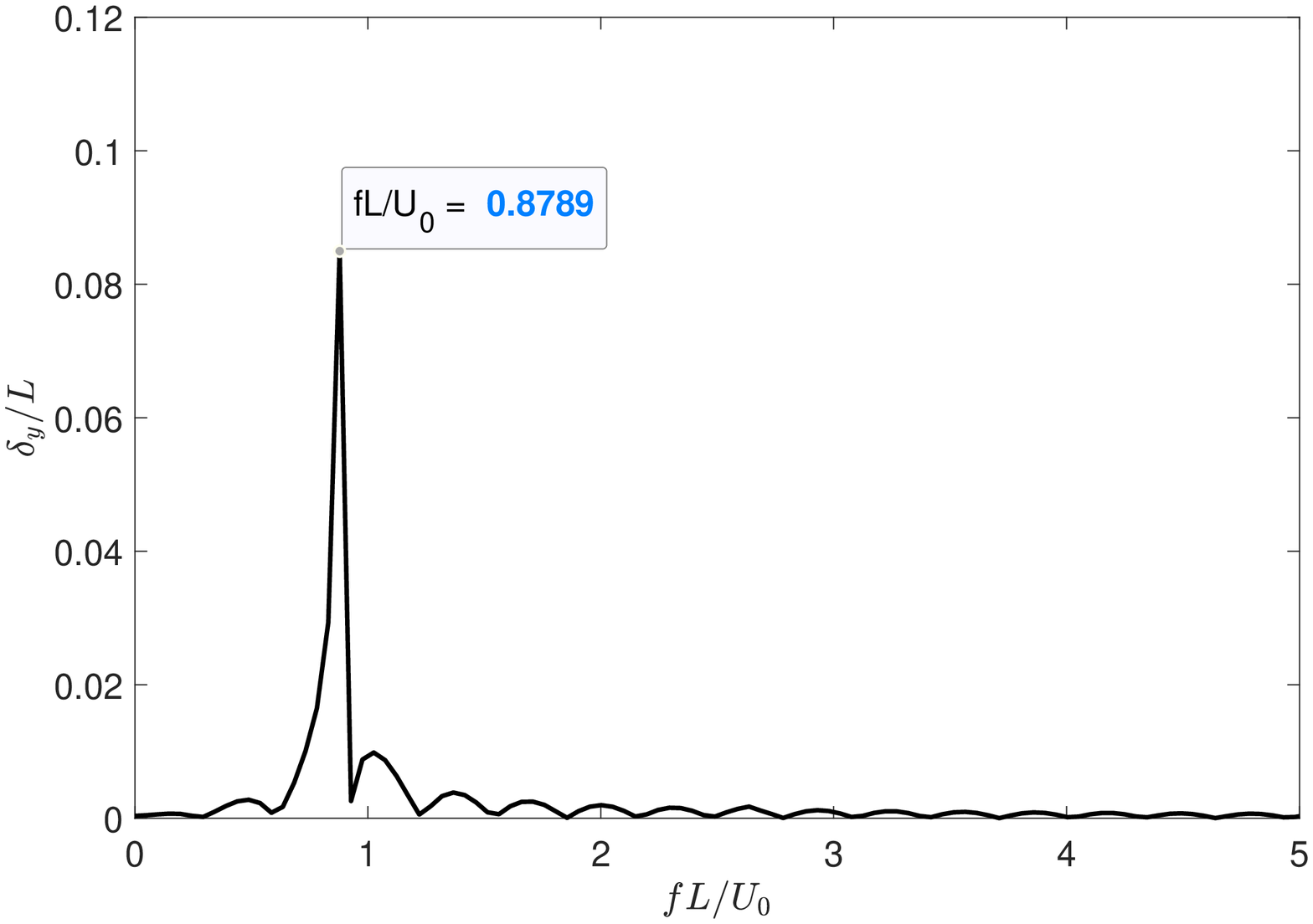}
			\caption{$ \Delta=2400 $}	
		\end{subfigure}\\
		\begin{subfigure}{0.49\textwidth}
			\includegraphics[width=0.99\columnwidth]{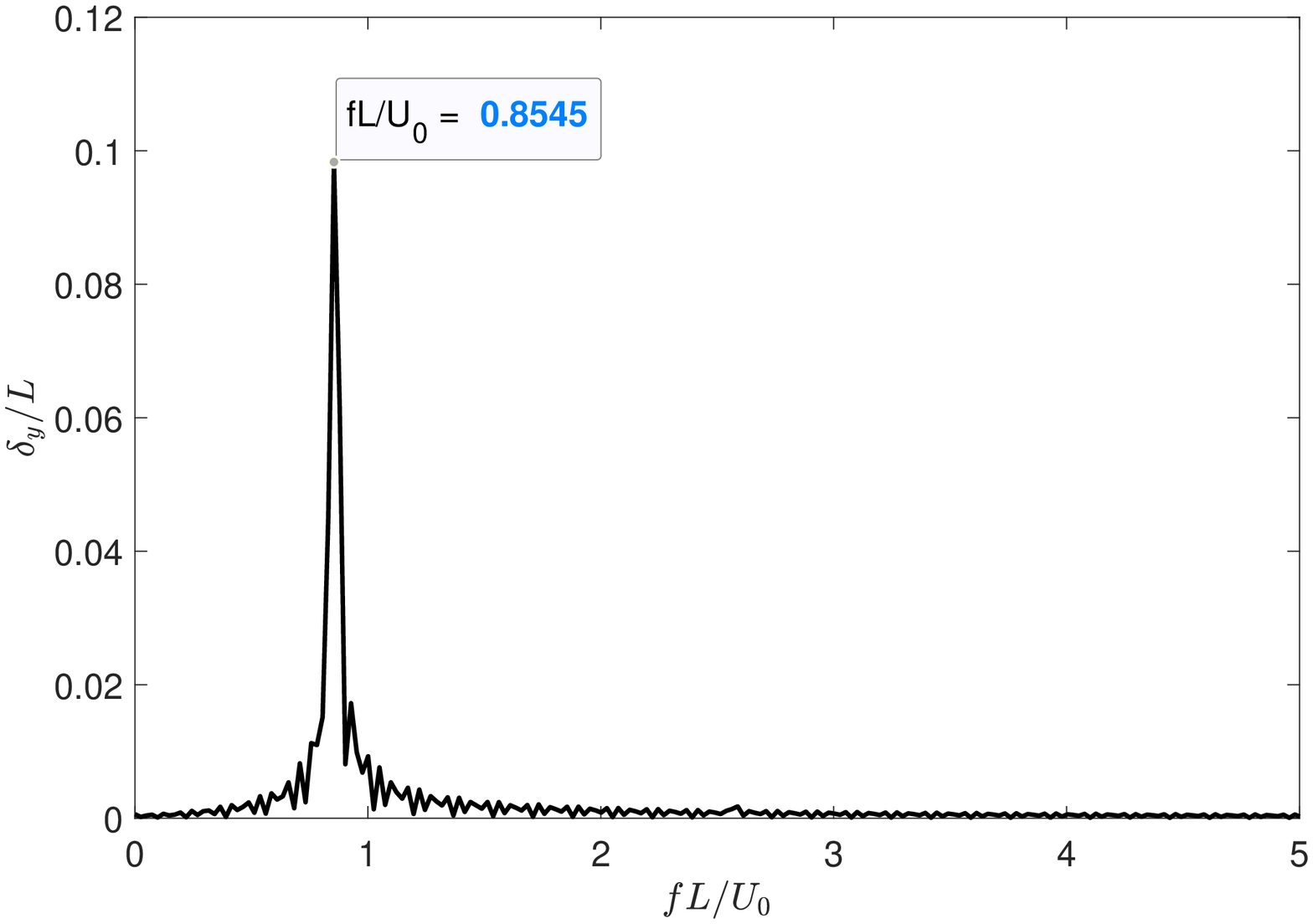}
			\caption{$ \Delta=4800 $}	
		\end{subfigure}
		\begin{subfigure}{0.49\textwidth}
			\includegraphics[width=0.99\columnwidth]{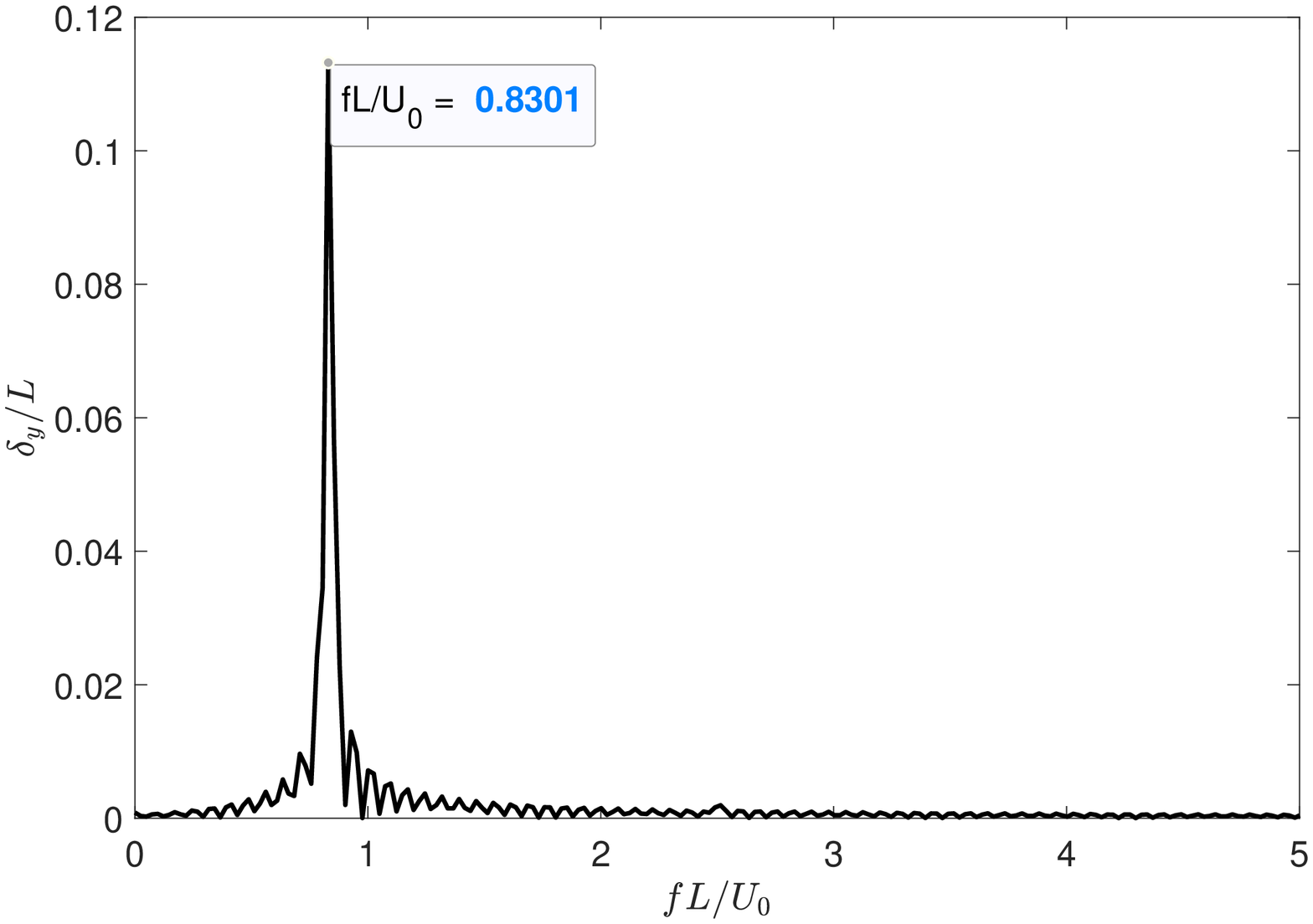}
			\caption{$ \Delta=7200 $}	
		\end{subfigure}
		\caption{Amplitude-frequency spectrum of the cross-stream displacement, with the dominant frequency marked, for: (a) $\Delta=0$, (b) $\Delta=2400$, (c) $\Delta=4800$ and (d) $\Delta=7200$ at a constant $Re=1000$, $m^*=0.1$, $H_\mathrm{t}/H=H_\mathrm{b}/H=0.5$ and $\beta_{\mathrm{avg}}=6000 $ ($ K_B=0.0005 $)}
		\label{fft_6000}
	\end{figure}
	
	\begin{figure}
		\centering	
		\begin{subfigure}{0.49\textwidth}
			\includegraphics[width=0.99\columnwidth]{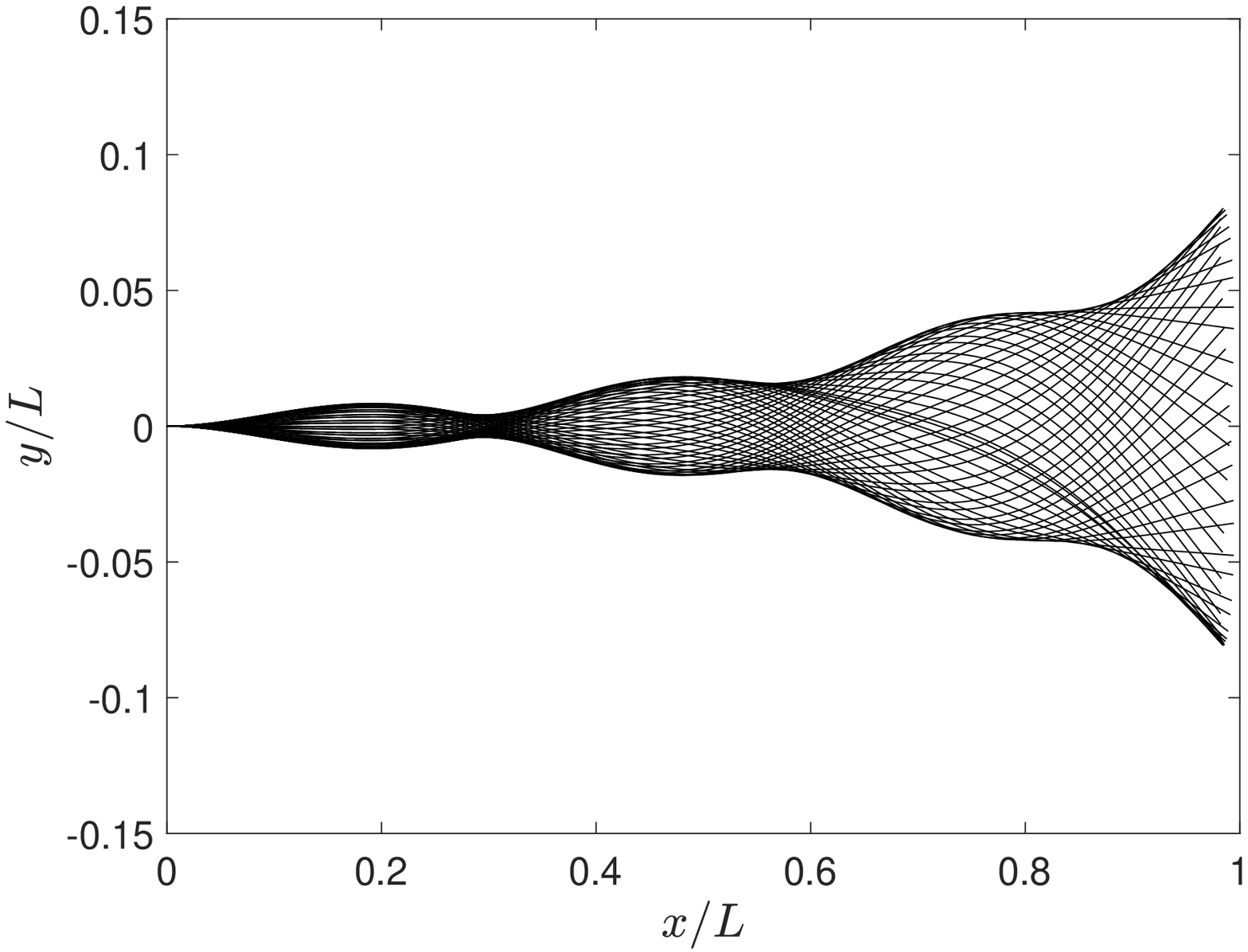}
			\caption{$\Delta=0$}
		\end{subfigure}
		\begin{subfigure}{0.49\textwidth}
			\includegraphics[width=0.99\columnwidth]{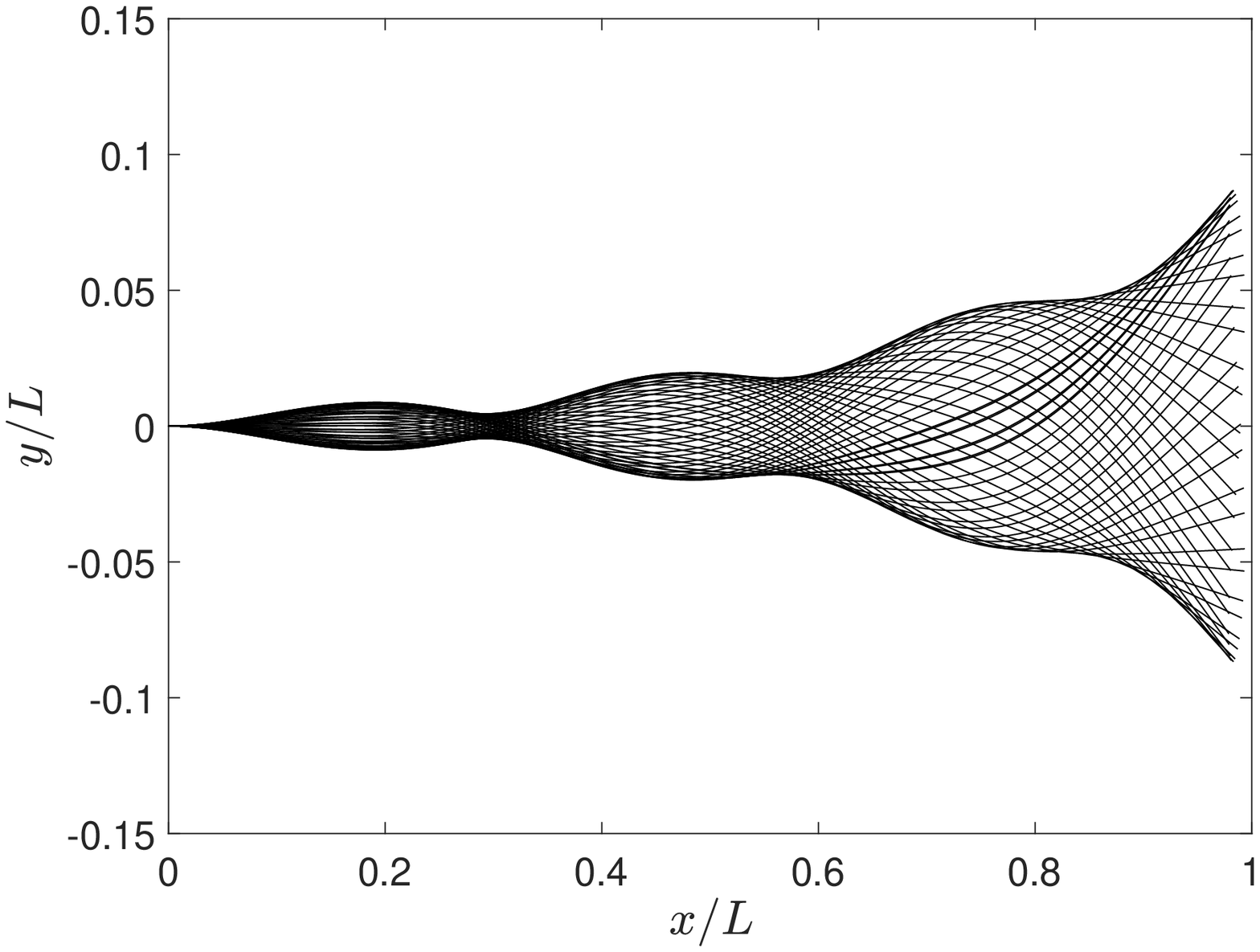}
			\caption{$\Delta=2400$}	
		\end{subfigure}\\
		\begin{subfigure}{0.49\textwidth}
			\includegraphics[width=0.99\columnwidth]{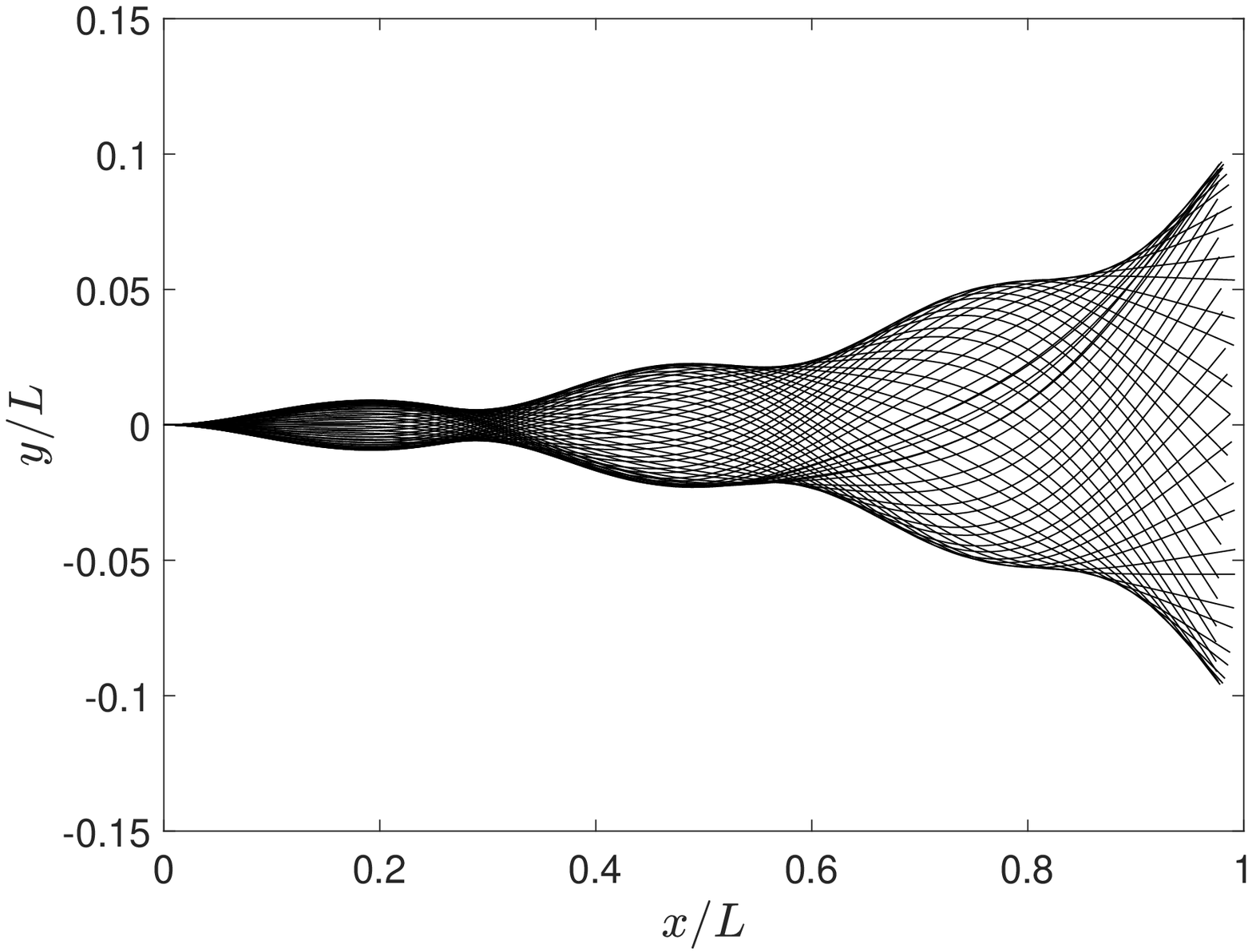}
			\caption{$\Delta=4800$}	
		\end{subfigure}
		\begin{subfigure}{0.49\textwidth}
			\includegraphics[width=0.99\columnwidth]{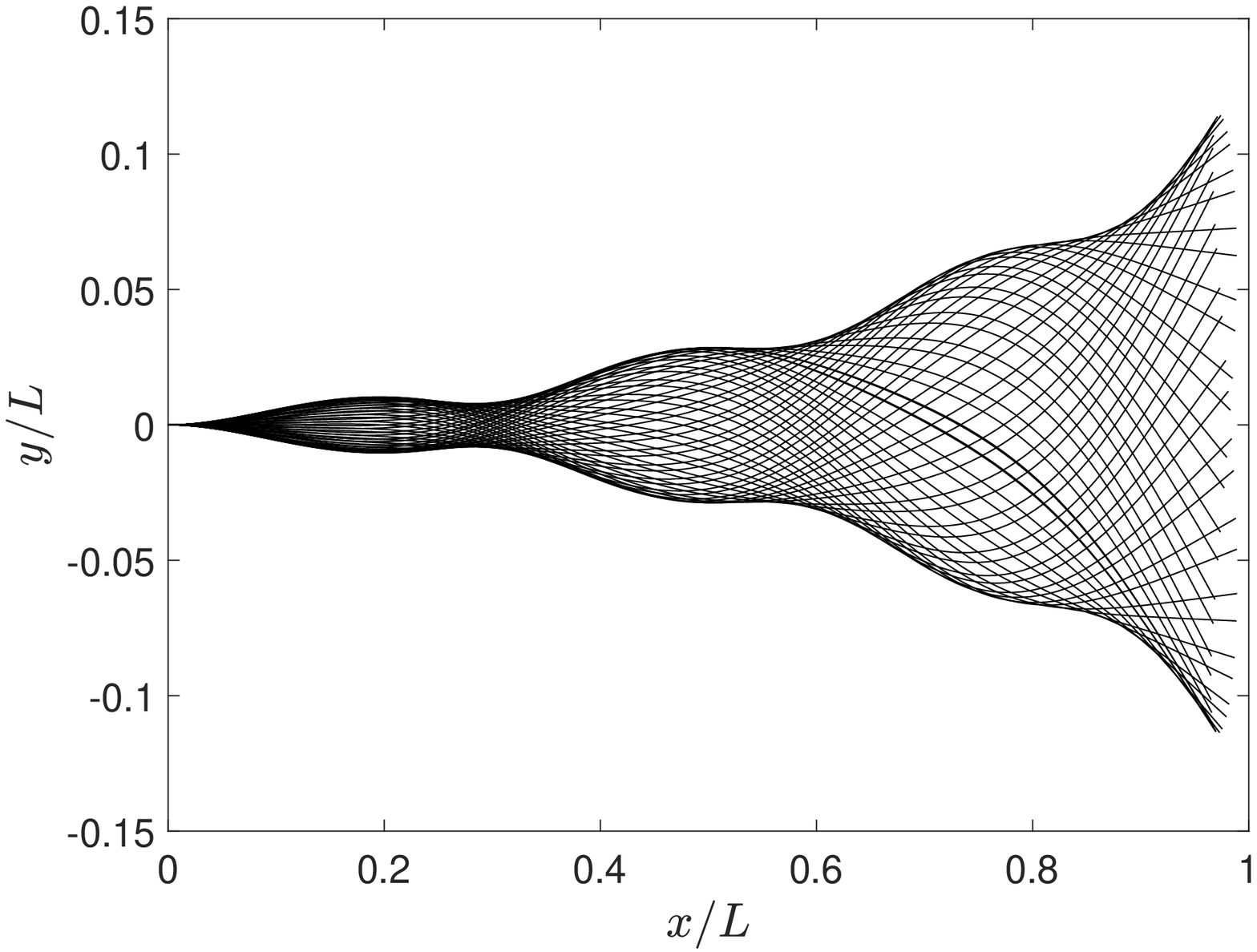}
			\caption{$\Delta=7200$}	
		\end{subfigure}
		\caption{Full-body response of the two-layered flexible plate over a complete oscillation cycle for : (a) $\Delta=0$, (b) $\Delta=2400$, (c) $\Delta=4800$ and (d) $\Delta=7200$ at a constant $Re=1000$, $m^*=0.1$, $H_\mathrm{t}/H=H_\mathrm{b}/H=0.5$ and $\beta_{\mathrm{avg}}=6000 $ ($ K_B=0.0005 $)}
		\label{profile_6000}
	\end{figure}
	
	Next, the frequency of flapping for each of these cases is investigated, and the corresponding FFT (Fast Fourier Transform) plots are shown in Fig.~\ref{fft_6000}. It can be seen here that the value of the dominant frequency shows a decreasing trend with increasing $ \Delta $. Following this, in Fig.~\ref{profile_6000}, the full-body response of the flexible plate over a complete cycle starting at an extremum is shown. These plate profile plots reinforce the increase in the amplitude of the trailing edge as $ \Delta $ is increased. Interestingly, it was noted that although the dominant frequency of flapping decreases, the mode of flapping remains constant across the four cases (Fig.~\ref{profile_6000}). This is in stark contrast to the flapping dynamics of a single-layered plate, where a change in dominant frequency is always accompanied by a change in the mode of oscillation, as illustrated in \cite{gurugubelli_JFM}.  
	
	\begin{figure}
		\centering
		\begin{subfigure}{0.49\textwidth}
			\includegraphics[width=0.99\columnwidth]{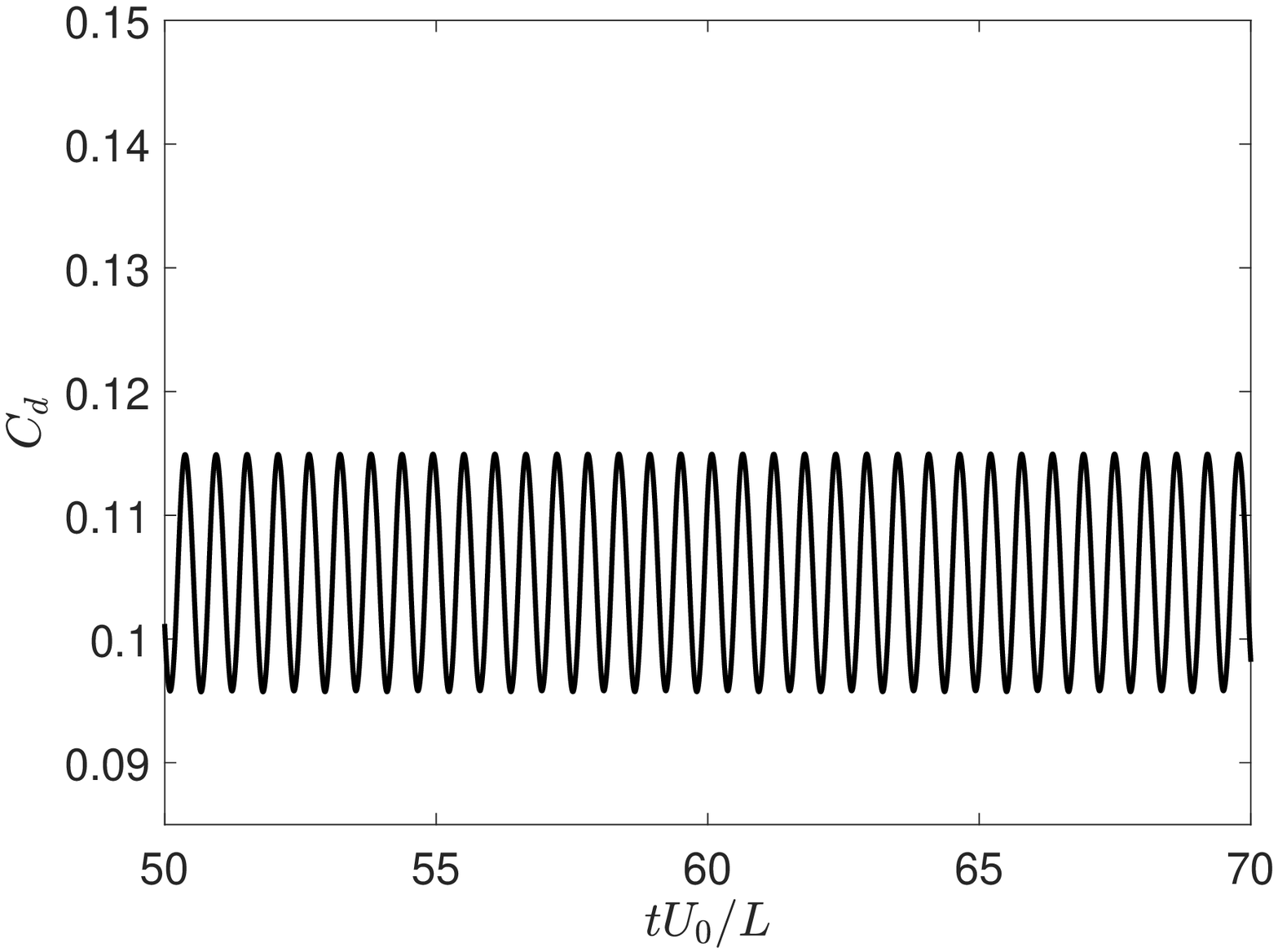}
			\caption{$ \Delta=0 $}	
		\end{subfigure}
		\begin{subfigure}{0.49\textwidth}
			\includegraphics[width=0.99\columnwidth]{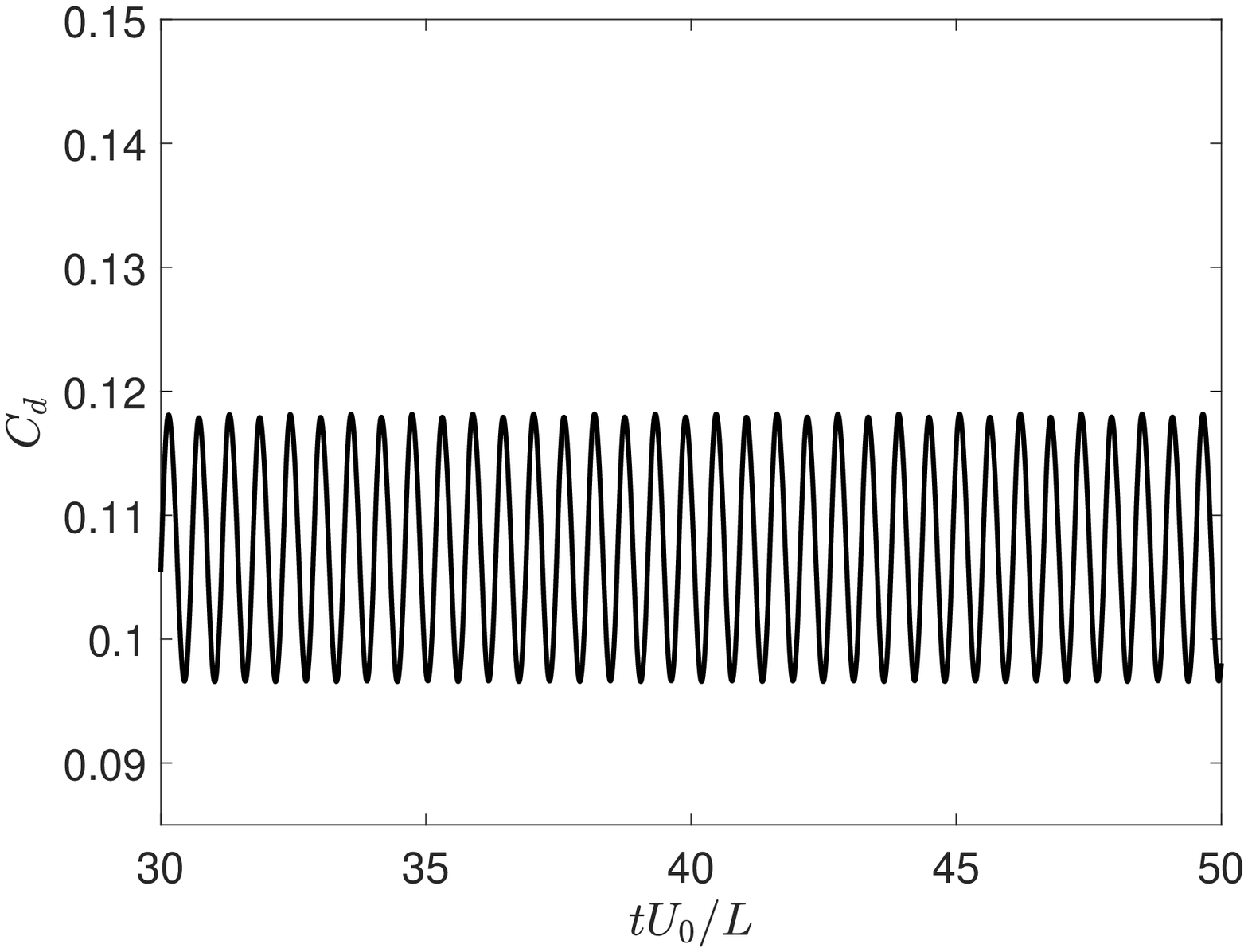}
			\caption{$ \Delta=2400 $}	
		\end{subfigure}\\
		\begin{subfigure}{0.49\textwidth}
			\includegraphics[width=0.99\columnwidth]{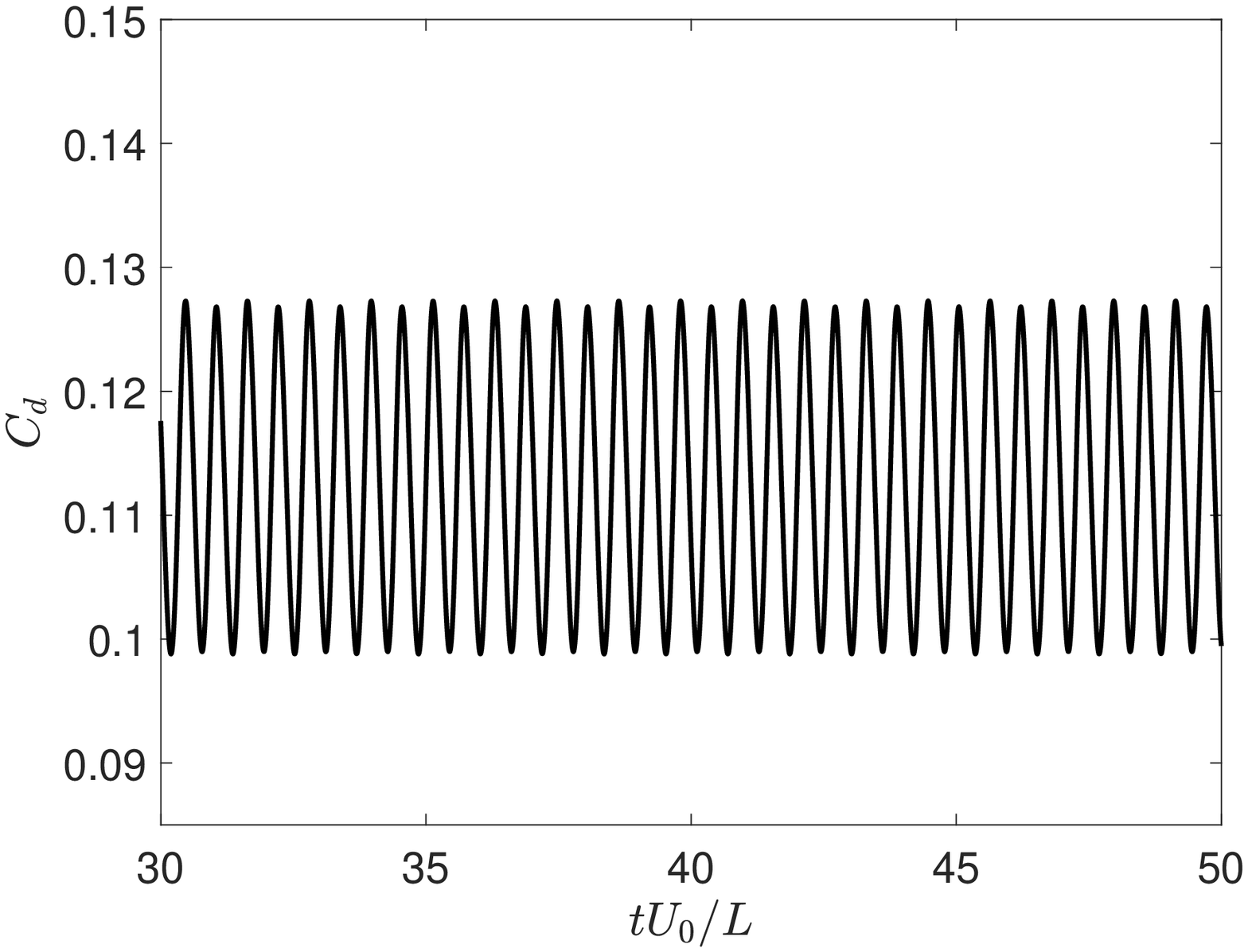}
			\caption{$ \Delta=4800 $}	
		\end{subfigure}
		\begin{subfigure}{0.49\textwidth}
			\includegraphics[width=0.99\columnwidth]{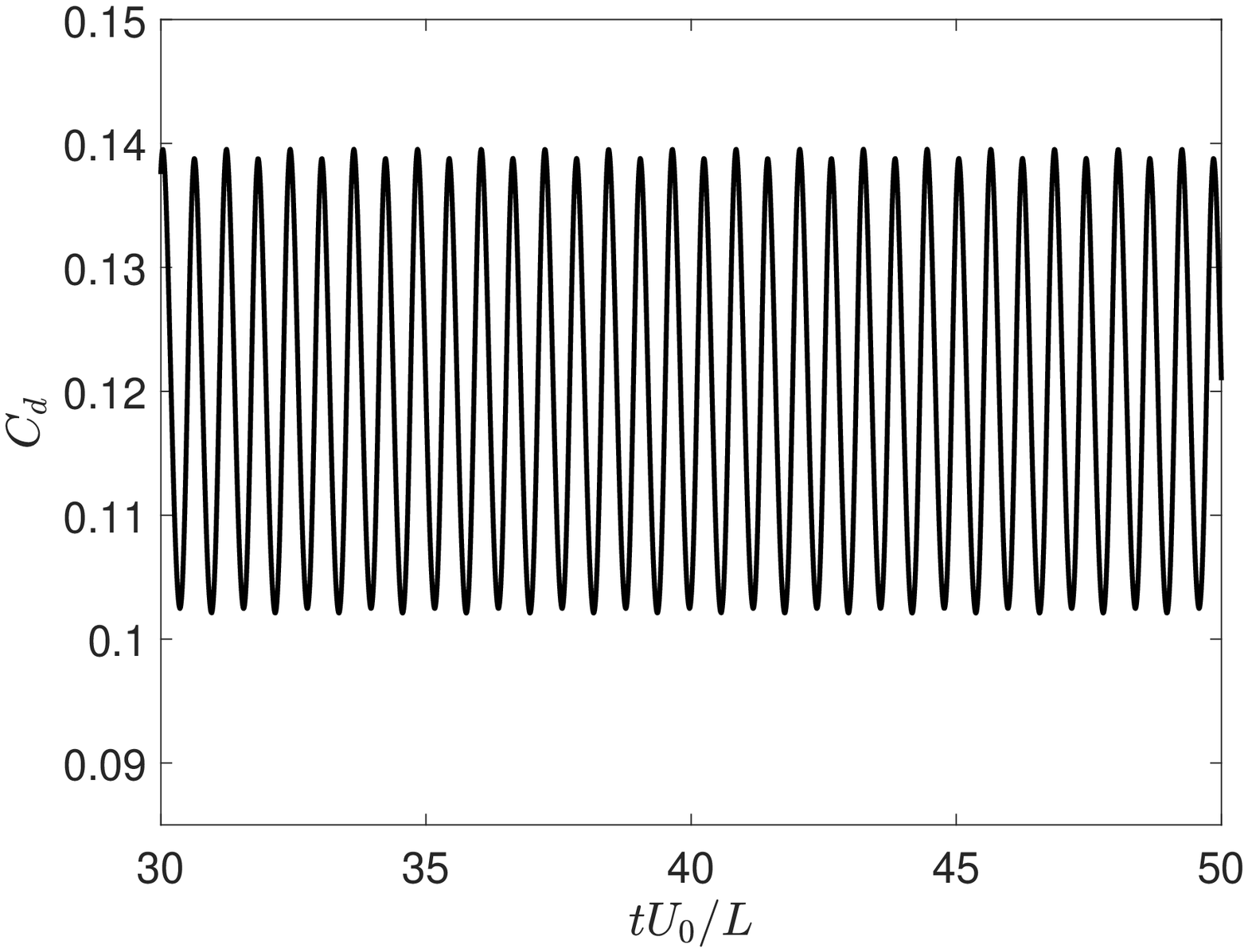}
			\caption{$ \Delta=7200 $}	
		\end{subfigure}
		\caption{Time histories of the drag coefficients over 20 non-dimensional time units for : (a) $\Delta=0$, (b) $\Delta=2400$, (c) $\Delta=4800$ and (d) $\Delta=7200$ at a constant $Re=1000$, $m^*=0.1$, $H_\mathrm{t}/H=H_\mathrm{b}/H=0.5$ and $\beta_{\mathrm{avg}}=6000 $ ($ K_B=0.0005 $)}
		\label{C_d_6000}
	\end{figure}
	
	\begin{figure}
		\centering
		\begin{subfigure}{0.49\textwidth}
			\includegraphics[width=0.99\columnwidth]{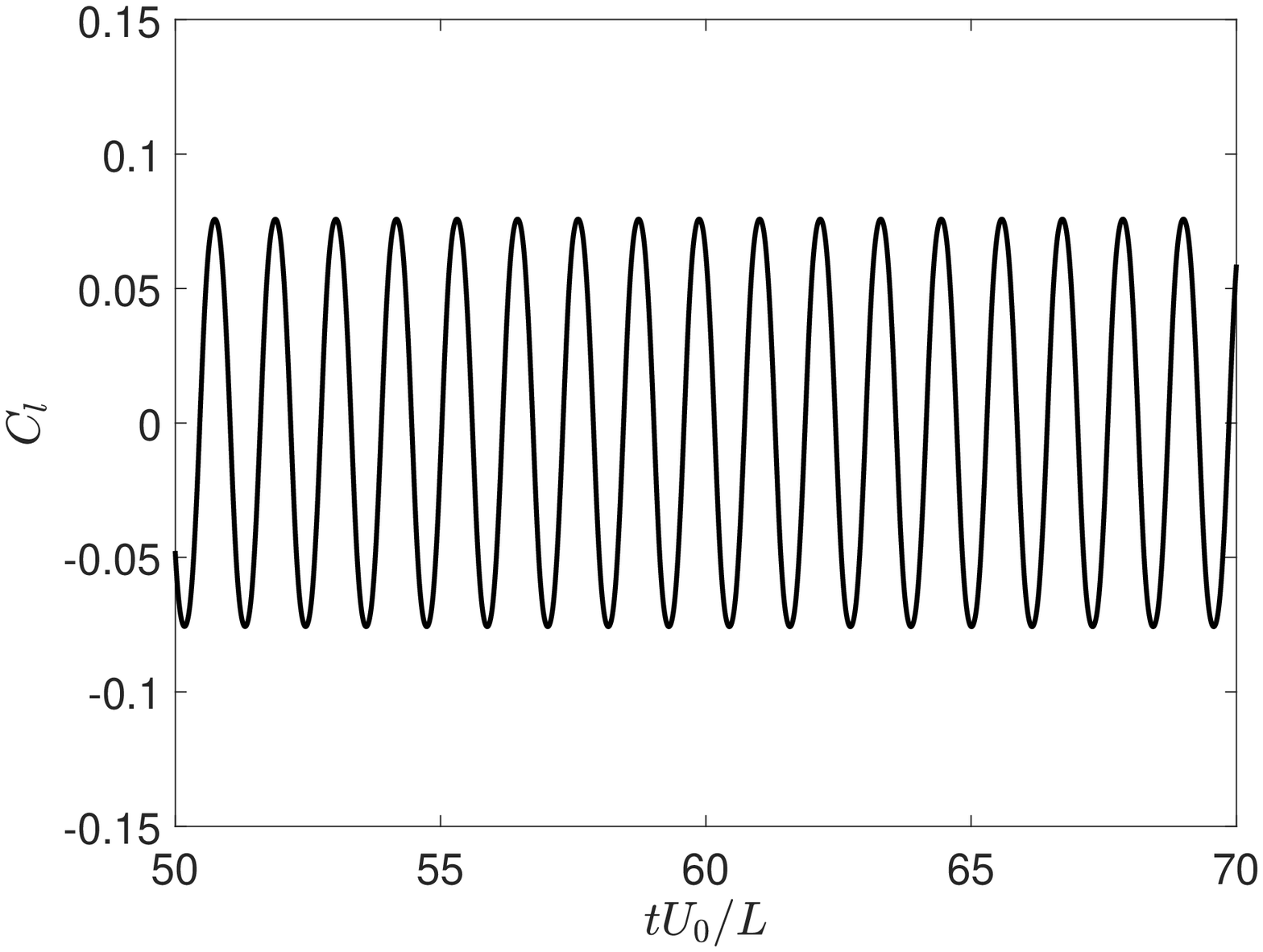}
			\caption{$ \Delta=0 $}	
		\end{subfigure}
		\begin{subfigure}{0.49\textwidth}
			\includegraphics[width=0.99\columnwidth]{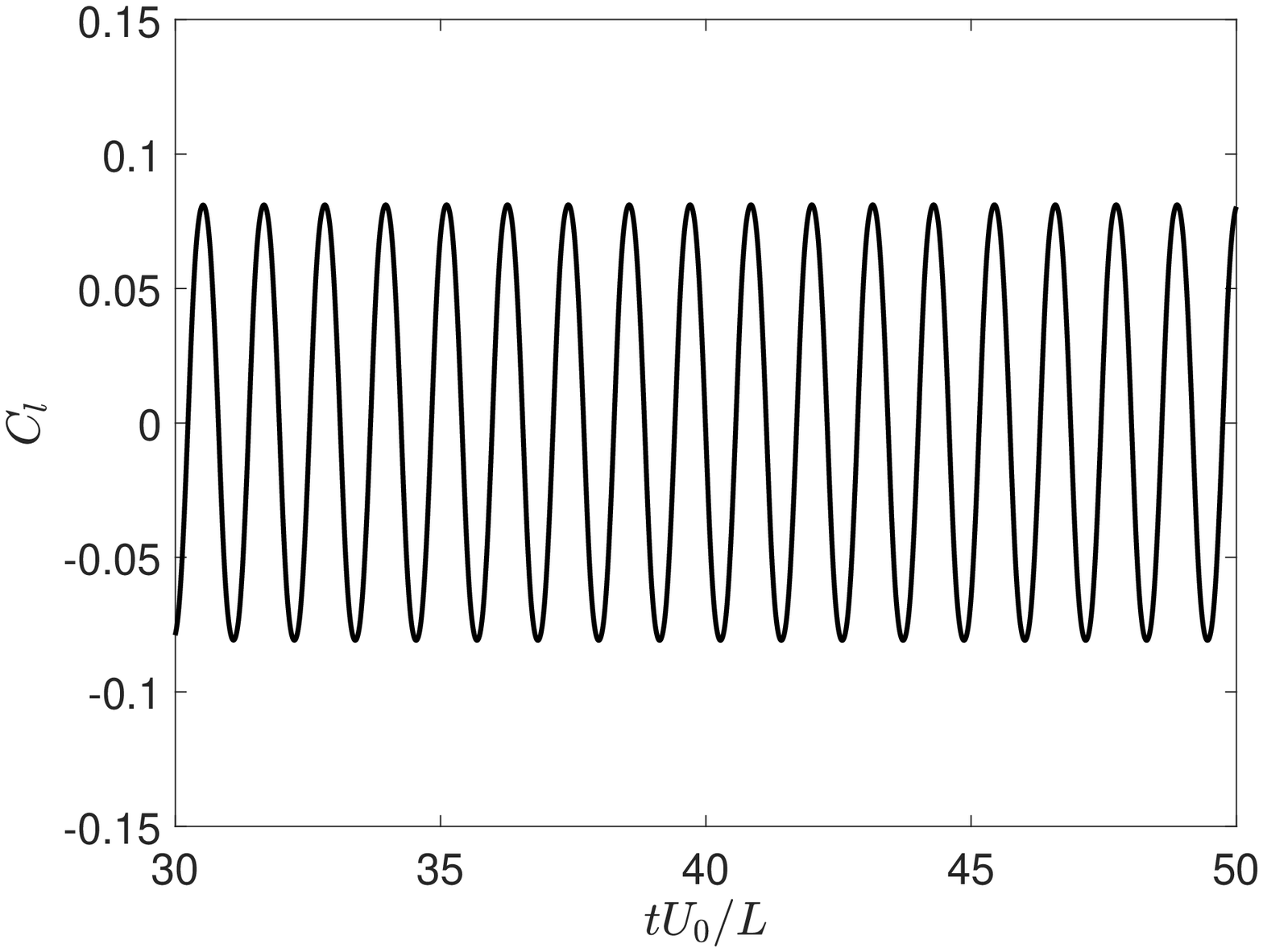}
			\caption{$ \Delta=2400 $}	
		\end{subfigure}\\
		\begin{subfigure}{0.49\textwidth}
			\includegraphics[width=0.99\columnwidth]{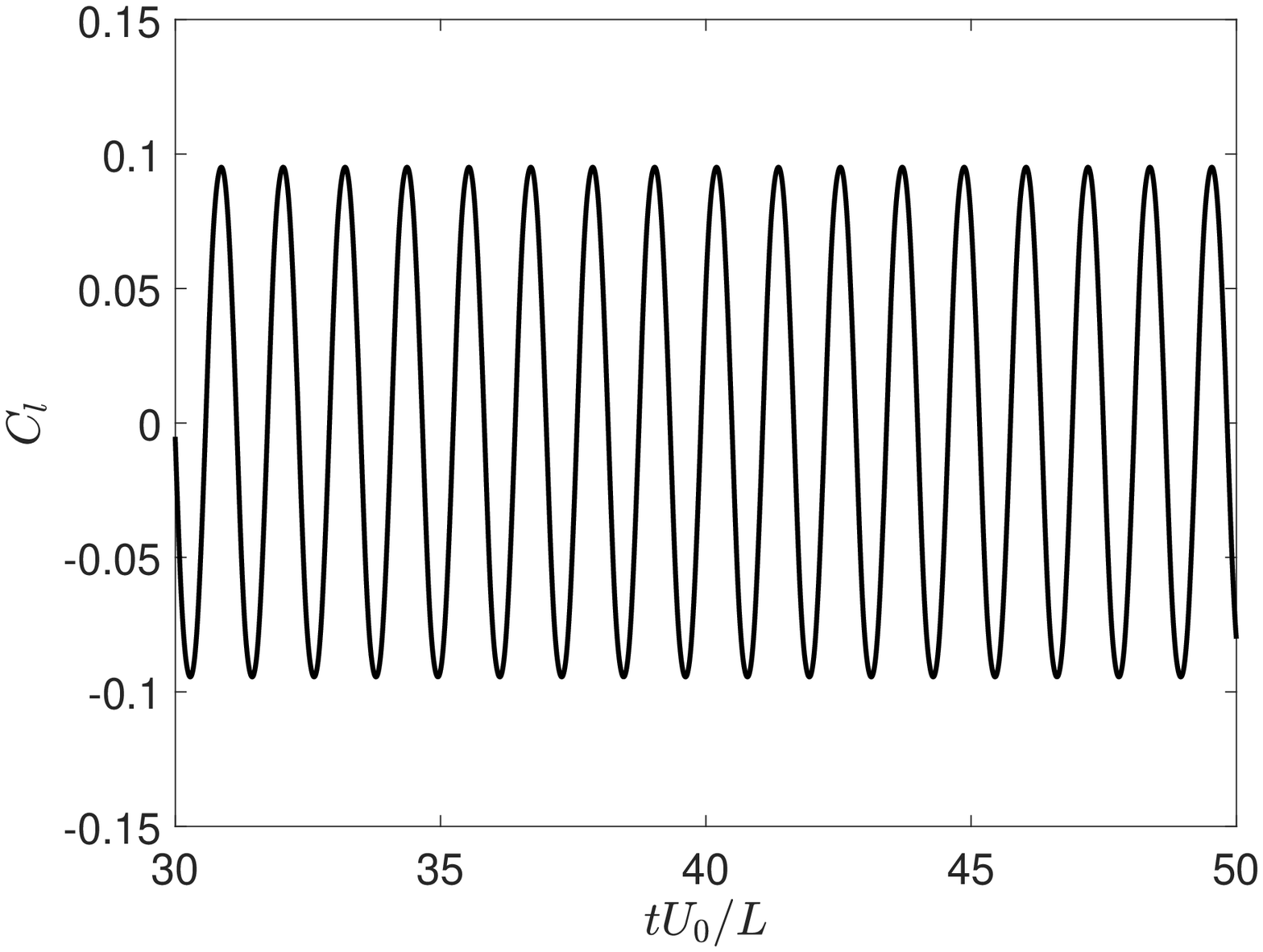}
			\caption{$ \Delta=4800 $}	
		\end{subfigure}
		\begin{subfigure}{0.49\textwidth}
			\includegraphics[width=0.99\columnwidth]{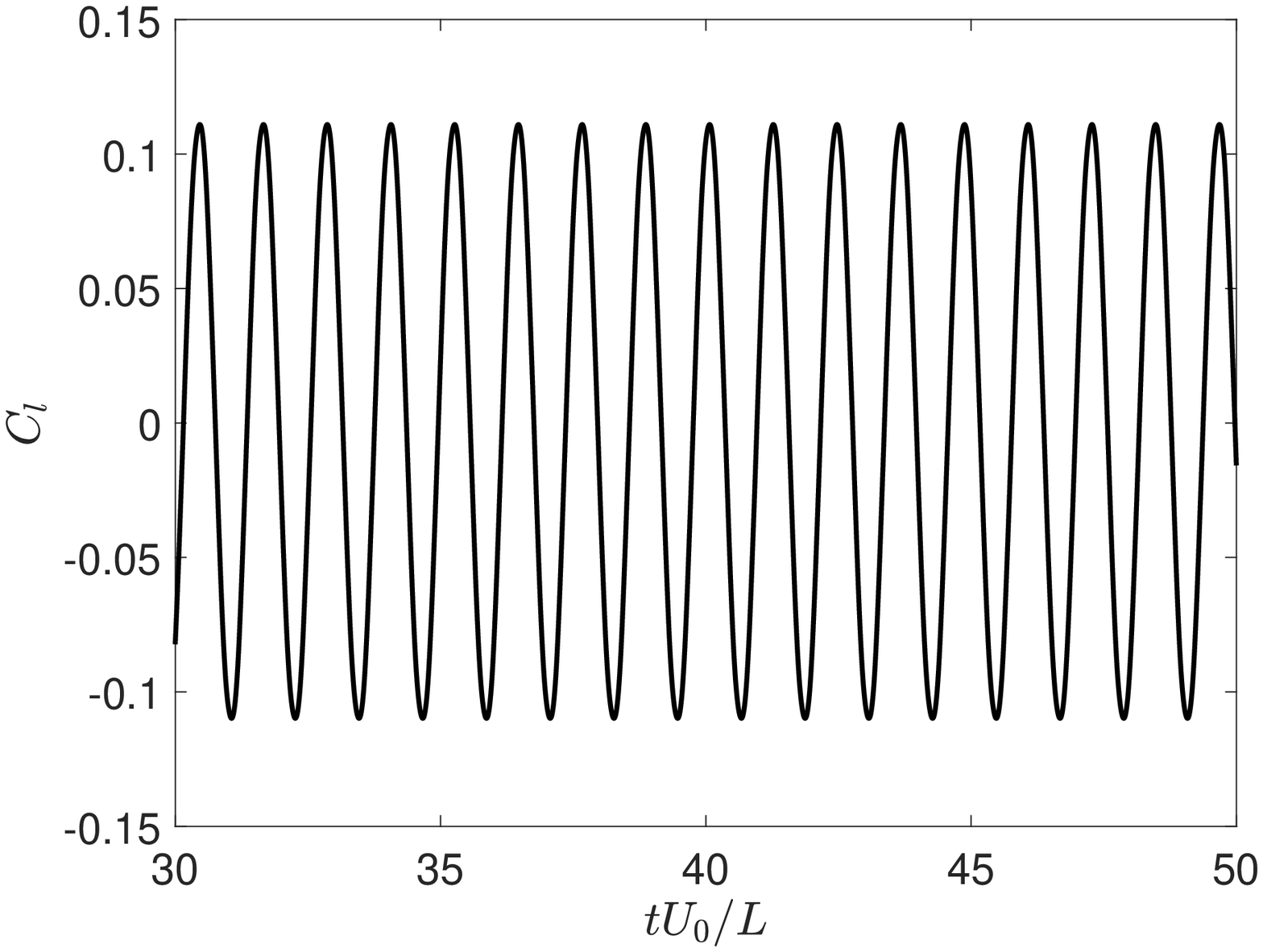}
			\caption{$ \Delta=7200 $}	
		\end{subfigure}
		\caption{Time histories of the lift coefficients over 20 non-dimensional time units for : (a) $\Delta=0$, (b) $\Delta=2400$, (c) $\Delta=4800$ and (d) $\Delta=7200$ at a constant $Re=1000$, $m^*=0.1$, $H_\mathrm{t}/H=H_\mathrm{b}/H=0.5$ and $\beta_{\mathrm{avg}}=6000 $ ($ K_B=0.0005 $)}
		\label{C_l_6000}
	\end{figure}
	
\begin{figure}
		\centering
		\includegraphics[width=0.75\columnwidth]{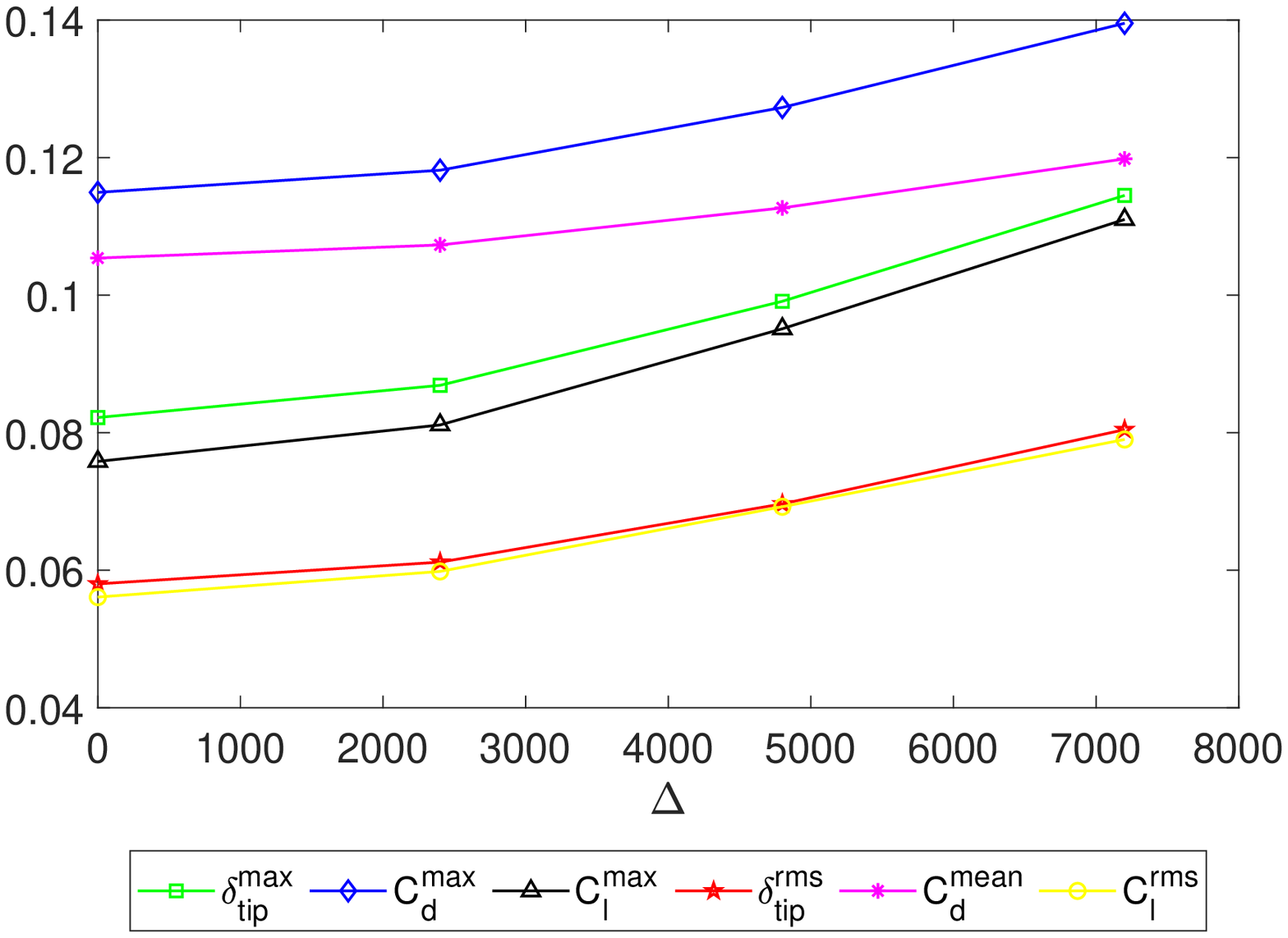}	
		\caption{Graphical representation of (a) maximum trailing edge displacement, (b) maximum drag coefficient, (c) maximum lift coefficient, (d) root mean square value of trailing edge displacement, (e) mean value of drag coefficient and (f) root mean square value of lift coefficient as a function of $ \Delta $ for the cases at a constant $Re=1000$, $m^*=0.1$, $H_\mathrm{t}/H=H_\mathrm{b}/H=0.5$ and $\beta_{\mathrm{avg}}=6000 $ ($ K_B=0.0005 $)}
		\label{summary_plot}
	\end{figure}

	The forces on the plate are studied by calculating the drag and lift coefficients, and these plots are shown in Figs.~\ref{C_d_6000}-\ref{C_l_6000}, respectively. Figure \ref{C_d_6000} clearly illustrates that both the mean value and the amplitude of the drag coefficient increases substantially with increasing $ \Delta $. It can also be noticed that the variation of drag coefficient is slightly less regular for the latter cases $ \left(\Delta=\left\{4800,7200\right\}\right) $. Figure~\ref{C_l_6000} shows that the lift coefficient increases with an increase in $ \Delta $. It is also worth noting that the RMS (Root Mean Square) value of the lift coefficient and the trailing edge displacement also follow an increasing trend with increasing $ \Delta $ as depicted in Fig.~\ref{summary_plot}.
	
	\begin{figure}
		\centering
		\begin{subfigure}{0.99\textwidth}
			\includegraphics[width=0.49\columnwidth]{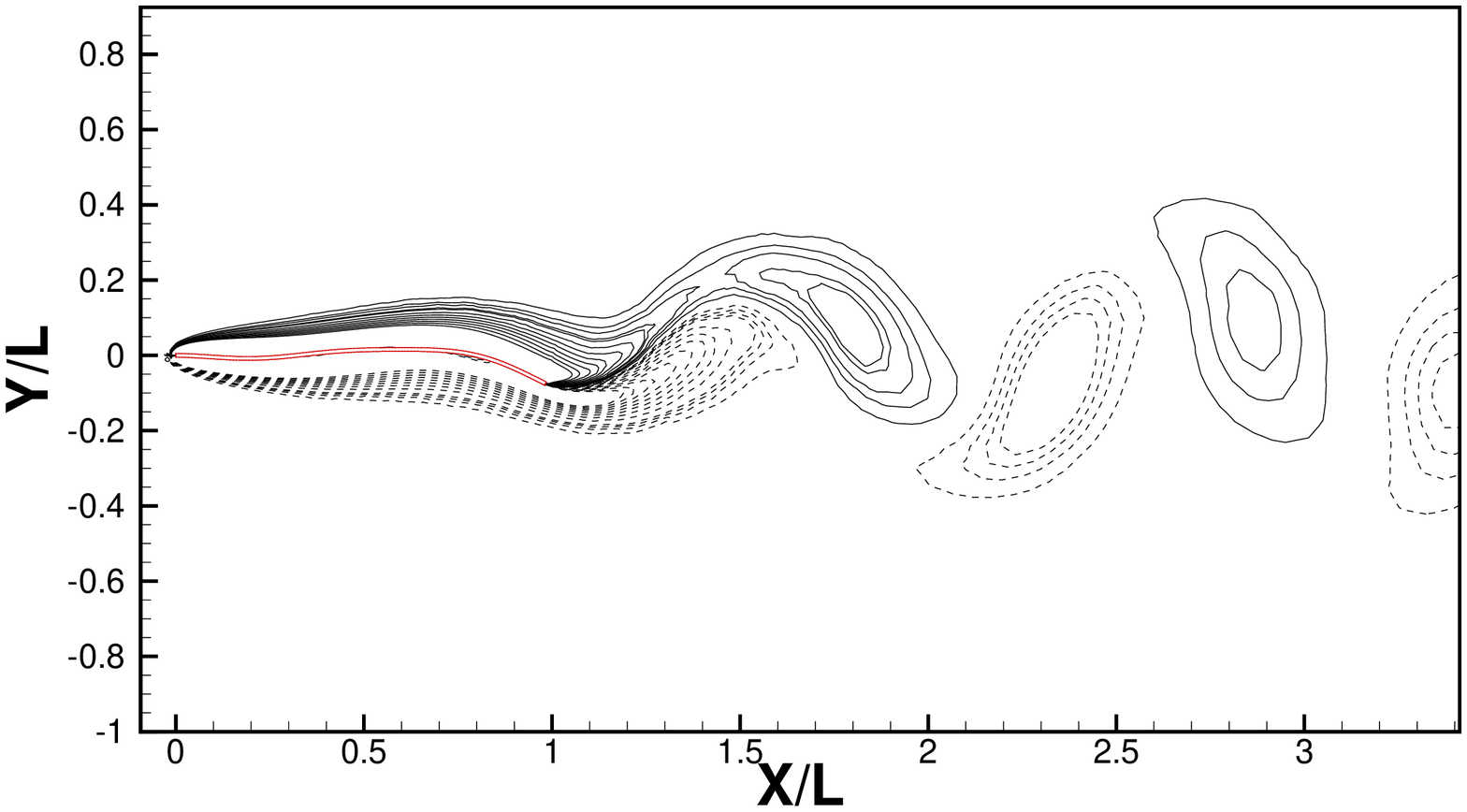}
			\includegraphics[width=0.49\columnwidth]{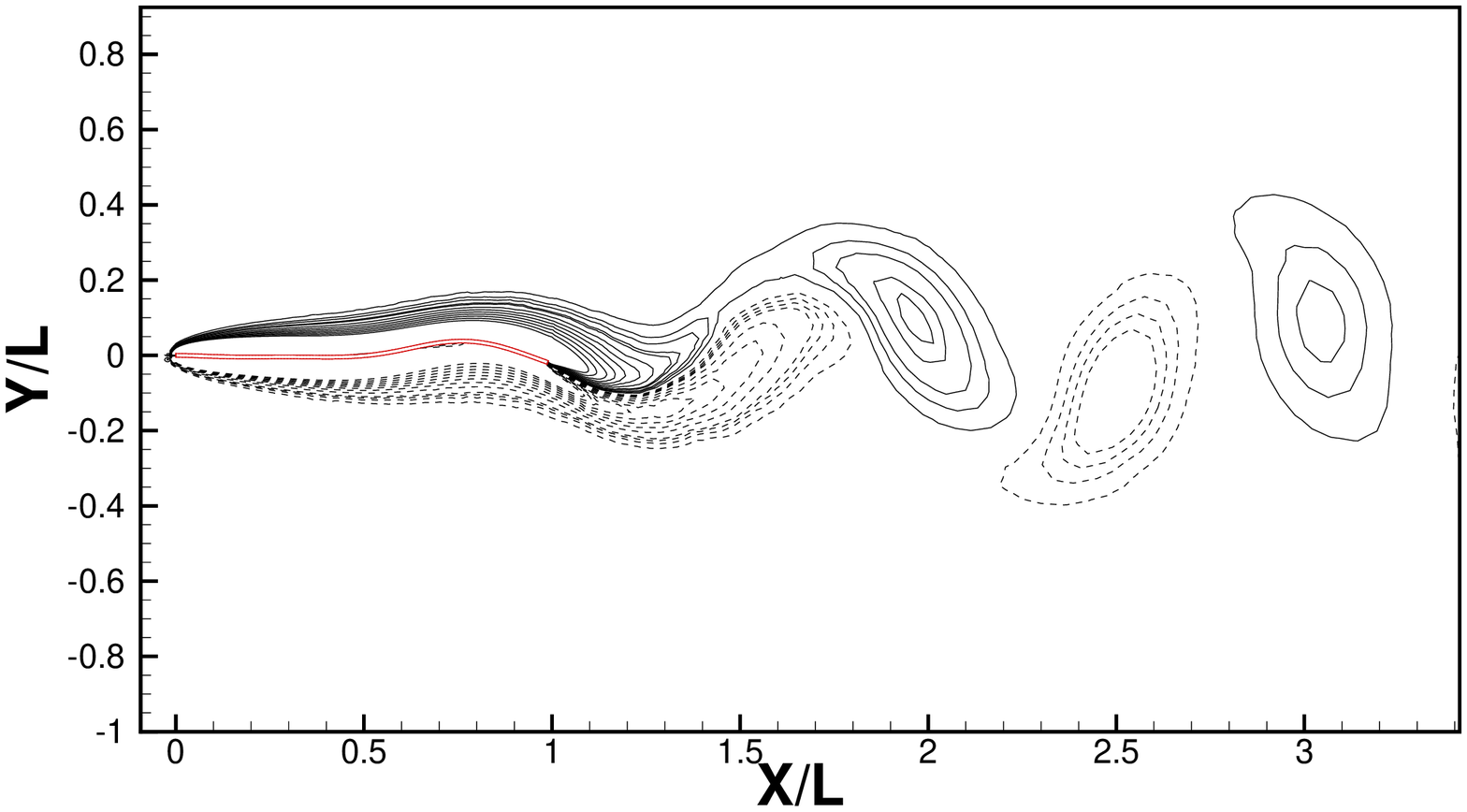}\\
			\includegraphics[width=0.49\columnwidth]{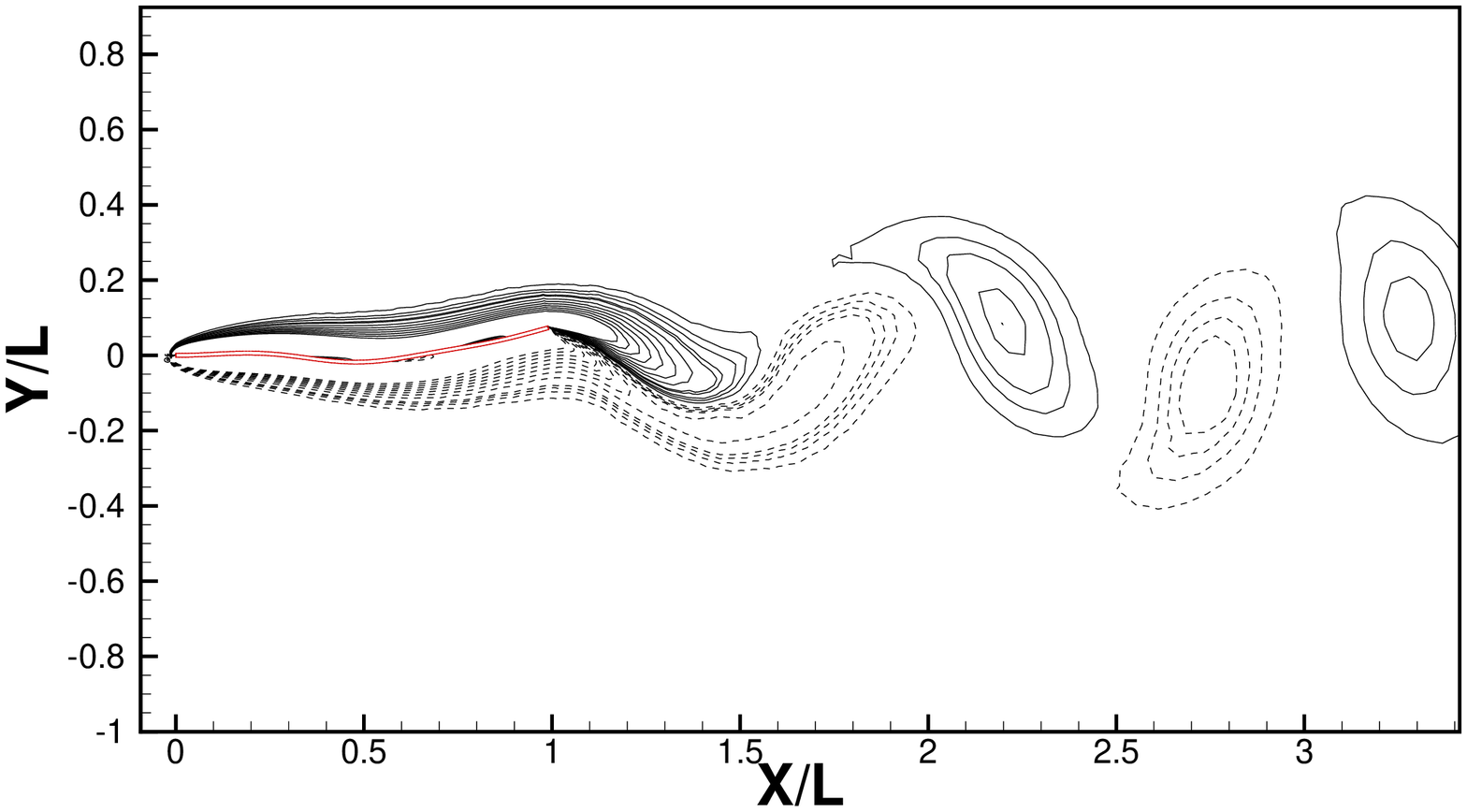}
			\includegraphics[width=0.49\columnwidth]{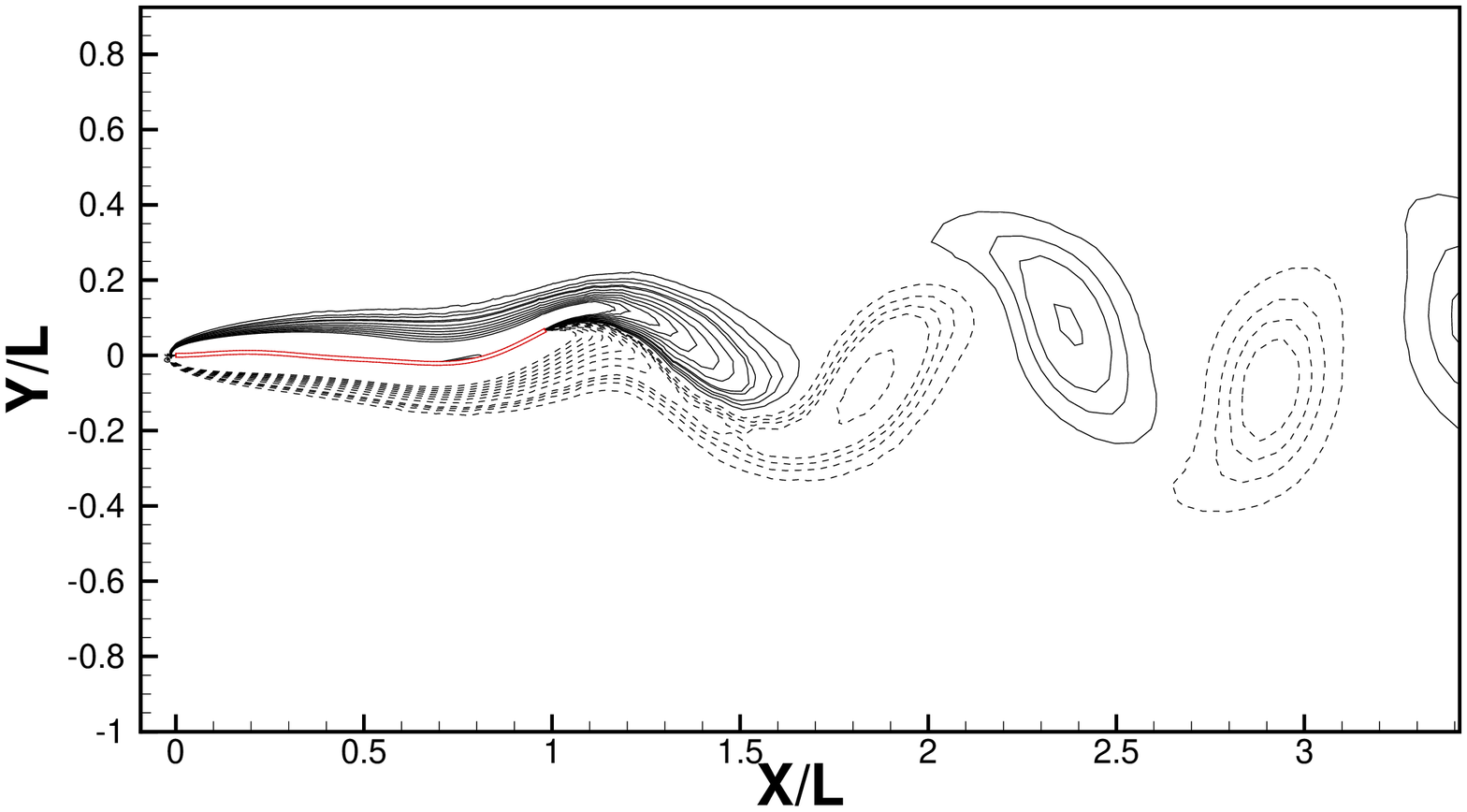}
			\caption{$ \Delta=0 $}	
		\end{subfigure}\\
		\begin{subfigure}{0.99\textwidth}
			\includegraphics[width=0.49\columnwidth]{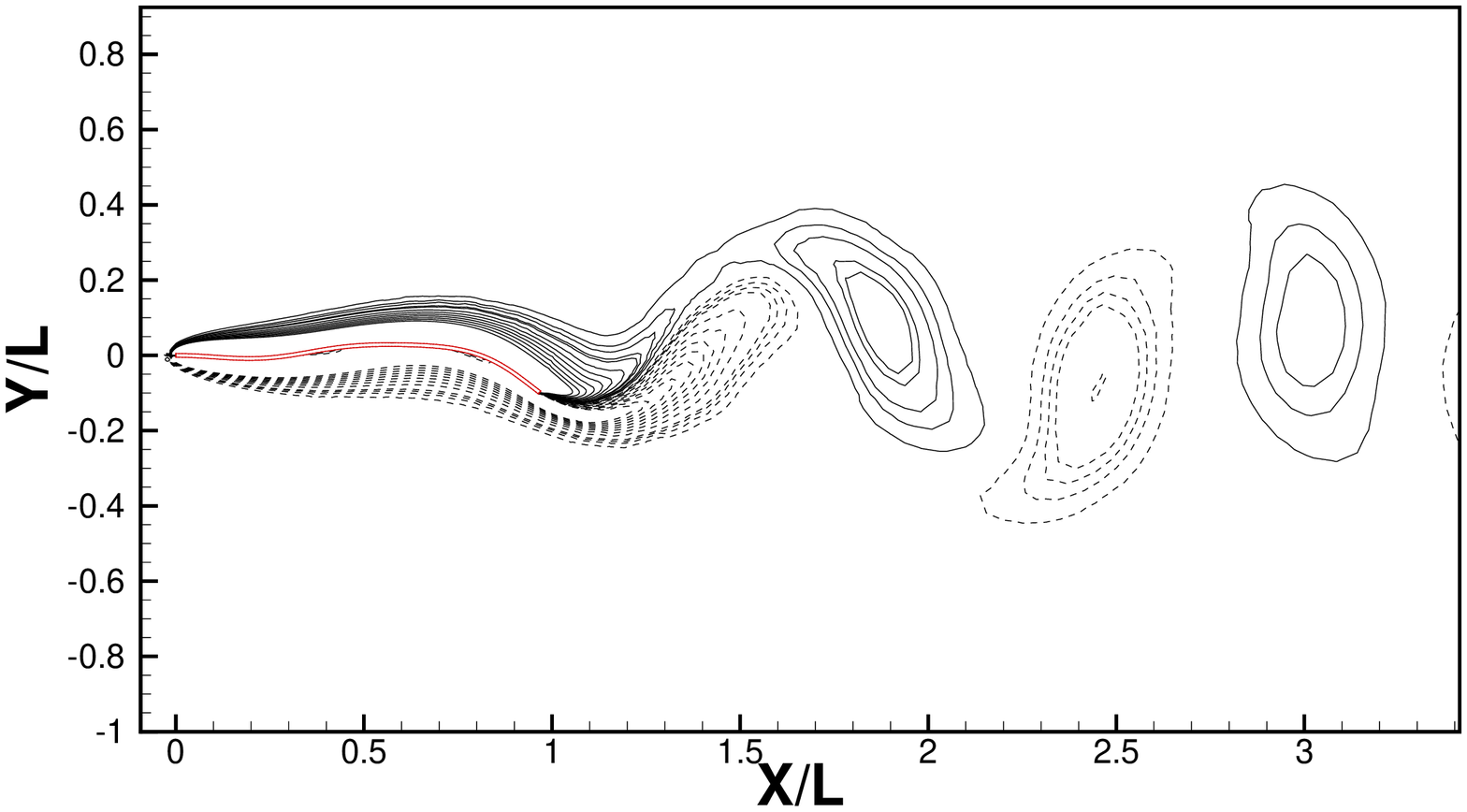}
			\includegraphics[width=0.49\columnwidth]{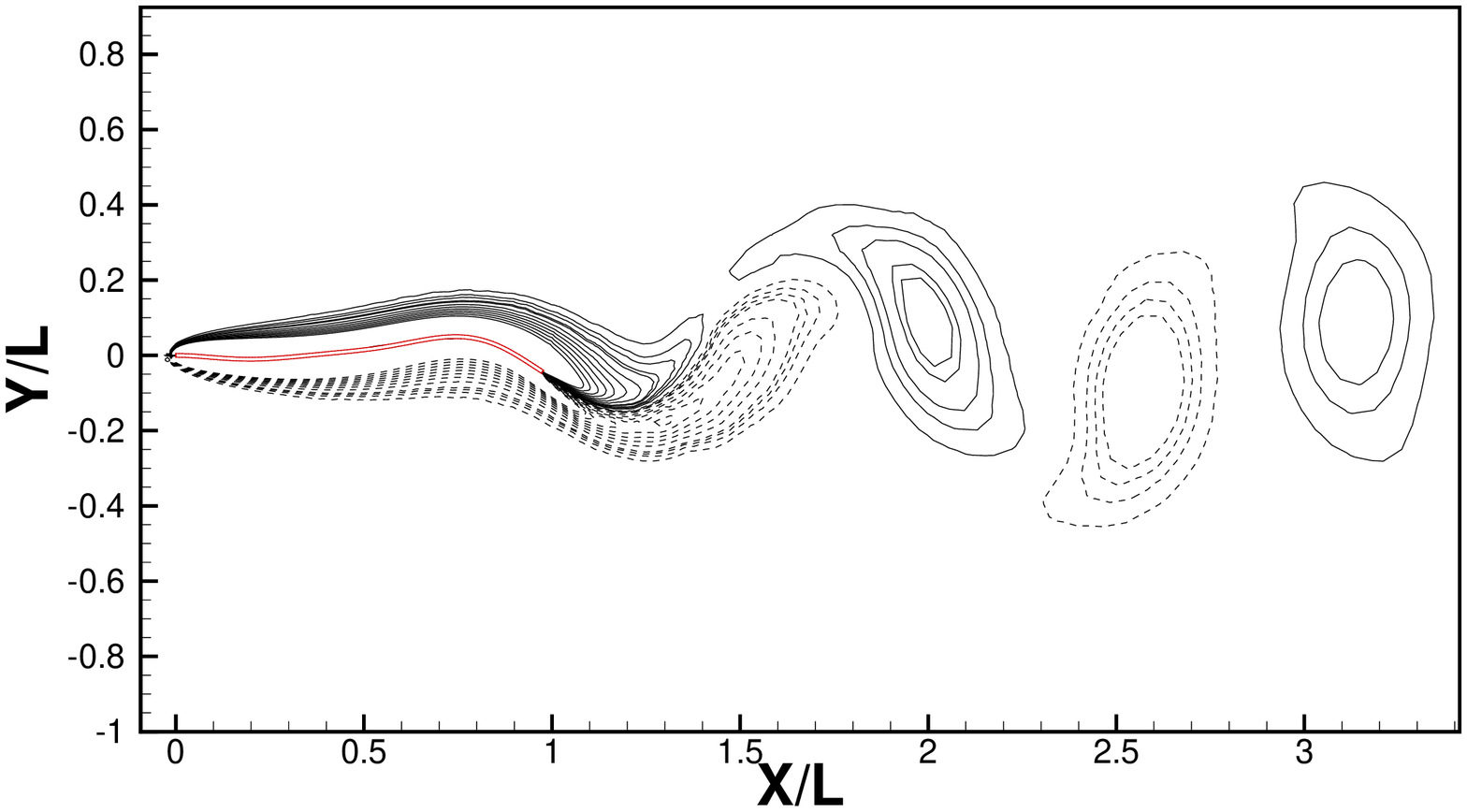}\\
			\includegraphics[width=0.49\columnwidth]{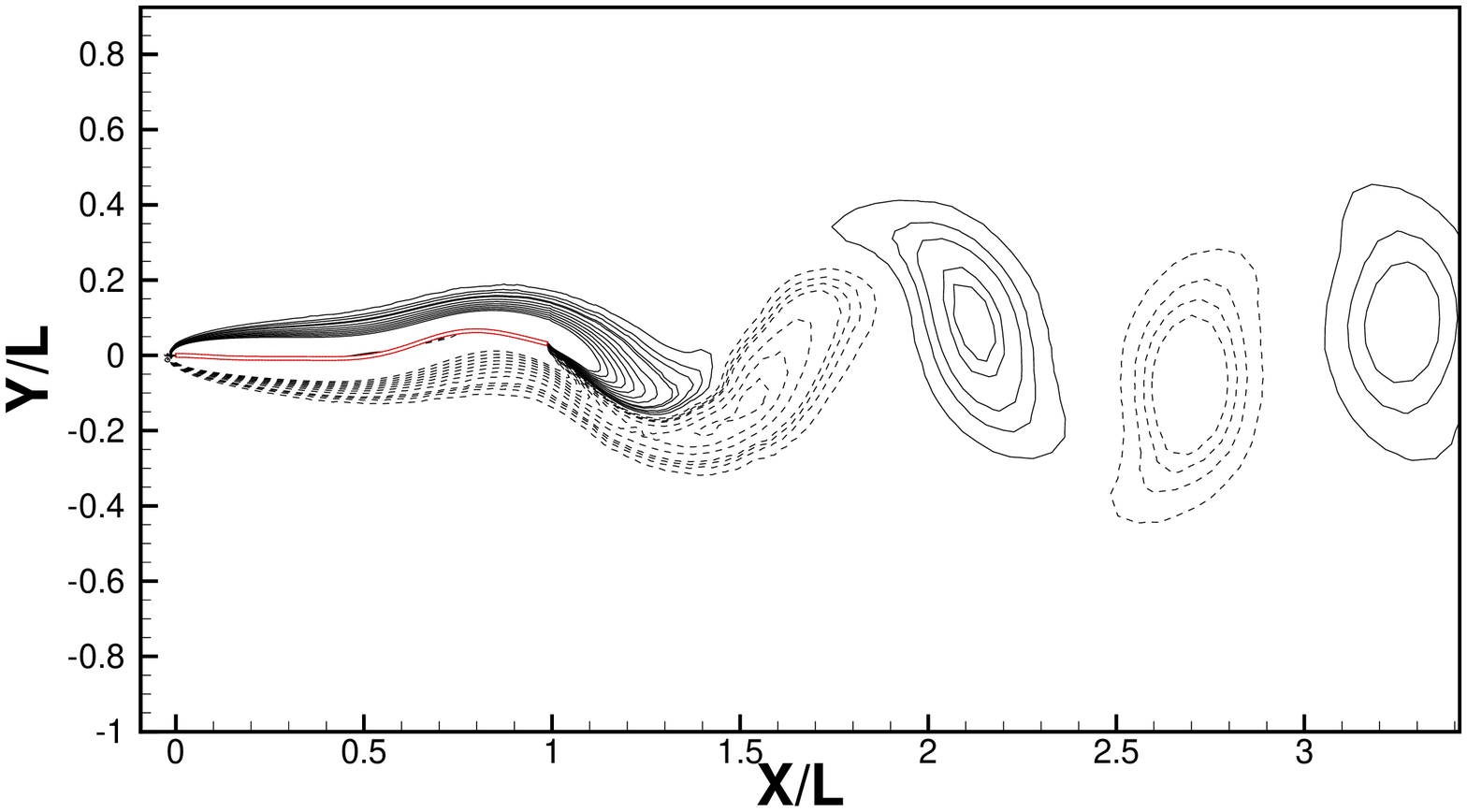}
			\includegraphics[width=0.49\columnwidth]{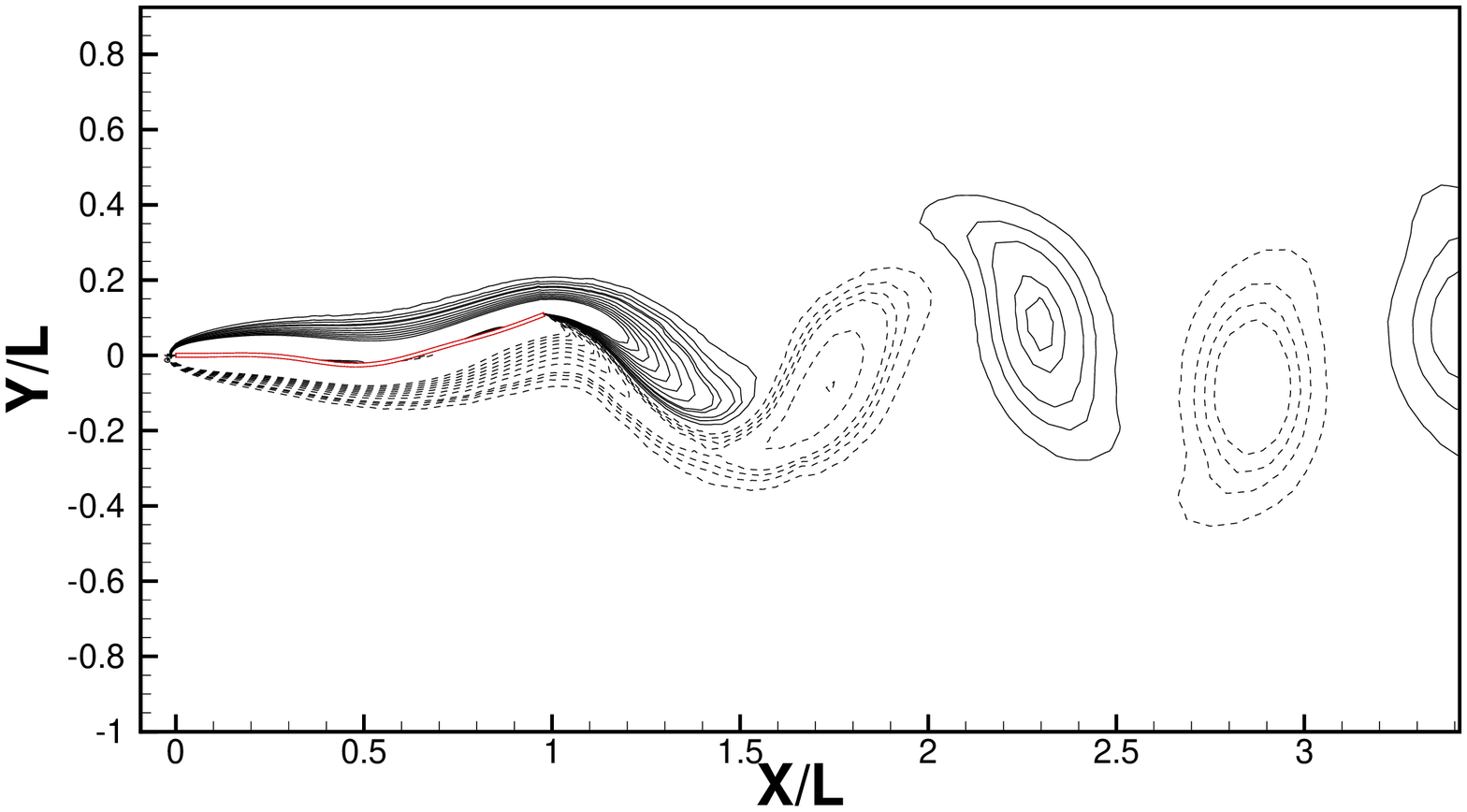}
			\caption{$ \Delta=7200 $}	
		\end{subfigure}
		\caption{Vorticity contours depicting vortex shedding patterns over an up-stroke motion of the plate for : (a) $\Delta=0$ and (b) $\Delta=7200$ at a constant $Re=1000$, $m^*=0.1$, $H_\mathrm{t}/H=H_\mathrm{b}/H=0.5$ and $\beta_{\mathrm{avg}}=6000 $ ($ K_B=0.0005 $). The solid and dashed lines represent positive and negative vorticity respectively}
		\label{vortex_6000}
	\end{figure}
	
	Finally, the vorticity contours are investigated to realize the influence of flapping motion on the vortex modes for the cases, $ \Delta=0 $ and 7200. Figure~\ref{vortex_6000} shows the snapshots of vortex shedding patterns over an up-stroke motion of the plate for $ \Delta=0 $ and 7200. In both cases, a strong clockwise rotating vortex can be seen spinning off from the trailing edge when it is at its extrema. As the trailing edge sweeps back to the mean position, the vortex is seen to be shed into the wake. Similarly, a vortex of the opposite sign is shed during the down-stroke. This cyclic process leads to a von Kármán vortex street with a continuous series of alternatively signed vortices. On careful observation of Fig.~\ref{vortex_6000}, it can be seen that the vortices are slightly more elongated along the cross-stream direction for the case with non-equal material properties as compared to the case with identical material properties. Since the Strouhal number was noted to increase with increasing $ \Delta $, this elongation in vortices is in accordance with \cite{connell2007}, where the vortices were seen to elongate in the cross-stream direction as the Strouhal number increased for the case of a traditional single-layered plate.
	
	\subsection{Effect of Elastic Modulus on the Onset of Flapping Instability}
	
	\begin{figure}
		\centering
		\begin{subfigure}{0.99\textwidth}
			\includegraphics[width=0.49\columnwidth]{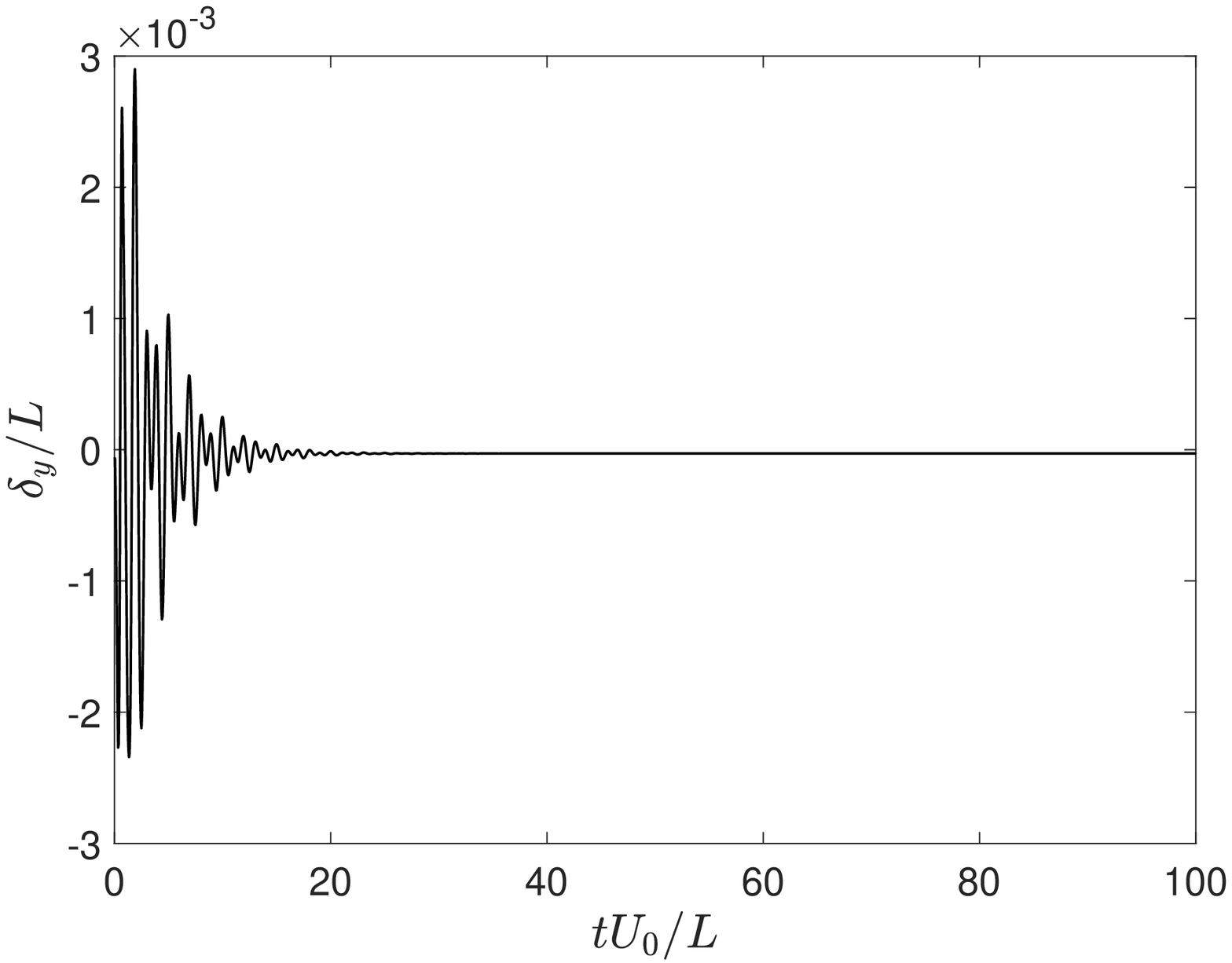}
			\includegraphics[width=0.49\columnwidth]{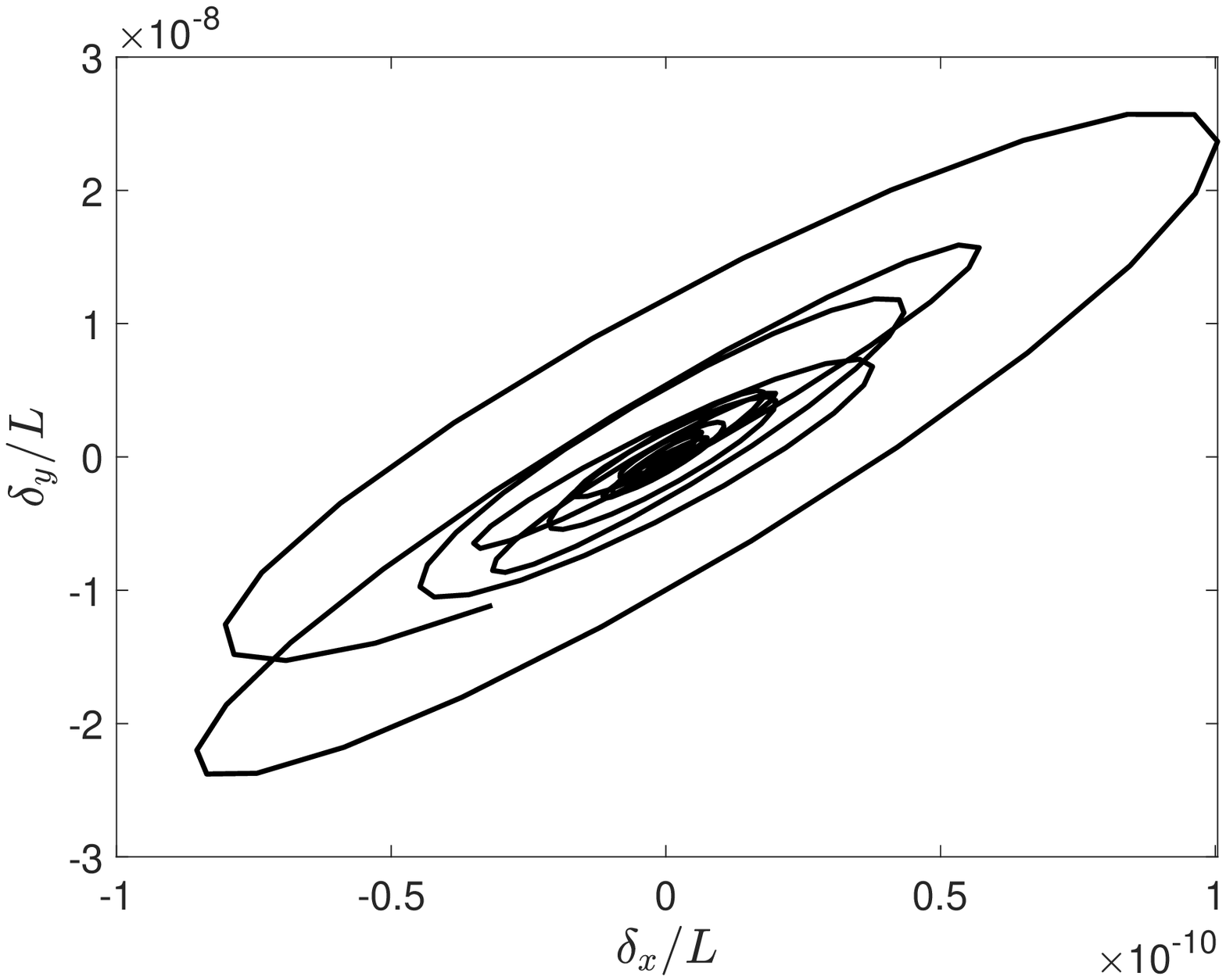} 
			\caption{$ \Delta=7200 $}
		\end{subfigure}\\    
		\begin{subfigure}{0.99\textwidth}
			\includegraphics[width=0.49\columnwidth]{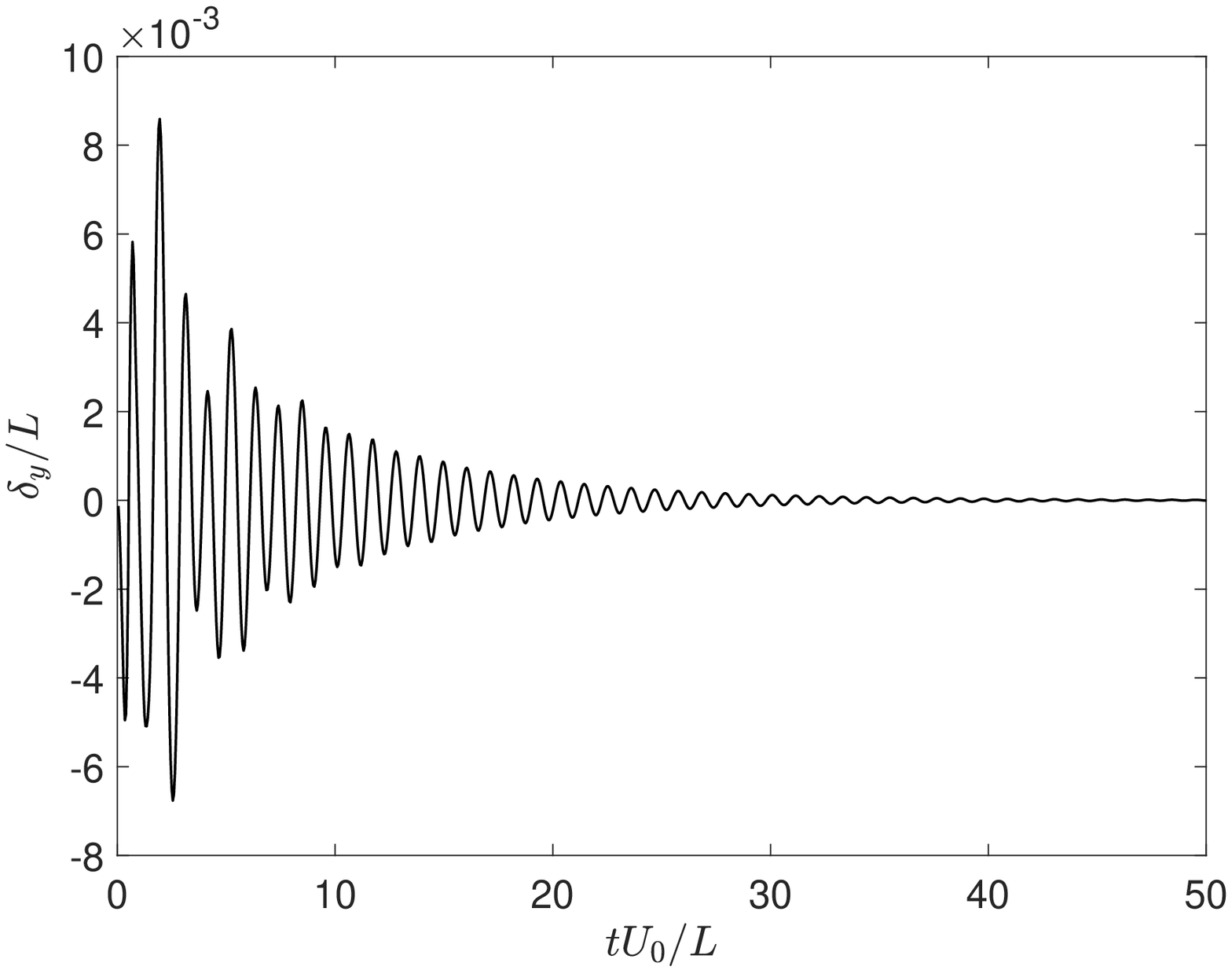}
			\includegraphics[width=0.49\columnwidth]{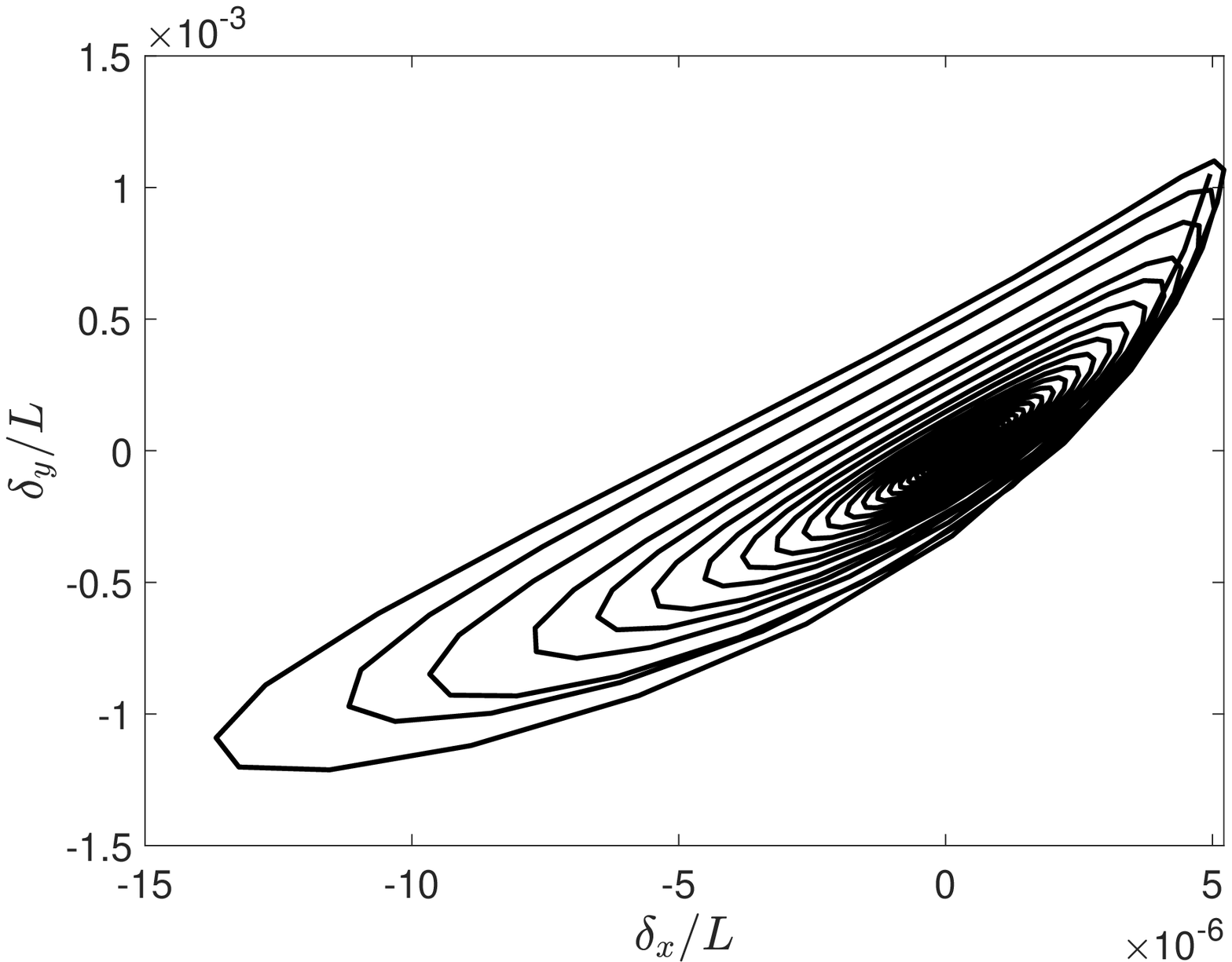} 
			\caption{$ \Delta=14400 $}
		\end{subfigure}\\
		\begin{subfigure}{0.99\textwidth}
			\includegraphics[width=0.49\columnwidth,trim=0mm 0mm 0mm 5mm,clip]{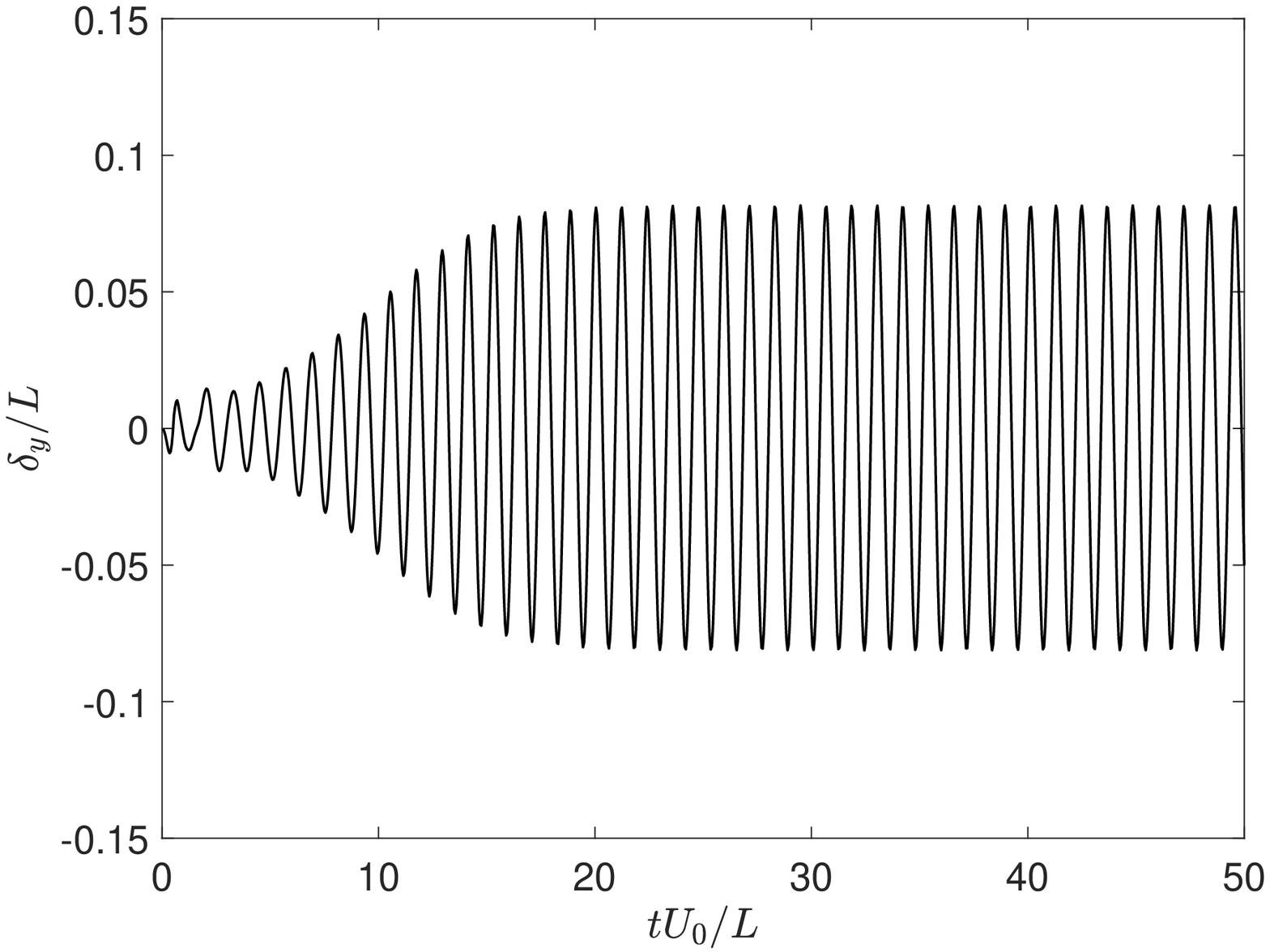}
			\includegraphics[width=0.49\columnwidth,trim=0mm 0mm 0mm 5mm,clip]{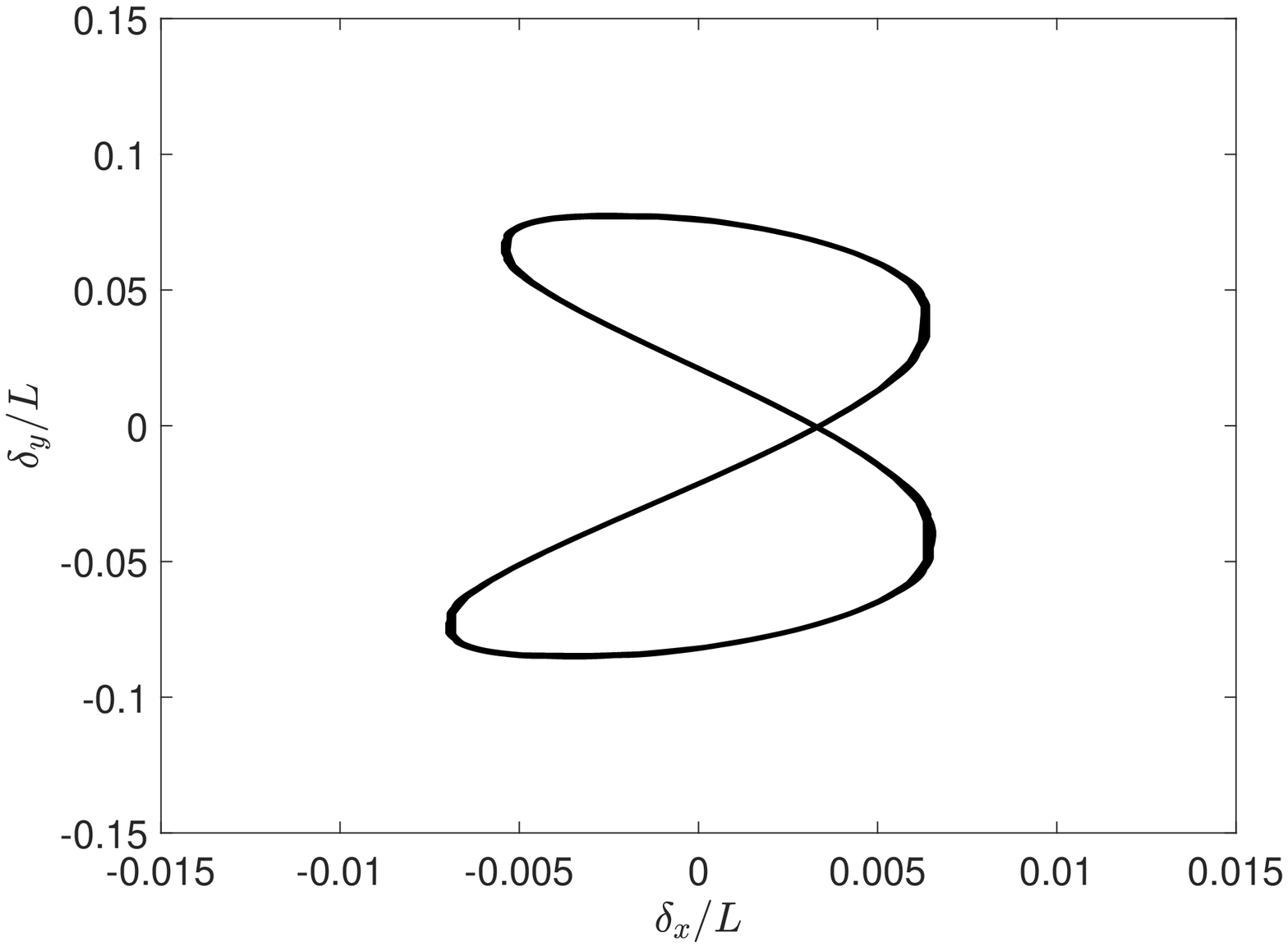} 
			\caption{$ \Delta=21600 $}
		\end{subfigure}\\
		\begin{subfigure}{0.99\textwidth}
			\includegraphics[width=0.49\columnwidth,trim=0mm 0mm 0mm 5mm,clip]{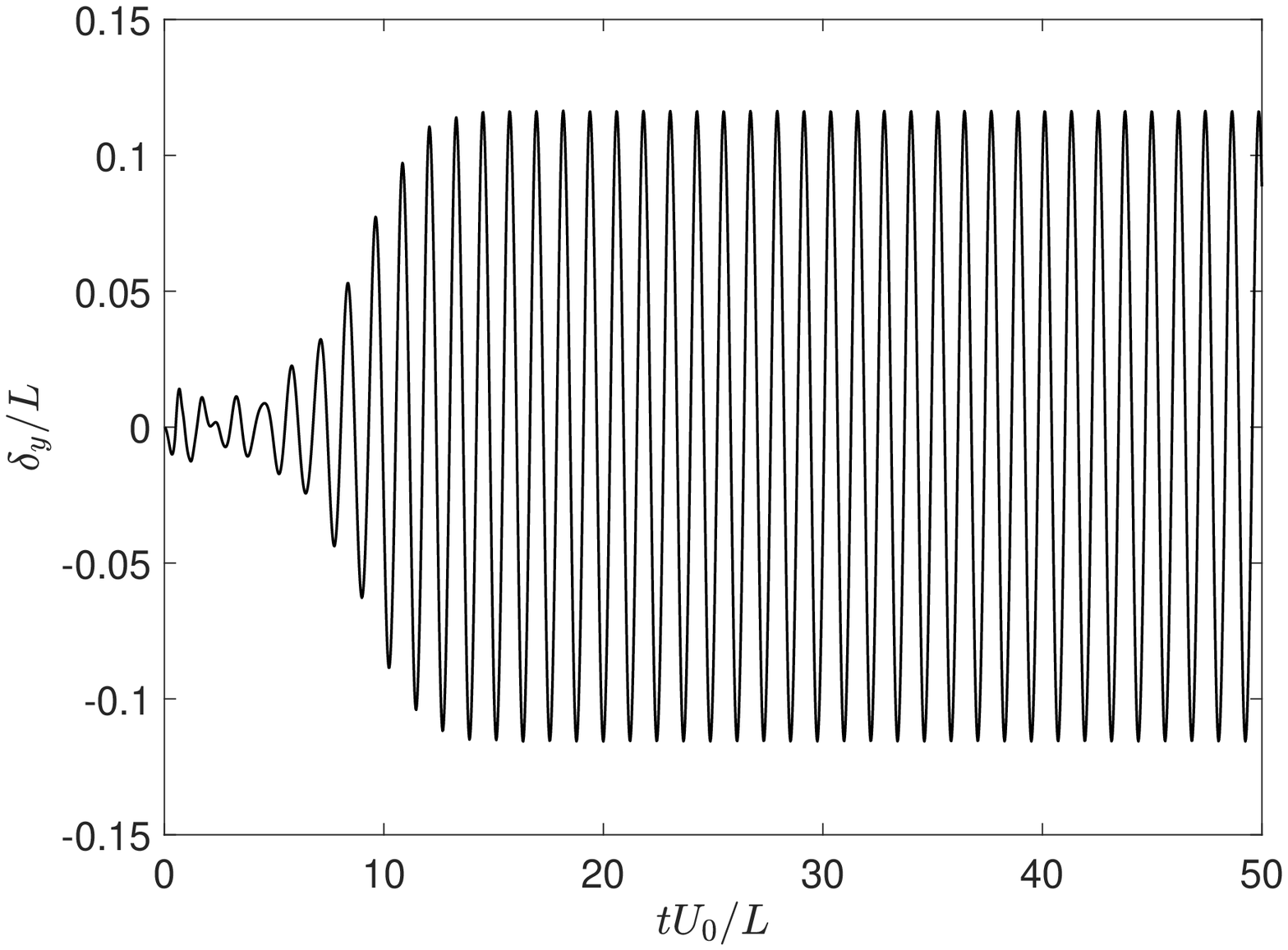}
			\includegraphics[width=0.49\columnwidth,trim=0mm 0mm 0mm 5mm,clip]{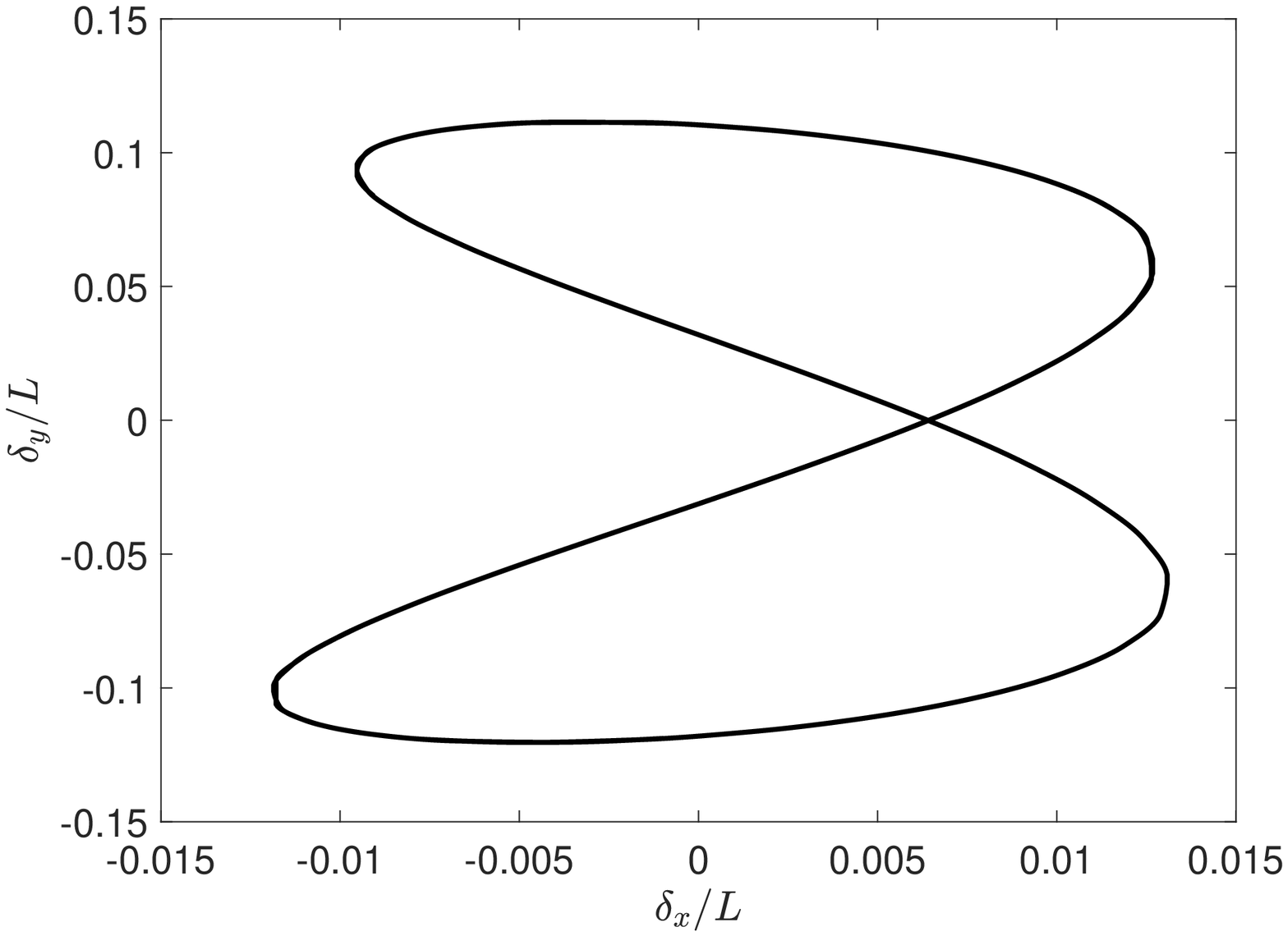} 
			\caption{$ \Delta=25200 $}
		\end{subfigure}    	
		\caption{Time history of trailing edge cross-stream displacement (left) and the corresponding Lissajous curves (right) for : (a) $\Delta=7200$, (b) $\Delta=14400$, (c) $\Delta=21600$ and (d) $\Delta=25200$ at a constant $Re=1000$, $m^*=0.1$, $H_\mathrm{t}/H=H_\mathrm{b}/H=0.5$ and $\beta_{\mathrm{avg}}=9600 $ ($ K_B=0.0008 $).}
		\label{onset_tip_disp}
	\end{figure}
	
	The onset of flapping instability has already been extensively investigated for the case of a single-layered plate. Here a similar investigation considering the two-layered plate is conducted in order to gain more insight about the effect of difference in elastic properties and see how it compares with the single-layered plate. A total of four cases $\Delta=\left\{7200, 14400, 21600, 25200\right\}$ with $\beta_{\mathrm{avg}} $ = 9600 ($ K_B=0.0008 $) for a constant Reynolds number of $ \mathrm{Re}=1000$, mass ratio $ \left(m^*\right)_{\mathrm{avg}}=0.1$ and $H_\mathrm{t}/H=H_\mathrm{b}/H=0.5$ are investigated.
	
	Although the average non-dimensional elastic modulus is constant across the five cases, a significant difference in the flapping response is observed. Figure~\ref{onset_tip_disp} summarizes the trailing edge displacement responses and the corresponding Lissajous curves for all four cases. From Fig.~\ref{onset_tip_disp}, the response dynamics can be characterized into two distinct response regimes as a function of $\Delta$: (I) fixed-point stable and (II) periodic LCO (Limit Cycle Oscillations). 
	The first two cases $ \left(\Delta=\left\{7200,14400\right\}\right) $ fall under regime (I), wherein, after the initial transients, the trailing edge of the plate settles to a steady straight configuration with no cross-stream displacement. On the other hand, the latter two cases $ \left(\Delta=\left\{21600,25200\right\}\right) $ fall under regime (II), where the response eventually settles into periodic LCO with a constant frequency and amplitude. This difference in response can also be distinctively seen from the Lissajous curves (Fig.~\ref{onset_tip_disp}, right). For regime (I), the phase plot spirals into a single point and contrastingly, for the regime (II), the phase plot exhibits a shape of figure eight, which indicates periodic LCO. Furthermore, the phase plots for the cases exhibiting LCO also reveal an asymmetry in the size of figure eight's upper and lower lobes, indicating that the trailing edge's stream-wise displacement is asymmetric during the upstroke and downstroke of the flapping motion. This is in accordance with the observations made in Section \ref{Delta_study}. 
	\subsection{Effect of Mass Ratio on Flapping Dynamics}
	
	\begin{figure}
		\centering
		\begin{subfigure}{0.99\textwidth}
			\includegraphics[width=0.49\columnwidth]{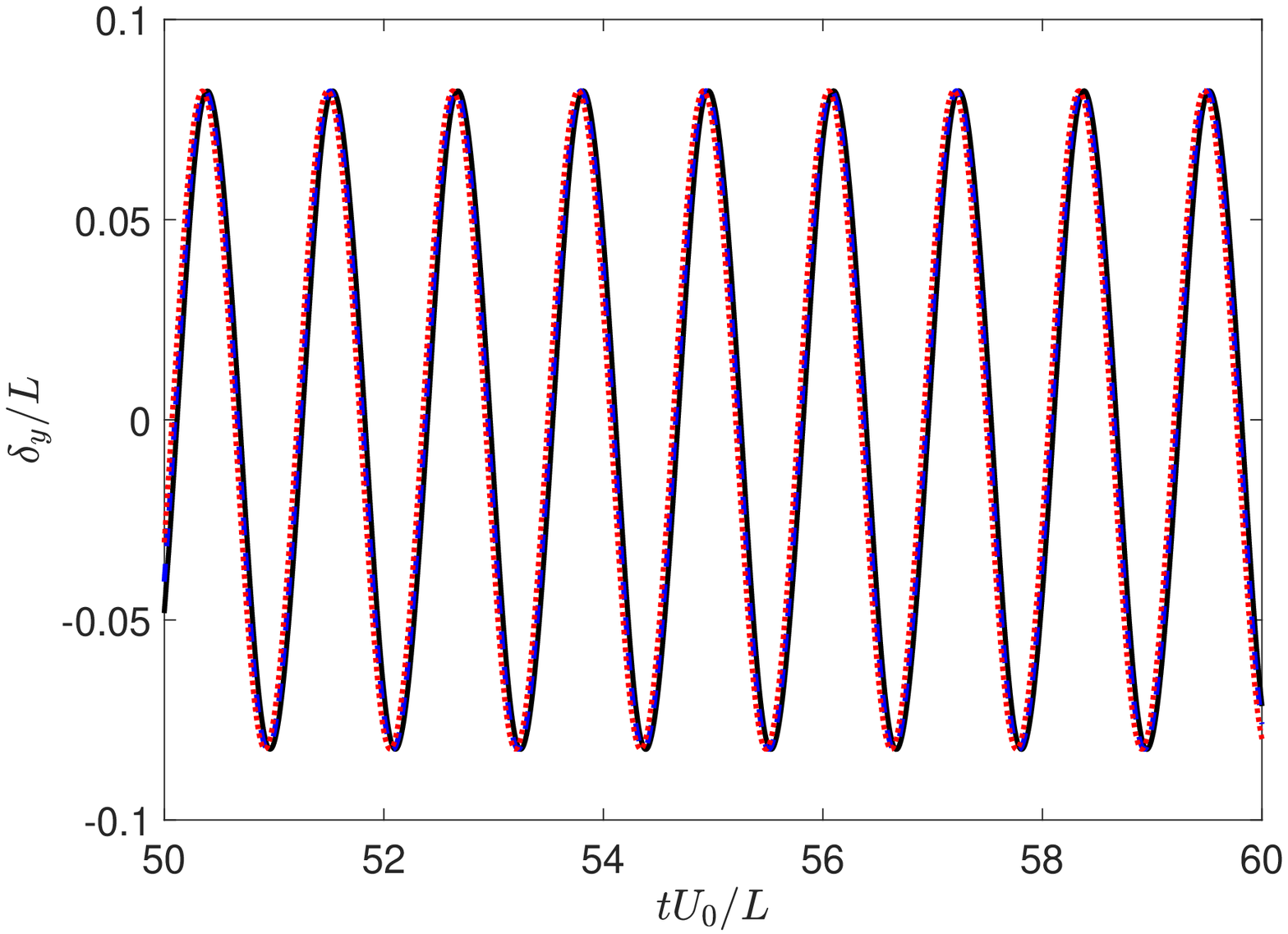}
			\includegraphics[width=0.49\columnwidth]{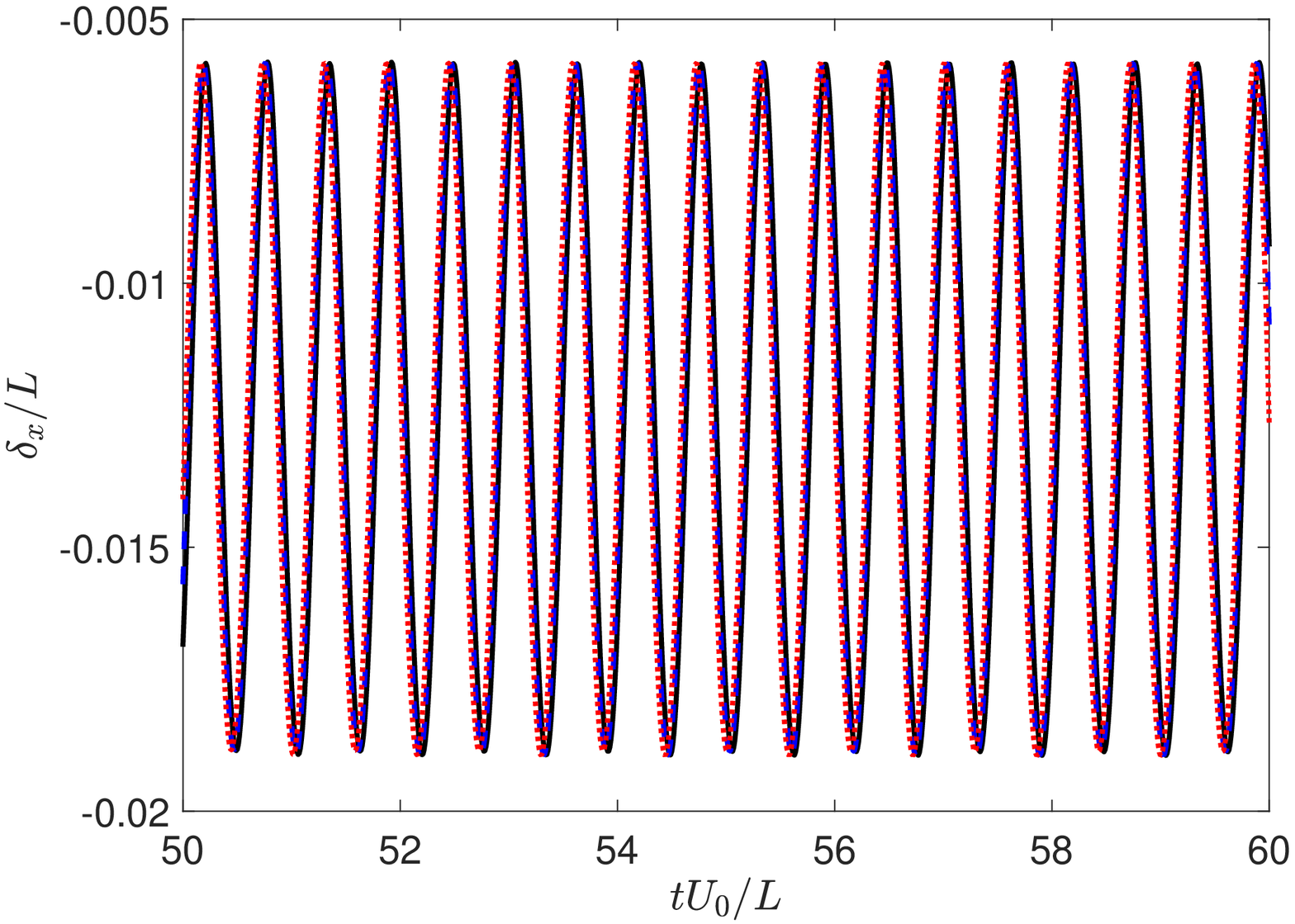} 
			\caption{Trailing edge displacement}
		\end{subfigure}\\    
		\begin{subfigure}{0.99\textwidth}
			\includegraphics[width=0.49\columnwidth]{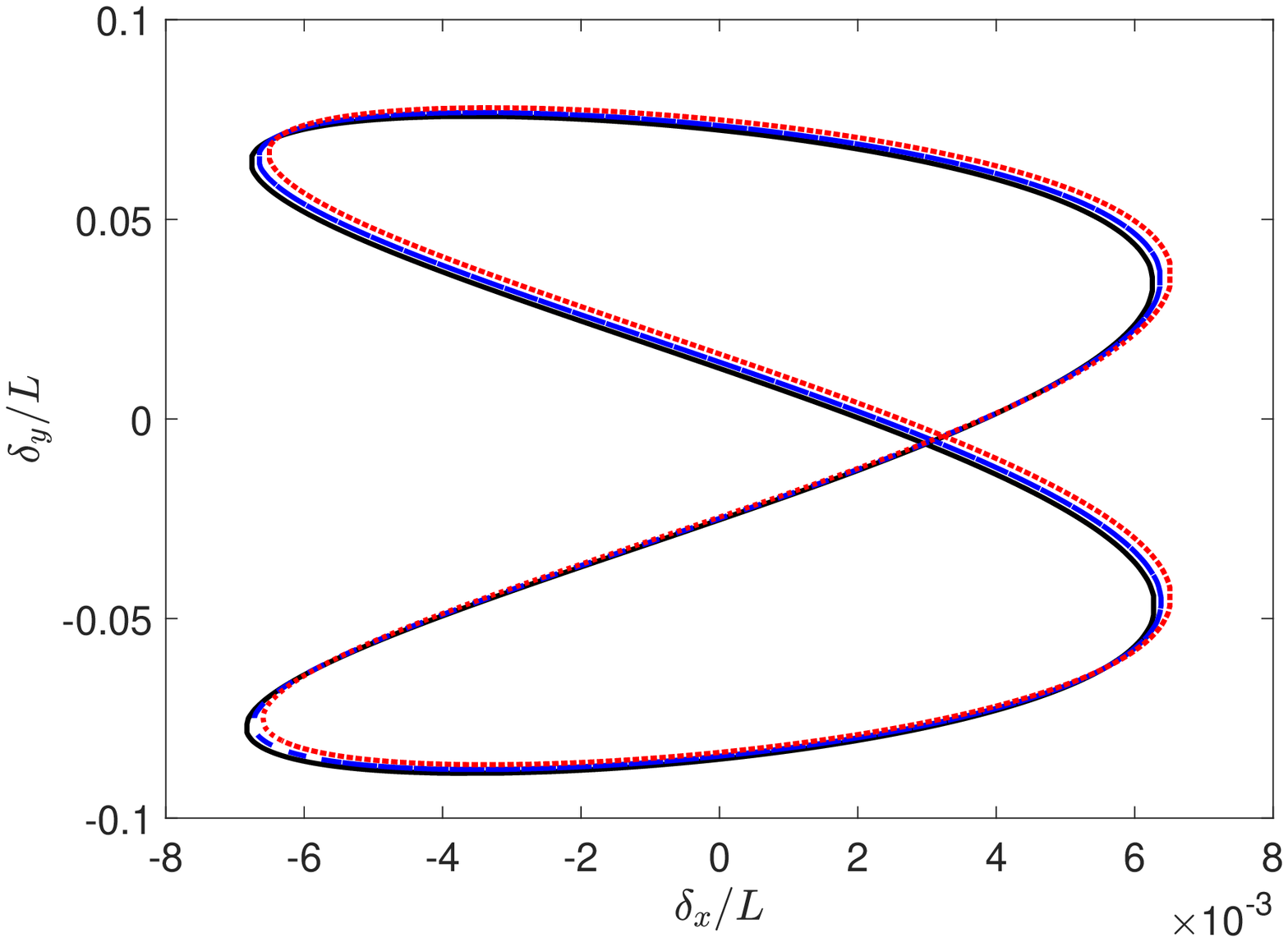}
			\includegraphics[width=0.49\columnwidth]{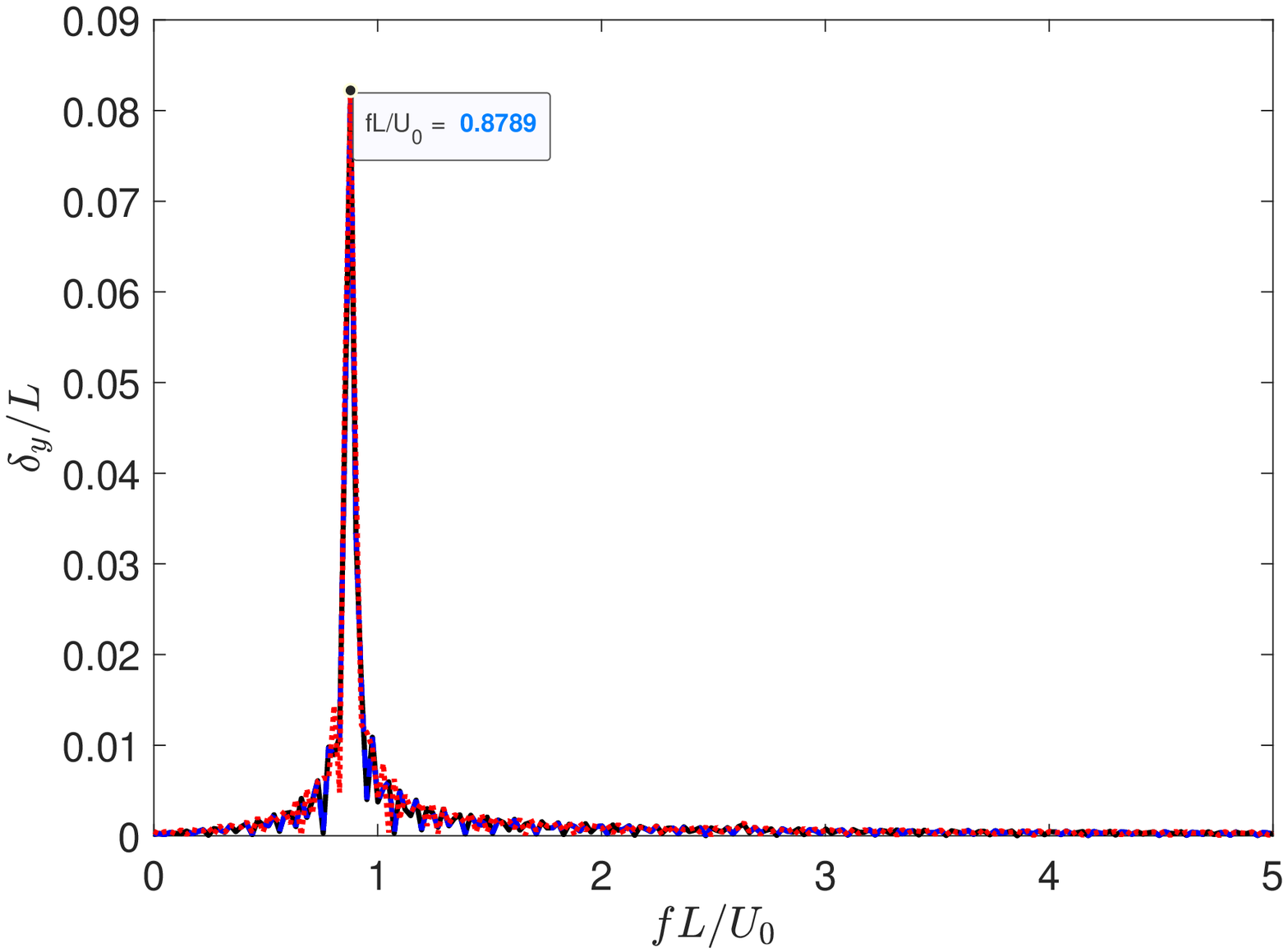} 
			\caption{Lissajous curves and amplitude-frequency spectrum}
		\end{subfigure}\\
		\begin{subfigure}{0.99\textwidth}
			\includegraphics[width=0.49\columnwidth]{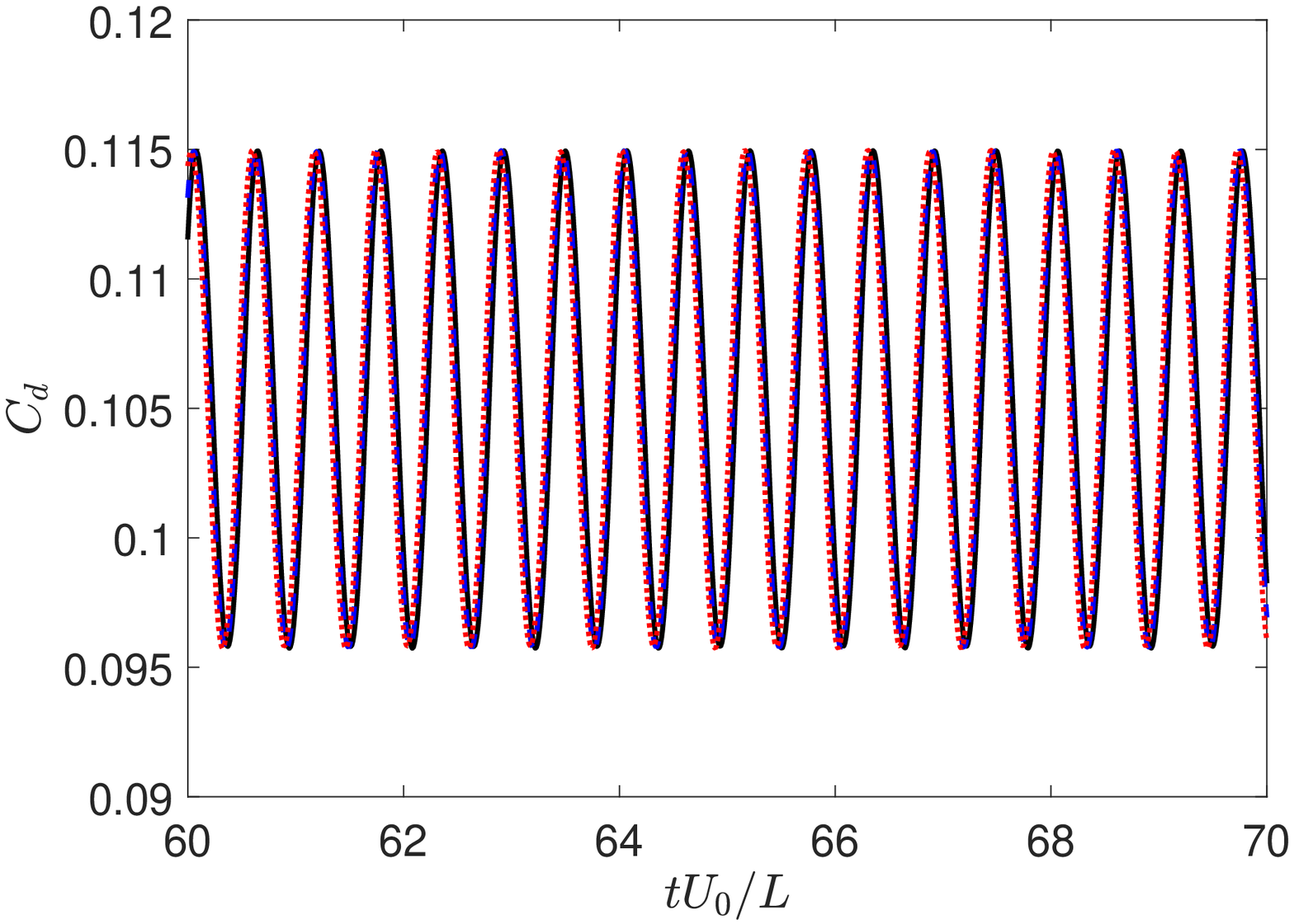}
			\includegraphics[width=0.49\columnwidth]{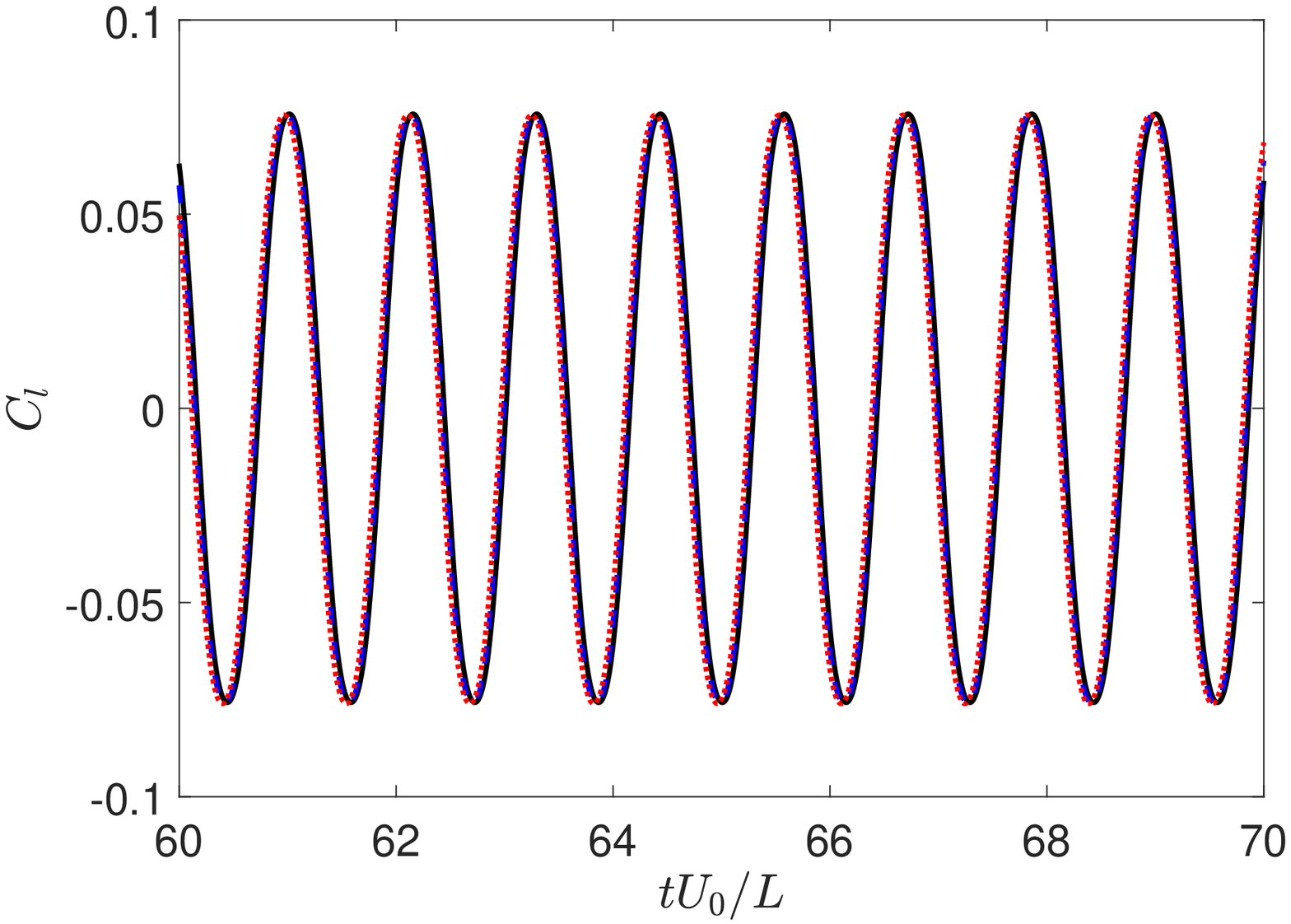} 
			\caption{$C_d$ and $C_l$ plots}
		\end{subfigure}	    	
		\caption{Plots depicting (a) the time history of trailing edge displacement, (b) Lissajous curves (left) and amplitude-frequency spectrum (right) and (c) the time history of drag and lift coefficient for the cases (i) $ \left(m^*_t,m^*_b\right) = \left(0.100,0.100\right)$ black solid lines (--------), (ii) $ \left(m^*_t,m^*_b\right) = \left(0.125,0.075\right)$ blue dashed lines $(---)$ and (iii) $ \left(m^*_t,m^*_b\right) = \left(0.150,0.050\right)$ red dotted lines $\left(\cdots\cdots\cdot\right)$ at a constant $Re=1000$ $H_\mathrm{t}/H=H_\mathrm{b}/H=0.5$ and $\beta_{\mathrm{avg}}=6000 $ ($ K_B=0.0005 $)}
		\label{mass_ratio}
	\end{figure}
	
	Investigations into the flapping dynamics of a single-layered plate have made it clear that the mass ratio is an important parameter in predicting the flapping dynamics of a plate. Here the influence of difference in the structural density between the two layers on the flapping dynamics of a two-layered plate is studied. Therefore, four cases with $ \left(m^*_{t},m^*_{b}\right) = \left\{\left(0.100,0.100\right),\left(0.125,0.075\right),\left(0.150,0.050\right)\right\} $ for a constant Reynolds number of $ \mathrm{Re}=1000$, non-dimensional elastic modulus  $\beta_{\mathrm{avg}}=6000$ ($K_B = 0.0005$) and $H_\mathrm{t}/H=H_\mathrm{b}/H=0.5$ are investigated.
	
	Interestingly, no significant differences are noticed across all three cases. Figure~\ref{mass_ratio}~(a) shows the time history of the cross-stream and stream-wise displacement of the trailing edge. It can be seen that all the three curves overlap one another indicating a similar flapping amplitude across the cases. The same can be inferred from the phase plots and FFT plots depicted in Fig.~\ref{mass_ratio}~(b). Finally, a comparison of the lift and drag coefficients (Fig.~\ref{mass_ratio}~(c)) reaffirms the fact that  the difference in density between the layers has next to no effect on the flapping dynamics of the plate. This observation is in stark contrast to the effects of elastic properties presented in Section \ref{Delta_study}.

	\section{Conclusion}\label{conc}
	
	We have numerically investigated the flapping dynamics of a two-layered flexible plate placed in a uniform flow. The two-layered flexible plate serves as a simplified model of two or more flexible plate-like structures mounted on top of each other as in the case of piezoelectric patches. 
	A robust, numerically stable, quasi-monolithic formulation based on finite element method is proposed to solve the non-linear fluid-flexible structure interaction of a multilayered plate. An in-house solver based on this formulation is developed to solve the problem of a two-layered flexible plate placed in a uniform flow with the leading edge fixed. This solver is then validated against the results for a single-layered plate by considering identical properties for the two layers. 
	
	The in-house solver has been used to conduct parametric studies to understand the effects of difference in the material properties such as Young's modulus and density between the top and bottom layers on the flapping dynamics of the plate. The material properties for the top and the bottom layer are selected such that the average material property of the combined two-layered plate is kept constant. 
	Initially, parametric studies are conducted for the case of $ Re=1000 $ and $ \left(m^*\right)_\mathrm{avg}=0.1 $, by choosing different values of Young's modulus for the top and bottom layers such that $ \left(K_B\right)_\mathrm{avg}=0.0005$. Interestingly, it is observed that as the difference between the elastic modulus of the layers is increased the flapping amplitude and forces on the plate also steadily increased. Simultaneously, a decrease in the dominant frequency of flapping is observed without any change in the mode of oscillation. It was also noticed that as the difference in elastic properties increased the two-layered plate exhibited an asymmetric stream-wise displacements of the trailing edge between the up-stroke and the down-stroke. This observation can be attributed to different degrees of bending in the up-stroke and down-stroke and is in stark contrast to the case of a single-layered plate, where the displacements are always symmetric.
	
	Following this, the case of $ Re=1000 $, $ \left(m^*\right)_\mathrm{avg}=0.1 $ and $ \left(K_B\right)_\mathrm{avg}=0.0008 $, for which, a single-layered plate does not undergo self-sustained flapping is considered, and the Young's modulus of the two layers are systematically varied. The results show that the introduction of such variation between the layers eventually leads to the onset of self-sustained flapping. Two distinct response regimes are observed, and they are described as the (I) fixed point stable regime and the (II) limit cycle oscillations (LCO) regime. In regime (I) the plate settles into a steady straight configuration after the initial transients whereas in regime (II) the plate exhibits periodic oscillations with a constant amplitude and frequency. Once again, asymmetry in the stream-wise displacement of the trailing edge between the up-stoke and down-stroke is observed as the difference in elastic modulus between the layers is increased. Finally, cases with unequal structural densities for the two layers are considered while keeping the average fluid mass ratio of the plate constant at $ \left(m^*\right)_\mathrm{avg}=0.1 $. It turns out that unlike the case with different elastic properties, the variation of structural densities have a negligible impact on the flapping dynamics of the two-layered plate. 
	
	In conclusion, the effects presented in this work clearly illustrate that the consideration of a multilayered model might be sometimes necessary to predict the flapping dynamics of plates accurately, especially in the cases when the layers of the plate possess different material properties. Therefore, more research considering such multilayered plates could be valuable in predicting the flapping dynamics of such multilayered flexible plates.
	
	\section*{Acknowledgment}
	The corresponding author would like to acknowledge the financial support from the OPERA award and Research Initiation Grant from Birla Institute of Technology and Science–Pilani.
	
	\bibliography{refs}
\end{document}